\documentclass[%
 reprint,
nofootinbib,
 amsmath,amssymb,
 aps,
]{revtex4-2}

\usepackage{mathtools,amssymb,lipsum}
\usepackage{graphicx}% Include figure files
\usepackage{dcolumn}% Align table columns on decimal point
\usepackage{bm}% bold math
\usepackage{makecell}
\usepackage[most]{tcolorbox}

\begin{document}

\preprint{APS/123-QED}

\title{Generalized Unruh Effect: A Potential Resolution to the Black Hole Information Paradox}% Force line breaks with \\

\author{Angela Chen}
 \altaffiliation[Email: ]{anqi.chen@ipmu.jp}%Lines break automatically or can be forced with \\
 \affiliation{Kavli Institute for the Physics and Mathematics of the Universe
(WPI), The University of Tokyo Institutes for Advanced Study,
The University of Tokyo, Kashiwa, Chiba 277-8583, Japan}

\date{\today}

\begin{abstract}
%The Unruh effect, first described in `Notes on black-hole evaporation' (W. Unruh 1976) gives an alternative interpretation for the Hawking radiation of an evaporating black hole. In the past several decades, it has been mainly investigated as an effect for the vacuum state expressed in a Bogoliubov transformed basis. 
We generalize the vacuum-Unruh effect to arbitrary excited states in the Fock space and find that the Unruh mode at the horizon induces coherent excitation on the canonical background ensemble measured by an accelerated observer. When there is only one type of Unruh mode in the system, for example, the ones outgoing from a black hole horizon, the mapping from an arbitrary density matrix on the maximal foliation to a vector space spanned by the pseudo-thermal density matrix on the partitioned spacetime wedge is one-to-one. Hence we propose that the information of the particles that is inside a collapsing shell, thus inside the asymptotic black hole horizon is at least partially retrievable by measuring the deviation of the Hawking radiation from the black body radiation spectrum. This work shows that the long-standing black hole information confusion might come from overlooking the possibility that the information could be preserved much better than we have expected in a seemingly non-unitary process when the partitions of the system are strongly entangled.
\end{abstract}

\maketitle	

\tableofcontents
\newpage

\section{Introduction}
\label{sec:introduction}

In ref \cite{Unruh:1976db}, W.G. Unruh first demonstrated a mechanism now called the Unruh effect, which states that a uniformly accelerating observer would detect a thermal bath from expressing the stationary vacuum state in terms of a different set of generator/annihilation basis defined along the time-like killing vector in their relatively accelerating coordinate system. The Unruh effect is a direct result of the non-unique canonical quantization of a field living in a Riemannian spacetime \cite{Fulling:1972md}. Following the original derivation on a vacuum state, the Unruh effect has been mainly quoted and discussed in the originally proposed scenario of a vacuum state, in the past decades \cite{Crispino:2007eb,krishnan2010quantum,Brenna:2013txa}. Most famously, the Unruh effect has helped understanding the Hawking radiation \cite{Hawking:1974rv, Gibbons:1977mu}, and is sometimes quoted as an alternative way for interpreting the black hole radiation/evaporation phenomenon. In recent years, the field has seen a growing number of works utilizing the Unruh effect to understand the entanglement generation and degradation in curved spacetime \cite{Fuentes-Schuller:2004iaz,Adesso:2007gm,martin2009fermionic,Landulfo:2009wg,Montero:2010jj,Martin-Martinez:2010yva,ramzan2012decoherence}. 

Schematically, the Unruh effect prescribes that under a Bougoliubov basis transformation, the Minkowski vacuum is transferred into a thermal ensemble on left for right Rindler wedge, likewise, the Kruskal vacuum is transferred into a thermal ensemble living outside or inside the horizon of a Schwartzschild black hole metric. However, when we scrutinize what lies in the core of the derivation of the Unruh effect, we can easily see how it can be applied to much more general scenarios. First of all, the mixture of positive and negative frequency modes is not only happening in Minkowski-Rindler observer pairs and Kruskal-Schwarzschild observer pairs. Although a unified theory has not been developed, in recent decades we have seen many examples of positive and negative frequency modes mixing in non-inertial frames, usually investigated by a tool called Bogoliubov transformation \cite{Bogolyubov:1958km}. A common feature those scenarios share is the local partition of the spacetime manifold by a particle horizon, which could be induced by acceleration, gravitation, or inflation \cite{Pannia:2021lso,Hemming:2000as,Chung:1998bt}. %An even more interesting observation is that one of the common features those physical process has is that they each involve the non-constant inner-product of a pair of time-like Killing vector fields. %(Inertial-Accelerating, Freefalling-Asymptotic, $t$-$t+\Delta t$ foliation under inflation) \footnote{Unlike in Lorentz transformation, the inner product between $\frac{\partial}{\partial t}$ and $\frac{\partial}{\partial t'}$ is a constant 1 or $\gamma$.}. 
The second direction of generalizing the Unruh effect is that Bogoliubov basis transformation is essentially a basis transformation that can be applied to arbitrary states in the Hilbert space, not just the vacuum state. In fact, given that in quantum field theory in curved spacetime framework the vacuum is defined based on the local time-like Killing vector, it is not that special from the point of view of quantum field theory in the curved spacetime. Vacuum is just one among infinitely many other states in the Fock states. The calculation in this paper is mainly dedicated to the generalization of the Unruh effect onto non-vacuum states. We start from Minkowski metric, then further into asymptotic Schwarzschild black hole scenarios depicted by collapsing shell metric. By the end of this paper, as we go into a discussion of the wider application of the generalized Unruh effect from the perspective of view of the first point made in this paragraph, we will see that a new definition of vacuum modulo any canonical ensemble might be in call for further development of quantum field theory in curved spacetime, and possibly for an inclusive description of quantum gravity.

%The following calculation that has been carried out on a vacuum state, by definition could also be carried out on an arbitrary state in the Fock space of the Unruh modes eigenstates. And because the Unruh modes connect to the conventional plane wave modes through a positive-to-positive frequency transformation, the results we obtained on the Fock space of the Unruh modes can also be applied to Fock space of the normal plane wave modes.

Unruh effect calculations are usually done in these three steps:

\begin{itemize}
\item Firstly, we secure two complete sets of orthonormal modes for the solution of classical field theory on a four-dimensional Lorentzian manifold. They each correspond to a maximal Cauchy surface foliation of the complete Lorentzian manifold. Each one of the orthonormal basis modes sets gives a complete expression of the solutions to the classical field equation of motion. Their positive and negative frequency modes would be mixed, in the mutual transformation due to different time foliations that only coincide on one slice. %As shown by \cite{Unruh:1976db}, it could happen when the time-like killing vector between the two reference frames has non-constant inner product (one is accelerating with respect to another).
\item Secondly, we canonically quantize the field and obtain the creation and annihilation operators corresponding to the two sets of modes which are the equation of motion solutions secured in the previous step. For a given state in the Fock space of one type of the quantized EoM eigenmodes, we carry out Bougoliubov transformation to map the state onto another set of the basis of the quantum field. 
\item Lastly, due to the existence of a horizon, for an observer living on one of the partial foliations of the spacetime manifold we contract out the states living on the other partition of the spacetime manifold.

\end{itemize}

Notice that the last step by its nature breaks the unitarity: we partition an entangled quantum system into two parts separated by a horizon, then trace out one of them. However, information is not necessarily lost, at least might not be at an as severe extent as thought before, in this non-unitary process, thanks to the strong entanglement across the horizon. This point has been made previously in the series of works by K. Lochan and T. Padmanabhan \cite{Lochan:2014xja,Lochan:2015oba}. We formulate this idea in a more explicitly quantum mechanical way by adopting the density matrix representation of countably infinite dimensional Hilbert space of the scalar field eigenmodes. We also would like to point out a more radical implication of this phenomenon on the approach to quantum gravity/unification of general relativity and quantum field theory.

The above summary might seem too abstract as a starting point. In the main body of this paper, we actually begin with this rather concrete question: given a pure state in the Fock space of Minkowski metric, what an observer accelerating along a parabolic worldline on a Rindler wedge would see. The answer is seemingly simple and has been treated straightforwardly so far, that they will see particles accelerating in the opposite direction, sitting in a thermal bath. Such approximation is adopted in, for example, \cite{Vanzella:2001ec} in relative scenarios. This straightforward picture works well in most cases, but this paper may reveal more of the story through a deeper contemplation and more detailed calculation. 

One of the highly intriguing interpretations of the generalized Unruh effect is its application to the `Black Hole Information Paradox' \cite{hawking2005information,giddings2012black,Calmet:2022swf}. By scrutinizing the evolution of plane wave modes inside the shell of a collapsing shell metric, we found a non-trivial mapping between arbitrary density matrix generated by the in-shell plane wave Fock states and the pseudo-thermal density matrix generated by the positive frequency modes living at the asymptotically flat region outside the Schwarzschild black hole horizon. We thus draw the following implication from our calculations: we may restore the information of the amplitudes and frequencies of the infalling particles collapsed into a black hole event horizon, that exists only for a finite time due to Hawking radiation, by measuring the deviation of the Hawking radiation spectrum from the perfect black body radiation spectrum. Here, we continue to refer to the generally featured black hole emission from the horizon as Hawking radiation. The concept of `stimulated Hawking radiation' has been studied for a long time \cite{Wald:1976ka,Bekenstein:1977mv,audretsch1992amplification,Weinfurtner:2010nu,Agullo:2021vwj}. Just like the original Unruh effect can be regarded as an alternative interpretation of the Hawking radiation without localized wavepacket assumption, the stimulated Hawking radiation could be seen as a similar precursor of the generalized Unruh effect in this work.

%For now, all of the calculation in this note happens in Minkowski space/Rindler space, following the convention of \cite{Unruh:1976db}. The calculation does not use the specific form of the field EOM solutions, so in principle if the three elements of the Unruh effect listed above can be satisfied, the conclusion can be generalized to other relatively non-inertial coordinate systems and field theories. But they do need further validation in the future.

This paper is organized in the following way. In section \ref{sec:definitions}, we go through the well-established prerequisite for this work, formatting the definitions and conventions. They include the specification of spacetime metric, equation of motion of the field, and canonical quantizations. In section \ref{sec:singlemode}, the calculation of a single Unruh particle state contracting R-(R+) states out to an R+(R-) density matrix is presented. In section \ref{sec:beyondsingle}, we generalize the results from the previous section to an arbitrary eigenstate with two types of Unruh modes. %A non-trivial mapping is found between the Minkowski Fock space and the R+ or R- pseudo-thermal density matrices, but whether it is an isomorphism remains as an open question, due to the complexity in a system residing two types of Unruh modes simultaneously. 
Up to this point, we work in Minkowski $\leftrightarrow$ Rindler scenario. Starting from section \ref{sec:blackhole}, we explore the Kruskal $\leftrightarrow$ Schwarzschild scenario. We investigate the evolution of an in-going mode living on a collapsing shell metric, its behavior at the horizon, and its density matrix measured by a future observer outside. %, and find that we only need one type of Unruh modes to represent the infalling plane wave modes inside the collapsing shell. %In this case, an unambiguous isomorphism can be found between the most general density matrix of a Kruskal-Unruh modes and the vector space spanned by the pseudo-thermal density matrices living outside the Schwarzschild black hole horizon. 
In section \ref{sec:spectrum}, we calculate the energy spectrum corresponding to the pseudo-thermal density matrices we obtained from sections \ref{sec:singlemode} and \ref{sec:beyondsingle}. In section \ref{sec:discussions}, we discuss the assumptions, caveats, and further implications of this work. The conclusion is in section \ref{sec:conclusions}. 

%We believe that our work could open a gate with fresh perspective into many long-existing questions on the interface between quantum mechanics, general relativity and thermodynamics, while using only fairly well-established, less-exotic presumptions in the traditional QFT in curved spacetime framework. The most aggressive assumption in our work is that we apply some modern quantum computation and communication concepts, for example density matrices, Von-Neumann entropy and positive operator valued measure (POVM) developed for finite Hilbert space to infinite dimensional Hilbert space in quantum field theory, without formally proving their validity in this paper.  %we outlook the countless future works that can be done along the direction this paper. For example, various information metrics can be calculated and compared; agreement of the radiation to the Page curve can be tested; spectrum features can be calculated for the black holes with different masses and at different stages of evaporation, etc. Lastly and most importantly, although expected to be very difficult, in principle this paper gives experimentally testable results in terms of the black hole evaporation process, or for less extreme curved spacetime generalized Unruh effects.

\section{Specification of the System}
\label{sec:definitions}

We start from Minkowski spacetime with $(+---)$ signature:
\begin{equation}
ds^2 = dt^2-dx^2-dy^2-dz^2
\end{equation}
After coordinate transformation:
\begin{align}
t = 2\sqrt{\rho}\sinh{\frac{1}{2}\tau}, \ \ \ \ z = 2\sqrt{\rho}\cosh{\frac{1}{2}\tau} \ \ \ \  (R+)\\
t = -2\sqrt{\rho}\sinh{\frac{1}{2}\tau}, \ \ \ \ z = -2\sqrt{\rho}\cosh{\frac{1}{2}\tau} \ \ \ \ (R-) \\
t = 2\sqrt{\rho}\cosh{\frac{1}{2}\tau}, \ \ \ \ z = 2\sqrt{\rho}\sinh{\frac{1}{2}\tau} \ \ \ \  (F)\\
t = -2\sqrt{\rho}\cosh{\frac{1}{2}\tau}, \ \ \ \ z = -2\sqrt{\rho}\sinh{\frac{1}{2}\tau} \ \ \ \  (P)
\end{align}
We get the Rindler metric:
\begin{align}
ds^2 = \rho d\tau^2-\frac{d\rho^2}{\rho}-dx^2-dy^2 \ \ \ \ (R\pm)\\
ds^2 = -\rho d\tau^2+\frac{d\rho^2}{\rho}-dx^2-dy^2 \ \ \ \ (F, P)
\label{rindlermetric}
\end{align}

The four wedges above cover the full Minkowski spacetime.

The constant proper acceleration worldlines in Minkowski spacetime are expressed by the hyperbola:
\begin{equation}
z^2-t^2 = 4\rho = \frac{1}{\alpha^2}
\end{equation}
where $\alpha$ is the constant proper acceleration. Hence for any modes expressed in terms of $\tau$, we can get the corresponding mode for $d\tau' = \sqrt{\rho}d\tau = \frac{1}{2\alpha} d\tau$ by multiplying the frequency with $2\alpha$ and deform the $\rho$ dependent part of the solution accordingly. %\footnote{We only point out the existence of $g(\rho)$ deformation without explicitly calculating it, as the space-like coordinate wave-forms are far less important than the time-like coordinate frequency of the modes. }. Bearing this in mind, we will continue the calculation with the metric \ref{rindlermetric}, so that we can directly quote the results from \cite{Unruh:1976db}. 

Next, we set up a free scalar field obeying the EoM:
\begin{equation}
(\nabla^2-\mu^2)\phi = 0
\label{eom}
\end{equation}
where $\nabla$ is the covariant derivative. The Hamiltonian of this scalar field only has the dynamical and the mass terms, with no interactions of any sort. Hence thermalization does not happen in this system except through gravitation. This massive real scalar field could serve as a minimal toy model for the investigation of the collapsing shell and asymptotically black hole metric later in Section \ref{sec:blackhole}. %Regardless of the initial condition, the collapsing to a black hole can be solely motivated by the gravity. This also means that, when we only consider one field through out the formation and evaporation of the black hole, 

Now we quantize the field $\Phi$ defined on the full manifold that can be charted by the Minkowski metric or the Rindler metric. First, we solve the equation of motion (\ref{eom}) for the classical field $\phi$, and find two sets of positive frequency basis for the solutions \cite{Unruh:1976db}:
\begin{align}
\phi^{M}_{\omega, \vec{k}}(x) &= \frac{e^{-i\omega t}}{[(2\pi)^3 2 \omega]^{1/2}} e^{i\vec{k}\cdot \vec{x}}\\
\phi^{R}_{\tilde{\omega}, \vec{q}}(\tilde{x}) &= \frac{e^{-i\tilde{\omega}\tau}}{[(2\pi)^3 2\tilde{\omega}]^{1/2}}g(\rho)e^{i(q_x x+q_y y)}
\label{eq:basis}
\end{align}
Where $g(\rho)$ satisfies:
\begin{equation}
[\rho\frac{d}{d \rho} \rho \frac{d}{d \rho} + \tilde{\omega}^2 - (\mu^2+q_x^2 +q_y^2)\rho]g(\rho) = 0
\label{rhoeom}
\end{equation}

and $\omega = \sqrt{\mu^2+\vec{k}^2}$.

%\begin{tcolorbox}[breakable]
%Remark: See also ref \cite{Crispino:2007eb} section II.F for another argument for the completeness of Rindler modes. However, also note that similar calculation cannot be trivially analogize to Kruskal-Schwardschild modes.

%Eq \ref{rhoeom} is similar to a Bessel equation but not exactly (negative $x^2$ term). We do not know the solution off hand. But presumably it is with some eigenvalues $q_{\rho}$ having an identity with $\tilde{\omega}, q_x, q_y$. 
%\end{tcolorbox}

Notice that, by its physical definition $\rho$ is positive, and we can duplicate $\phi^{R}_{\tilde{\omega}, \vec{q}}(\tilde{x})$ to cover the modes on $R^+$ and $R^-$. On future and past wedges, things are more complicated. Because $\frac{d}{d\rho}$ is the timelike direction there, and it is not a killing vector, we do not have solution modes that can be expressed as constant frequency waves in terms of the time ($\rho$). However, we can analytically continue the solutions $\phi^{R+}_{\tilde{\omega}, \vec{q}}(\tilde{x})$ to future wedge, and $\phi^{R-}_{\tilde{\omega}, \vec{q}}(\tilde{x})$ to past wedges. This way of continuation can label the EoM solutions on $F, P$ with $\tilde{\omega}, \vec{q}$, as a complete set. Because the combination of $\phi^{R+}_{\tilde{\omega}, \vec{q}}(\tilde{x})$ and $\phi^{R-}_{\tilde{\omega}, \vec{q}}(\tilde{x})$ on the spacelike surface $t=0$ can fully represent $\phi^{M}_{\omega, \vec{k}}(x)$, and as long as a field is fully determined by the initial condition on certain spacelike slice, the decomposition of that field on this specific slice can be applied to the full foliated spacetime even if the explicit analytical form of the solution remains unknown, as the time evolution of the field is completely unitary. 

In the following text, $\phi^{R+}_{\tilde{\omega}, \vec{q}}(\tilde{x})$ is by definition non-zero on $R+, F$, and $\phi^{R-}_{\tilde{\omega}, \vec{q}}(\tilde{x})$ is by definition non-zero on $R-, P$ (even though we have saved exploring the analytical expression of them on $P, F$). Combining the two, we can get a full piece of $\phi^{M}(t_0,\vec{x})$ on any $t_0$ slice. Thus we define the creation and annihilation operators through the integral of each solution mode with the quantum field $\Phi$, both living on the spacetime manifold charted by certain coordinate systems. %\footnote{We did not define explicitly the bracket here, any conventional definition should work. But one point to make is that we prefer our bracket to be an integral in 4D spacetime, with reasons stated in the Remark on this page.}:

\begin{align}
a_{\omega, \vec{k}} & = \langle \phi^{M}_{\omega, \vec{k}}(x), \Phi(x)\rangle
\label{eq:operatora}\\
%\int_M dx^4 \phi^{*M}_{\omega, \vec{k}}(x) \Phi(x)\\
a^{\dagger}_{\omega, \vec{k}} & = \langle \phi^{*M}_{\omega, \vec{k}}(x), \Phi(x) \rangle
\label{eq:operatoradagger}\\
%\int_M dx^4 \phi^{M}_{\omega, \vec{k}}(x) \Phi(x)\\
b_{\tilde{\omega},\vec{q}} & = \langle \phi^{R+}_{\tilde{\omega}, \vec{q}}(\tilde{x}), \Phi(\tilde{x}) \rangle
\label{eq:operatorb}\\
%\int_{R+,F} d\tilde{x}^4 \phi^{*R+}_{\tilde{\omega}, \vec{q}}(\tilde{x}) \Phi(\tilde{x})\\
b^{\dagger}_{\tilde{\omega},\vec{q}} & = \langle \phi^{*R+}_{\tilde{\omega}, \vec{q}}(\tilde{x}), \Phi(\tilde{x})\rangle
\label{eq:operatorbdagger}\\
%\int_{R+,F} d\tilde{x}^4 \phi^{R+}_{\tilde{\omega}, \vec{q}}(\tilde{x}) \Phi(\tilde{x})\\
d_{\tilde{\omega},\vec{q}} & = \langle \phi^{R-}_{\tilde{\omega}, \vec{q}}(\tilde{x}), \Phi(\tilde{x})\rangle
\label{eq:operatord}\\
%\int_{R-,P} d\tilde{x}^4 \phi^{R-}_{\tilde{\omega}, \vec{q}}(\tilde{x}) \Phi(\tilde{x})\\
d^{\dagger}_{\tilde{\omega},\vec{q}} & = \langle \phi^{*R-}_{\tilde{\omega}, \vec{q}}(\tilde{x}), \Phi(\tilde{x}) \rangle
\label{eq:operatorddagger}
%\int_{R-,P} d\tilde{x}^4 \phi^{*R-}_{\tilde{\omega}, \vec{q}}(\tilde{x}) \Phi(\tilde{x})\\
\end{align}
%\begin{tcolorbox}[breakable]
%Remark: %Why only $d_{\tilde{\omega},\vec{q}}$ on $R^-$ is contracting with the complex conjugate function (unlike the other two annihilator)? Because $\phi^{R\pm}_{\tilde{\omega}, \vec{q}}(\tilde{x})$ are mathematically the same except for $e^{\pm i \tilde{\omega}\tau}$ factor, and while the positive time flow $\frac{d}{d\tau}$ in $R+$ has positive inner product with $\frac{d}{dt}$ (flowing in the semi-aligned direction), $\frac{d}{d\tau}$ in $R-$ flows in the opposite direction with $t$. It is this fact caused the positive and negative frequency modes mixing in Rindler space with respect to the Minkowski positive frequency modes.
%The continuation to future and past wedges here is different from \cite{Unruh:1976db}. Our definition of non-zero $\phi^{R+}_{\tilde{\omega}, \vec{q}}(\tilde{x})$ on $R+, F$ and non-zero $\phi^{R-}_{\tilde{\omega}, \vec{q}}(\tilde{x})$ on $R-, P$ ensures a unique expansion of the field in terms of the Rindler bases, on the full Minkowski spacetime. 
%\end{tcolorbox}

\begin{tcolorbox}[breakable]
Remark: The quantum field $\Phi(x)$, or $\Phi(\tilde{x})$, is a scalar field defined on the full spacetime manifold $\mathcal{M}$ that is an identical collection of Lorentzian manifold points for Minkowski or Rindler coordinate system. When there is a boundary of $\mathcal{M}$, $\partial \mathcal{M}$ is transferred accordingly under two coordinate systems. We assume in our set up the boundary terms are vanishing. %Though each classical solution mode is defined non-zero on different casual wedge, when we define the inner product to be an integral over the full bulk of 4D spacetime manifold, Minkowski and Rindler operators can share the same definition of inner product (the bracket). 4D instead of 3D inner product is redundant in the usual QFT problems in inertial frames, because the initial space slice can determine every other slice given the unitary time evolution in a single inertial frame. However, in such non-inertial frames problems where we have two sets of different time foliation, the inner product should in 4D to take care of the complexity of a eigenstate evolution in an alternative time-flow direction.

Minkowski and Rindler foliation structures have a shared Cauchy surface $t=\tau=0$. The canonical quantization should be done on this specific shared time slice because only this hyper-surface sets the initial condition along every time-flow (forward or backward) direction. So the bracket in the above creator/annihilator definitions should be an integral either on the shared Cauchy surface $t=\tau=0$, or by default over the full chart of 4D spacetime manifold if the shared time slice was not identified. Doing otherwise will obtain an incomplete set of creator/annihilator to represent the quantum field states living on the entire manifold, under one of the reference frames in the pair that we are concern about for the problem.
%It will not enlarge the Hilbert space; For each way of complete Cauchy surface foliation, inner product defined in 4D is redundant. 4D definition of the inner product is only necessary for addressing the space-\textbf{time} varying decomposition of the vectors in the Hilbert space, between two bases.
\end{tcolorbox}

The quantum field can thus be decomposed in two ways: 

\begin{align}
    \Phi(x) &= \int dk^3 [a_{\omega, \vec{k}}\phi^{M}_{\omega, \vec{k}}(x) + a^{\dagger}_{\omega, \vec{k}}\phi^{*M}_{\omega, \vec{k}}(x)]\\
    & = \int dq^3 [b_{\tilde{\omega},\vec{q}}\phi^{R+}_{\tilde{\omega}, \vec{q}}(\tilde{x}) + d^{\dagger}_{\tilde{\omega},\vec{q}}\phi^{R-}_{\tilde{\omega}, \vec{q}}(\tilde{x}) \\
    & +d_{\tilde{\omega},\vec{q}}\phi^{*R-}_{\tilde{\omega}, \vec{q}}(\tilde{x})+ b^{\dagger}_{\tilde{\omega},\vec{q}}\phi^{*R+}_{\tilde{\omega}, \vec{q}}(\tilde{x})]
\end{align}

with canonical quantization \cite{Srednicki:2007qs}:
\begin{align}
    [a_{\omega, \vec{k}},a^{\dagger}_{\omega', \vec{k'}}] & = \delta^3(\vec{k}-\vec{k'})\\
        [b_{\omega, \vec{q}},b^{\dagger}_{\omega', \vec{q'}}] &= \delta^3(\vec{q}-\vec{q'})\label{eq:bcommu}\\
            [d_{\omega, \vec{q}},d^{\dagger}_{\omega', \vec{q'}}] &= \delta^3(\vec{q}-\vec{q'})\label{eq:dcommu}
\end{align}
All other $a$ or $b,d$ commutators are zero. $a, a^{\dagger} $ does not necessarily commute with the Rindler creating and annihilating operators. \footnote{Here we referred to Srednicki eq (3.19) and (3.29). The normalization of $\Phi(x)$ and Minknowski modes here should give the same result as (3.19) in Srednicki, just moving the $\omega$ normalizing factor into $a_{\omega, \vec{k}}$ and $\phi^M_{\omega,\vec{k}}(x)$. As for the normalization of the Rindler modes, we swipe them under the definition of $g(\rho)$.}

In the following sections, we will first calculate how a single frequency mixed Unruh mode is observed by an R+ or R- observer, then proceed to the multi-particle case. 

\section{Unruh Effect Generalized to Single Positive Minkowski Frequency State}
\label{sec:singlemode}
We now carry out a quantum mechanical way of calculation from a single particle state generated by Unruh mode to a density matrix tracing out one of the Rindler wedges. This process has been done in \cite{Unruh:1976db} for a Minkowski vacuum, and we do it for an Unruh wavepacket excited state  $| \tilde{\omega}(\vec{q})\rangle_U$. 

\subsection{An Unruh Wavepacket Excited on Minkowski Vacuum}
\label{subsec:unruhwavepacket}
 Skipping the standard Bogoliubov transformation of the classical field solutions in Minkowski/Rindler metrics, we start from equation (2.19a) in \cite{Unruh:1976db}. It gives the relationship between creation and annihilation operators as the result of the Bogoliubov transformation of the classical field modes:
\begin{align}
    (e^{\pi \tilde{\omega}}b_{\tilde{\omega},\vec{q}}-e^{-\pi \tilde{\omega}}d^{\dagger}_{\tilde{\omega},\vec{q}})| 0\rangle_M &= 0 \\
    \label{eq:unruhannihilator}
    (e^{-\pi \tilde{\omega}}b^{\dagger}_{\tilde{\omega},\vec{q}}-e^{\pi \tilde{\omega}}d_{\tilde{\omega},\vec{q}})| 0\rangle_M &= 0
\end{align}
Let us define the following annihilators for Minkowski vacuum:
\begin{align}
    u_{\tilde{\omega},\vec{q}} &= \frac{1}{\sqrt{2 \sinh{2\pi \tilde{\omega}}}} (e^{\pi \tilde{\omega}}b_{\tilde{\omega},\vec{q}}-e^{-\pi \tilde{\omega}}d^{\dagger}_{\tilde{\omega},\vec{q}})\\
    v_{\tilde{\omega},\vec{q}} &= \frac{1}{\sqrt{2 \sinh{2\pi \tilde{\omega}}}} (e^{-\pi \tilde{\omega}}b^{\dagger}_{\tilde{\omega},\vec{q}}-e^{\pi \tilde{\omega}}d_{\tilde{\omega},\vec{q}})
\end{align}
that satisfies the normal commutating relationships:
\begin{align}
        [u_{\omega, \vec{q}},u^{\dagger}_{\omega', \vec{q'}}] &= \delta^3(\vec{q}-\vec{q'}),\\
        [v_{\omega, \vec{q}},v^{\dagger}_{\omega', \vec{q'}}] &= \delta^3(\vec{q}-\vec{q'}),
\end{align}
with all other commutators in the above operators set equal to zero. We call these operators Unruh creators and annihilators type I and II, and call the mode they generate Unruh mode/excitation.
%\begin{tcolorbox}[breakable]
%Remark: $u^{\dagger}_{\omega, \vec{q}}$ and $v^{\dagger}_{\omega, \vec{q}}$ creates a single corresponding particle on top of the Minkowski vacuum, yet it is obviously not the Minkowski plane wave mode corresponding to $\phi^{M}_{\omega, \vec{k}}(x)$. For $\tilde{\omega}>>1$, we can see that $u^{\dagger}_{\tilde{\omega},\vec{q}}$ generates quanta that is predominant by a $R+$ mode ($b^{\dagger}$), and vice versa. That is why in some of the literature $u^{\dagger}_{\omega, \vec{q}}$ modes are called positive frequency and $v^{\dagger}_{\omega, \vec{q}}$ negative frequency modes, which is wrong. They are never 'purely' positive or negative frequency, and this sits in the core of Unruh effect: the mixture between positive and negative frequency mode, or, the mixture of modes flowing forward and backward with respect to a specific timelike Killing vector $\zeta^{\mu}$.
%\end{tcolorbox}

\subsection{A Single Unruh excitation Viewed by a Rindler Observer, on R+ and R-}
\label{subsec:singleparticle}

Given the transformation between creators and annihilators, in principle, we have obtained a transformation between the Fock states under two basis. To investigate how a single Unruh excitation on Minkowski vacuum is viewed by a Rindler observer on R+ or R-, we will act $u^{\dagger}_{\tilde{\omega},\vec{q}}$ on both sides of the equation connecting Minkowski and Fulling-Rindler vacuum, then trace out the Hilbert space for the states confined on $R-$ or $R+$. We are expected to obtain a highly stochastic density matrix due to the entanglement between two wedges, and it is distinguishable from the canonical thermal density matrix that a Minkowski vacuum would resolve into. By measuring the observables of this density matrix, for example, the energy spectrum, an observer on either $R+$ or $R-$ could resume the Unruh mode generated on the Minkowski vacuum that we started from. This result applies for both type I and type II Unruh modes, and they each are detectable on either $R+$ or $R-$.

The traditional Unruh effect for Minkowski vacuum is formally expressed by equation (2.19b) in \cite{Unruh:1976db}:
\begin{equation}
    |0\rangle_M = Z \left[ \prod_{q} \exp{(e^{-2\pi \tilde{\omega}} b^{\dagger}_{\tilde{\omega},\vec{q}} d^{\dagger}_{\tilde{\omega},\vec{q}})} \right] |0 \rangle_F
    \label{eq:unruhvacuum}
\end{equation}
where $|0 \rangle_F$ is the Fulling-Rindler vacuum vanished by $b,d$ annihilators, and the normalization constant $Z^{-2} = \sum_{\mathcal{N}}e^{-4\pi E_{\rm tot}}$ is the canonical ensemble partition function. $\mathcal{N}$ are all the possible configurations of the eigenmodes, which will be explained further in equation (\ref{eq: Ndefinition}), and $E_{\rm tot} = \sum_i n_i \tilde{\omega}_i$ is the total energy corresponding to the eigenmode. %Notice that as long as we have done our quantization on the shared Cauchy surface in both ways of foliating the spacetime, we are allowed to work in a shared Hilbert space for $| \Phi \rangle$ living on the full Minkowski/four-wedge extended Rindler manifold. \footnote{The shared Hilbert space here is including the states evolving both-ways in time, advanced and retarded. In a physics problem involving observations later, the word `Hilbert space' that is sufficient to an observer is usually only including those states only traveling one-way in time. In a sense, the lower-bounded algebra defined by `annihilation operators' vanishing the `vacuum' on a Cauchy surface is already secretely marginalizing the states traveling in opposite time-flow that immediately fall out the future lightcone of an observer in certain time-flow. This is why non-inertial frames have such complexity -- the Hilbert spaces of one-way time-traveling states do not coincide completely in different frames.}. Hence the equality is exact in equation (\ref{eq:unruhvacuum}). %The size of the Hilbert space only changes when we partition and trace out a part of the system cut by the event horizon. 

Acting a type-I and type-II combined Unruh creator $A_I u^{\dagger}_{\tilde{\omega},\vec{q}}+A_{II}v^{\dagger}_{\tilde{\omega},\vec{q}}$, where $A_{I}^2+A_{II}^2=1$, on both side of equation (\ref{eq:unruhvacuum}):
\begin{align}
    | \tilde{\omega}(\vec{q})\rangle_U &= Z A_I\left( \frac{e^{\pi \tilde{\omega}}}{\sqrt{2\sinh{2\pi\tilde{\omega}}}}b^{\dagger}_{\tilde{\omega},\vec{q}}\hat{S} -  \frac{e^{-\pi \tilde{\omega}}}{\sqrt{2\sinh{2\pi\tilde{\omega}}}}d_{\tilde{\omega},\vec{q}}\hat{S} \right) \\ 
    & + Z A_{II}\left( \frac{e^{\pi \tilde{\omega}}}{\sqrt{2\sinh{2\pi\tilde{\omega}}}}d^{\dagger}_{\tilde{\omega},\vec{q}}\hat{S} -  \frac{e^{-\pi \tilde{\omega}}}{\sqrt{2\sinh{2\pi\tilde{\omega}}}}b_{\tilde{\omega},\vec{q}}\hat{S} \right)
    \label{eq:singleunruhstate}
\end{align}
Where 
\begin{equation}
    \hat{S} = \left[ \prod_{q} \exp{(e^{-2\pi \tilde{\omega}} b^{\dagger}_{\tilde{\omega},\vec{q}} d^{\dagger}_{\tilde{\omega},\vec{q}})} \right].
\end{equation}
Using
\begin{align}
    [b^{\dagger}_{\tilde{\omega},\vec{q}},\hat{S}] & =  [d^{\dagger}_{\tilde{\omega},\vec{q}},\hat{S}] =0, \label{eq:vcommu1}\\
    [d_{\tilde{\omega},\vec{q}},\hat{S}] & = e^{-2\pi\tilde{\omega}}\hat{S}b^{\dagger}_{\tilde{\omega},\vec{q}},     \label{eq:vcommu2}\\
    [b_{\tilde{\omega},\vec{q}},\hat{S}] & = e^{-2\pi\tilde{\omega}}\hat{S}d^{\dagger}_{\tilde{\omega},\vec{q}}
    \label{eq:vcommu3}
\end{align}
We get 
\begin{align}
    | \tilde{\omega}(\vec{q})\rangle_U & = Z e^{-\pi \tilde{\omega}}\sqrt{2\sinh{2\pi \tilde{\omega}}}\\
    & \times (A_I b^{\dagger}_{\tilde{\omega},\vec{q}} + A_{II} d^{\dagger}_{\tilde{\omega},\vec{q}}) 
    \hat{S} | 0\rangle_F
\end{align}

Next, we trace out the states generated by the R- creators, $d^{\dagger}$, acting on $| 0 \rangle_F$, to get the density matrix on R+, $\rho_U^{R+}(q)$. The result consists of four terms:
\begin{align}
    \rho_U^{R+}(q) &= %Z^2e^{-2\pi \tilde{\omega}}2\sinh{2\pi \tilde{\omega}}
    |A_I|^2\hat{\rho}^{b^{\dagger}b}  + |A_{II}|^2\hat{\rho}^{d^{\dagger}d} + A_I A_{II}^*\hat{\rho}^{b^{\dagger}d} + A_I^*A_{II}\hat{\rho}^{d^{\dagger}b}
    \label{eq:rhorplus}
\end{align}
where $q$ has energy component $\tilde{\omega}$, and $\hat{\rho}^{d^{\dagger}b} = \hat{\rho}^{b^{\dagger}d \dagger}$. All four terms have $q$ dependence through $b/d$ creator/annihilators. 

Before we demonstrate the detailed expressions of $\hat{\rho}$ in the above equation, let us introduce some convenient notations. We denote the configuration of momentum $q$ of an eigenstate in the following way:
\begin{equation}
    | \mathcal{N}\rangle = \{q_1, n_1,...q_i, n_i\}, n_i = 0,1,2,...
    \label{eq: Ndefinition}
\end{equation}
where $q$ is the four momentum, including $\tilde{\omega}$ as a component.
With respect to the configuration $| \mathcal{N}\rangle$,  a state missing one particle of momentum $q$ is denoted by:
\begin{equation}
    |\mathcal{N}, n_{q}-1 \rangle,
\end{equation}
and similarly for other number modifications of $n_q$. The states in this notation are normalized, in the sense that:
\begin{align}
    |\mathcal{N}, n_{q}-1 \rangle &= \frac{a_{\omega,q}}{\sqrt{n_{q}}} |\mathcal{N} \rangle, \\
    |\mathcal{N}, n_{q}+1 \rangle &= \frac{a^{\dagger}_{\omega,q}}{\sqrt{n_{q}+1}} |\mathcal{N} \rangle. 
\end{align}

This notation can be applied to any set of eigenstates and corresponding creators and annihilators. The vacuum that $| \mathcal{N}\rangle$ corresponds to will be labeled in the subscripts later on.

%The necessity of keeping track of the full momentum configuration, instead of only noting the number count of a specific frequency like in most of the past works should show up later in section \ref{sec:spectrum}. %There, the energy spectrum as a function of frequencies will be calculated for the pseudo-thermal ensemble. 
%We will see that the information engraved in $n_q$ of Kruskal Unruh eigenmodes is transferred into $n_q-n_{q'}$ of Outer-horizon Schwarzschild eigenmodes. Namely, on partitioned spacetime manifold, the relative number counting of the special momentum with respect to others is the operator that plays the role of the absolute number counting operator in the maximal foliation Hilbert space, thus the deviation of the spectrum from the canonical ensemble could carry some or even majority the information while the unitarity is broken.

With $| \mathcal{N}\rangle$ notation, we can denote 
\begin{align}
    \hat{S}|0\rangle_F & = \left[ \prod_{q} \exp{(e^{-2\pi \tilde{\omega}} b^{\dagger}_{\tilde{\omega},\vec{q}} d^{\dagger}_{\tilde{\omega},\vec{q}})} \right] | 0\rangle_F\\
    & = \sum_{\mathcal{N}}e^{-2\pi E_{\rm tot}} |\mathcal{N}\rangle_{R+}|\mathcal{N}\rangle_{R-},
\end{align}
where $E_{\rm tot} = \sum_i \tilde{\omega_i}n_i$, and $\mathcal{N}$ runs over all the possible configurations of the eigenstates of $\Phi$ generated by the quantization of equation (\ref{eq:basis}) as indicated by $R+/R-$ subscripts.

We start from the calculation of $\hat{\rho}^{b^{\dagger}b}$.
\begin{align}
    \hat{\rho}^{b^{\dagger}b}(q)  =& Z^2e^{-2\pi \tilde{\omega}}2\sinh{2\pi \tilde{\omega}}\\
    & \times \sum_{\mathcal{N}} \sum_{\mathcal{N'}}e^{-2\pi E'_{\rm tot}} \ _{R-}\langle \mathcal{N}|\mathcal{N'}\rangle_{R-}  b^{\dagger}_q|\mathcal{N'}\rangle_{R+} \\
    & \bigotimes \sum_{\mathcal{N''}}e^{-2\pi E''_{\rm tot}} \ _{R+}\langle \mathcal{N''}|b_q \ _{R-}\langle\mathcal{N''} | \mathcal{N}\rangle_{R-}\\
     =& Z^2e^{-2\pi \tilde{\omega}}2\sinh{2\pi \tilde{\omega}} \\ 
    &\times \sum_{\mathcal{N}} e^{-4\pi E_{\rm tot}} (n_q + 1)  |\mathcal{N},n_q+1\rangle_{R+} \ _{R+}\langle \mathcal{N},n_q+1|
    \label{eq:rhobb}
\end{align}
Next, the result for $\hat{\rho}^{d^{\dagger}d}$ is:
\begin{align}
    \hat{\rho}^{d^{\dagger}d}(q) = & Z^2e^{-2\pi \tilde{\omega}}2\sinh{2\pi \tilde{\omega}} \\
    & \times \sum_{\mathcal{N}} \sum_{\mathcal{N'}}e^{-2\pi E'_{\rm tot}} \ _{R-}\langle \mathcal{N}|d^{\dagger}_q|\mathcal{N'}\rangle_{R-} |\mathcal{N'}\rangle_{R+} \\
    & \bigotimes \sum_{\mathcal{N''}}e^{-2\pi E''_{\rm tot}} \ _{R+}\langle \mathcal{N''}| \ _{R-}\langle\mathcal{N''} |d_q| \mathcal{N}\rangle_{R-}\\
    = & Z^2e^{-2\pi \tilde{\omega}}2\sinh{2\pi \tilde{\omega}} \\
    & \times \sum_{\mathcal{N}} e^{-4\pi E_{\rm tot}} (n_q + 1)  |\mathcal{N}\rangle_{R+} \ _{R+}\langle \mathcal{N}|
    \label{eq:rhodd}
\end{align}

Lastly, the result for $\hat{\rho}^{b^{\dagger}d}$ is:
\begin{align}
    \hat{\rho}^{b^{\dagger}d}(q)  = & Z^2e^{-2\pi \tilde{\omega}}2\sinh{2\pi \tilde{\omega}}\\
    & \times \sum_{\mathcal{N}} \sum_{\mathcal{N'}}e^{-2\pi E'_{\rm tot}} \ _{R-}\langle \mathcal{N}|\mathcal{N'}\rangle_{R-} b^{\dagger}_q|\mathcal{N'}\rangle_{R+} \\
    & \bigotimes \sum_{\mathcal{N''}}e^{-2\pi E''_{\rm tot}} \ _{R+}\langle \mathcal{N''}| \ _{R-}\langle\mathcal{N''} |d_q| \mathcal{N}\rangle_{R-}\\
    = & Z^2e^{-2\pi \tilde{\omega}}2\sinh{2\pi \tilde{\omega}} \\
    & \times \sum_{\mathcal{N}} e^{-2\pi \tilde{\omega}} e^{-4\pi E_{\rm tot}} \sqrt{(n_q+2)(n_q + 1)} \\
    & \times |\mathcal{N}, n_q+2 \rangle_{R+} \ _{R+}\langle \mathcal{N}|
    \label{eq:rhobd}
\end{align}

Substituting the above terms back to equation (\ref{eq:rhorplus}), we get:
\begin{widetext}

\begin{align}
    \rho_U^{R+}(q) &= Z^2e^{-2\pi \tilde{\omega}} 2 \sinh{2\pi \tilde{\omega}} \left[|A_I|^2\sum_{\mathcal{N}} e^{-4\pi E_{\rm tot}} (n_q + 1) |\mathcal{N},n_q+1\rangle_{R+} \ _{R+}\langle \mathcal{N},n_q+1| \right.\\
    & + |A_{II}|^2\sum_{\mathcal{N}} e^{-4\pi E_{\rm tot}} (n_q + 1) |\mathcal{N}\rangle_{R+} \ _{R+}\langle \mathcal{N}| \\
    &  + \left( A_I A_{II}^*\sum_{\mathcal{N}} e^{-2\pi \tilde{\omega}} e^{-4\pi E_{\rm tot}} \sqrt{(n_q+2)(n_q + 1)}\right.  \Biggl. \Biggl. |\mathcal{N}, n_q+2 \rangle_{R+} \ _{R+}\langle \mathcal{N}| + H.C. \Biggr) \Biggr]
\end{align}
\end{widetext}

For an R- observer, $\rho_U^{R-}(q)$ switches $A_I$ and $A_{II}$, and changes the subscript $R+$ to $R-$ for the states.

The first thing we would notice is that $\hat{\rho}^{b^{\dagger}b}$ and $\hat{\rho}^{d^{\dagger}d}$ only involves diagonal terms, while $\hat{\rho}^{b^{\dagger}d}$ only has non-diagonal terms. Remembering that $Z^{-2}$ is partition function, we find that ${\rm Tr}({\hat{\rho}^{b^{\dagger}b}}) = 1$ and $\hat{\rho}^{b^{\dagger}b} \geq 0$. Namely, both diagonal matrices $\hat{\rho}^{b^{\dagger}b}$ and $\hat{\rho}^{d^{\dagger}d}$ could be normalized density matrices, assuming such notation can be applied to infinite dimensional Hilbert space. As a result, the trace of $\rho_U^{R+}(q)$ is also safely equal to 1. 

When the Unruh mode is purely type I, $\rho_U^{R+}(q)$ only have $\hat{\rho}^{b^{\dagger}b}$ term, and $\rho_U^{R-}(q)$ only have $\hat{\rho}^{d^{\dagger}d}$ term. Comparing $\hat{\rho}^{b^{\dagger}b}$ and $\hat{\rho}^{d^{\dagger}d}$, we notice that although the normalized states have different labels, the matrix components are equal to each other term by term. Hence the Von-Neumann entropy $S = -{\rm tr}{\rho \log{\rho}}$ of $\rho_U^{R+}(q)$ and $\rho_U^{R-}(q)$ for type I Unruh mode must be the same. It is expected, because $| \tilde{\omega}(\vec{q})\rangle_U$ is a pure state. The bipartite of a pure state must have equal entropy \cite{nielsen2002quantum}. 

Our notation in this section has already implicitly assumed the quantum mechanical linear algebraic representation of the infinite-dimensional Hilbert space in QFT. Tracing back to the origin where this assumption is introduced, the notation $|\mathcal{N} \rangle$ in equation (\ref{eq: Ndefinition}) is actually not as innocent as it seems -- at this step, we implicitly assumed countability of the momentum defined with the eigenmodes of the field equation of motion solutions. It means that throughout this paper the maximal spacetime manifold we consider should be bounded, or equivalently we should have a lower cut-off on the acceleration $\alpha$ (minimum deviation from the inertial frame) so that the energy levels of the scalar field are discrete. Without losing generality, we assume the minimal interval between energy levels to be a constant $h_0$. The set of Fock states spanning the Hilbert space of this QFT system thus could be countably infinite. One of the strategies to count the states is to continuously increase $E_{\rm tot}$ from minimum possible value $\mu$ and to attribute a natural number to each eigenstate in ascending $E_{\rm tot}$ sequence. 

%Similar calculation has also been carried out before in a series of literature mostly on the topic of entanglement in curved spactime \cite{Landulfo:2009wg,bruschi2010unruh,Martin-Martinez:2010yva}. They usually only focus on the number states of a single frequency. Despite the difference in this setup, the similar expressions of the results here comparing to those previous literature could serve as a cross-check of our maths.

\section{Beyond single Unruh Excitation}
\label{sec:beyondsingle}
Now we generalize the calculation in the above section into the full Fock space generated by the Unruh creators $\{u^{\dagger},v^{\dagger}\}$. This Unruh Fock space should be a basis transformation with respect to the one generated by the Minkowski Klein-Gordon wave creators $a^{\dagger}$ since they share the same vacuum. Actually, the classical field solutions have shown that the Unruh modes are plane wave decomposition of the field in $\log{U}$ space, instead of in the natural lightcone coordinate, $U = t-x$ space. Without proof here, we quote the results from \cite{Unruh:1976db,Crispino:2007eb} that the transformation between plane-wave modes and the Unruh modes is positive frequency to positive frequency. Thus our conclusion in this section for the Unruh Fock space is in general applicable to the original Minkowski plane wave Fock space, with a basis transformation from the plane waves to the plane waves in $\log{x}$ space.

A general Fock state generated by the Unruh modes above Minkowski vacuum can be expressed by:
\begin{align}
    | \mathcal{N}_I \mathcal{N}_{II}\rangle_U & = \frac{1}{\sqrt{\alpha_1!...\alpha_N!\beta_1!...\beta_N!}}  u_{q_1}^{\dagger, \alpha_1}...u_{q_N}^{\dagger, \alpha_N}v_{q_1}^{\dagger, \beta_1}...v_{q_N}^{\dagger, \beta_N} \hat{S}| 0 \rangle_F
\end{align}

where $\alpha_i,\beta_i$ are the number counts for certain momentum modes, and they can be zero. 

Utilizing the commutators between $b,d,b^{\dagger},d^{\dagger}$ and $\hat{S}$ in equations (\ref{eq:bcommu}), (\ref{eq:dcommu}), (\ref{eq:vcommu1}), (\ref{eq:vcommu2}), (\ref{eq:vcommu3}), we notice that, 
\begin{equation}
    |\mathcal{N}_I \mathcal{N}_{II} \rangle_U = \prod_{q_i}\sum_{m} B(\tilde{\omega}_i|m+\alpha_i-\beta_i,m) b^{\dagger,m+\alpha_i-\beta_i}_{q_i} d^{\dagger, m}_{q_i} \hat{S} | 0\rangle_F
    \label{eq:unruhfockstate}
\end{equation}

Here, assuming $\alpha_i>\beta_i$, $m = 0,...\beta_i$, and $B(\tilde{\omega}_i|m+\alpha_i-\beta_i,m)$ are positive coefficients that can be analytically calculated from the commutators between $\hat{S}, b, d, b^{\dagger}, d^{\dagger}$. Thus for an arbitrary eigenstate in the Minkowski vacuum Unruh mode Fock space, the building block for the density matrix on $\rho^{R+}$ or $\rho^{R-}$ is given by:
\begin{widetext}
\begin{align}
    \hat{\rho}_{R+}^{b^{\dagger,l}d^{\dagger,m}b^n d^s}(q)  = & B(\tilde{\omega}|l,m) B^*(\tilde{\omega}|n,s) \sum_{\mathcal{N}} \sum_{\mathcal{N'}}e^{-2\pi E'_{\rm tot}} \ _{R-}\langle \mathcal{N}|d^{\dagger,m}|\mathcal{N'}\rangle_{R-} b^{\dagger,l}_q|\mathcal{N'}\rangle_{R+} \sum_{\mathcal{N''}}e^{-2\pi E''_{\rm tot}} \ _{R+}\langle \mathcal{N''}|b_q^{n} \ _{R-}\langle\mathcal{N''} | d^s| \mathcal{N}\rangle_{R-}\\
     = & B(\tilde{\omega}|l,m) B^*(\tilde{\omega}|n,s) \sum_{\mathcal{N},n_{q}^{\rm min}} e^{-4\pi E_{\rm tot}} e^{2\pi(m+s)\tilde{\omega}} \frac{n_q! }{(n_q-m)!(n_q-s)!}\sqrt{(n_q-m+l)!(n_q-s+n)!} \\ 
     & \times |\mathcal{N},n_q-m+l\rangle_{R+} \ _{R+}\langle \mathcal{N},n_q-s+n|
    \label{eq:rhogeneral}
\end{align}
\end{widetext}
where the configuration runs over $\mathcal{N}$, with the minimum number of $q$ momentum modes $n_q^{\rm min} = \max{[m,s]}$. $l,m,n,s$ runs over non-negative integers. The normalization factor $B(\tilde{\omega}_i|l,m)$ ensures the traces of diagonal matrices  $\hat{\rho}_{R+}^{b^{\dagger,l}d^{\dagger,m}b^l d^m}(q)$ equal one. An arbitrary $\rho_{R+}(\Omega)$ tracing out R- states can be expressed by a linear combination of   $\hat{\rho}_{R+}^{b^{\dagger,l}d^{\dagger,m}b^n d^s}$, with the modulation of diagonal components $\hat{\rho}_{R+}^{b^{\dagger,l}d^{\dagger,m}b^l d^m}(q)$ being unity.

The expression above is only for a single Unruh mode frequency; Because the creators and annihilators of different momentum commute with each other, the generalization to multiple momenta is straightforward. We just need to manipulate $n_{q_i}$ with respect to the mode number eigenvalue configuration $\mathcal{N}$, for each momentum $q_i$. Notice that the choice of $\mathcal{N}$ is meaningful and non-trivial under the relabeling of $n_q$, because the Boltzmann factor $e^{-4\pi E_{\rm tot}}$ non-trivially weights a specific block $|\mathcal{N},n_q-m+l\rangle_{R+} \ _{R+}\langle \mathcal{N},n_q-s+n|$ in the density matrix.

Equation (\ref{eq:rhogeneral}) is all one needs to calculate the contracted density matrix on a partitioned spacetime wedge from an original density matrix living on the fully accessible spacetime. Thus we have a non-trivial mapping from the Minkowski-Unruh Fock states \footnote{They stand for the Fock states of the Unruh wave modes number eigenstates excited on Minkowski vacuum. Similarly, we would have Minkowski-Plane wave Fock states, etc.} to the pseudo-thermal density matrix on R+ and R-. We give the name `pseudo-thermal' density matrix to those infinite term density matrices like $ \hat{\rho}_{R+}^{b^{\dagger,l}d^{\dagger,m}b^n d^s}(q)$, that slightly deviates from the canonical ensemble. Also, notice that the pseudo-thermal density matrix cannot be constructed as the linear combination of a canonical ensemble and the excitation of the required number of certain momentum. The physical meaning of this fact is that the pseudo-thermal density matrix originated from the entanglement of excited states across the event horizon is distinguishable from the excitation on one of the Rindler wedges backlighted by the thermal bath generated by the vacuum at the horizon.

In section \ref{subsec:singleparticle}, we calculated the special cases for $lmns = 1010, 0101,$ and $1001$. They each agree with the general form equation (\ref{eq:rhogeneral}).

 It is obvious that the matrices $\hat{\rho}_{R+}^{b^{\dagger,l}d^{\dagger,m}b^n d^s}(q)$ are linearly independent in terms of the matrix summation algebra.%, because there is no coefficient for fixed $(lmns)$ combinations that could eliminate all the $n_q=n_q^{min},...\infty$ terms. 
 What is more, it is impossible for different Fock states, i.e. the eigenstates corresponding to Klein-Gordon equation solutions, $|\mathcal{N}_I \mathcal{N}_{II} \rangle_U$ to be mapped into the same $\hat{\rho}_{R+}^{b^{\dagger,l}d^{\dagger,m}b^n d^s}(q)$ linear combination. Because the $lmns$ maximum coefficient term for certain configuration $|\mathcal{N}_I \mathcal{N}_{II} \rangle_U$ is given by $l=n=\alpha, m=s=\beta$. With the always positive coefficient $B(\tilde{\omega}|l,m)$, the $\hat{\rho}_{R+}^{b^{\dagger,l}d^{\dagger,m}b^n d^s}(q)$ linear combination obtained from a distinct $|\mathcal{N}_I \mathcal{N}_{II} \rangle_U$ state is guaranteed to be different from the pseudo-thermal density matrices linear combination corresponding to other Unruh mode eigenstates. 

However, it needs further discussion to clarify whether each unique density matrix of the form $\sum_{i,j} A_{ij}|\mathcal{N}^i_I \mathcal{N}^i_{II} \rangle_U \ _U \langle\mathcal{N}^j_I \mathcal{N}^j_{II} |$ would be contracted into a unique $\hat{\rho}_{R+}^{b^{\dagger,l}d^{\dagger,m}b^n d^s}(q)$ linear combination. The answer is likely yes, but we leave it as an open question without carrying out a concrete mathematical proof in the scope of this paper. The complication comes from the contribution to the same $\hat{\rho}_{R+}^{b^{\dagger,l}d^{\dagger,m}b^n d^s}(q)$ term from different $|\mathcal{N}^i_I \mathcal{N}^i_{II} \rangle_U \ _U \langle\mathcal{N}^j_I \mathcal{N}^j_{II} |$ terms, when two types of Unruh modes exist simultaneously. %When there is only one type of Unruh mode, the question above is much more approachable. Hence we will come back to the issue of one-to-one mapping between maximal foliation density matrices and partitioned spacetime manifold pseudo-thermal density matrices in section \ref{sec:blackhole}, for Kruskal/Schwarzschild black hole metric. Because there is only one outgoing direction for the Unruh modes in black hole scenario, we only need to deal with one type of Unruh excitations there.  

The toy model of our massive scalar experiences no interaction other than gravity. Hence in principle, any slight deviation from the canonical ensemble in the density matrix would be preserved during its propagation on a Rindler wedge, without further thermalization from collisions. %Nor do we need to worry about the grey-body factor later for Kruskal/Schwarzschild black hole, under the Unruh effect set up with the diffused field EoM eigenmodes (no configuration space localization assumption to begin with).  
An observer sitting away from the horizon on a Rindler wedge can then measure the observables of this density matrix, then infer the original state living across the horizon on the Minkowski vacuum. When the acceleration is small, we asymptotically go back to the inertial frame case, where the pseudo-thermal density matrix is dominated by a peak corresponding to the Minkowski vacuum Unruh mode. %We have not proven the isomorphism under the case where both types of Unruh modes simultaneously exist, as discussed previously, however we have at least demonstrated that such mapping is nontrivial for a pure state excited on Minkowski vacuum. So even if not all the information could be retrieved, due to the potential degeneracy of $\sum_{i,j} A_{ij}|\mathcal{N}^i_I \mathcal{N}^i_{II} \rangle_U \ _U \langle\mathcal{N}^j_I \mathcal{N}^j_{II} |$ density matrices contracting to a single Rindler wedge density matrix in the form of $ \hat{\rho}_{R+}^{b^{\dagger,l}d^{\dagger,m}b^n d^s}(q)$ vector, features of these Minkowski systems can still be inferred from the Rindler observer measurements. Namely the horizon is at least never a firewall blocking all the information.

We are skipping the formal setups of the detectors and the formal definitions of the observable operators here \footnote{Something way more complicated than Unruh-DeWitt detector is required to measure the frequency space details of the density matrix.}. We believe the measurement can be done in a fairly standard way in the asymptotically flat region far from the Rindler horizons. %The Hilbert space we have been working in corresponds to the quantized equation of motion solutions to the classical theory, and it is (at least to our understanding) straightforward to setup momentum or number counting operators in the asymptotically flat region far from the horizon on a Rindler wedge. 
The measurement in the asymptotically flat region is sufficient for learning the property of the ensemble on any equal-time slice, because in our non-interacting massive scalar field toy model, the density matrix evolves trivially in energy-momentum eigenstates. %The infinite amount of measurements required to perfectly identify a pseudo-thermal density matrix is a separate issue that threatens the statements that we are making for the infinite dimensional Hilbert space. %however no exact measurement can be done on a pure state in quantum theory anyways. In fact, on quite the contrary, the impossibility of measuring a pseudo-thermal density matrix might serve as an explanation of the uncertainty principle. This is only a far-fetched proposal beyond the scope of this paper though.

%Notice that we cheated here to throw out the Rindler thermal density matrix terms that are redundant for the inverse determination of a Minkowski pure state. This means that the isomorphism does not come trivially when we take the Minkowski mixed states into account. The structure is still preserved, because ${\rm tr}_{R-}(\rho_M^A + \rho_M^B) = {\rm tr}_{R-}(\rho_M^A)+{\rm tr}_{R-}(\rho_M^B)$, but some discussion might be needed to prove the surjectivity and the injectivity. Things are more complicated when taking all the possibility into consideration, but I suspect that the isomorphism, or at least an injective mapping, still holds from Minkowski density matrix to Rindler thermal density matrices. In any case, to avoid uncertainty, we will confine to the Minkowski/Unruh (and later Kruskal) pure states in the rest of the note and leave the proof for density matrix to the mathematicians.

\begin{tcolorbox}[breakable]
Remark: The Boltzmann factor $e^{-4\pi E_{\rm tot}}$ should be treated exactly, carefully, without any approximation in the rest of this paper, where $E_{\rm tot}$ is the total energy for the configuration $\mathcal{N}$, not the physical total energy of the energy eigenstate. Relabeling $n_q$ will only change the reference zero energy point for every component in the density matrix $\hat{\rho}_{R+}^{b^{\dagger,l}d^{\dagger,m}b^n d^s}(q)$. In some previous work, for example, the Horowitz-Maldacena conjecture \cite{horowitz2004black} paper, people regarded this seemingly boring factor as an insignificant algebraic label without tracking the details in it. This might exactly be the approximation erasing all the information that a contracted Rindler density matrix carries. $E_{\rm tot}$ depends on everything in the number eigenvalue configuration $\mathcal{N}$. The coherent mismatch between the Boltzmann factor of $\mathcal{N}$ and the physical state $|\mathcal{N}, n_{q}-1 \rangle \langle \mathcal{N}, n_{q}-1|$, for example, is telling the stories about the Minkowski pure state before contracting a Rindler wedge out.

\end{tcolorbox}

For an arbitrary Unruh mode eigenstate simultaneously generated by both types of the creators, $u^{\dagger}$ and $v^{\dagger}$, those three observations hold:

\begin{itemize}
    \item The relationship between R- and R+ observer density matrix is:
\begin{equation}
    \hat{\rho}_{R+}^{b^{\dagger,l}d^{\dagger,m}b^n d^s}(q) = \hat{\rho}_{R-}^{b^{\dagger,m}d^{\dagger,l}b^s d^n}(q)
    \label{eq:rhohatsymmetry}
\end{equation}
    \item All the $\hat{\rho}_{R+}^{b^{\dagger,l}d^{\dagger,m}b^n d^s}(q)$ components that appear in the representation of a Rindler wedge contraction for Unruh mode eigenstate obey the identity $l-m=n-s\equiv{\delta}$. Thus $\rho_{R\pm}$ is always diagonal for Unruh mode eigenstates.
    \item Regardless of the Hilbert space on which they are built, pseudo-thermal matrices have component-level symmetry 
    \begin{equation}
        \hat{\rho}^{b^{\dagger,l}d^{\dagger,m}b^n d^s}_{ii}(q) = \hat{\rho}^{b^{\dagger,m}d^{\dagger,l}b^s d^n}_{i+\delta,i+\delta}(q)
    \end{equation} 
    when the identity $l-m=n-s$ holds. Without losing generality $i$ starts from $0$, and the components with indices smaller than $\delta$ of the matrix on the right-hand side vanish. An illustration of this symmetry is $\hat{\rho}^{b^{\dagger}b}(q)$ and $\hat{\rho}^{d^{\dagger}d}(q)$ in the single particle case in section \ref{sec:singlemode}. 

\end{itemize}

The above features of the pseudo-thermal density matrices grant us the same conclusion as the single particle case in section \ref{sec:singlemode}, for an arbitrary Unruh mode eigenstate: The Von-Neumann entropy $S=-{\rm tr}(\rho \log{\rho})$ on R+ or R- wedge partition of an Unruh mode eigenstate has equal value. 

%For the indices that obey $l-m=n-s$, which are the only terms that appears in R+ or R- contraction of an Unruh mode eigenstate $|\mathcal{N}_I \mathcal{N}_{II} \rangle_U$, we have the value of the elements in $\hat{\rho}_{R+}^{b^{\dagger,l}d^{\dagger,m}b^n d^s}(q)$ matrix equal to $\hat{\rho}_{R+}^{b^{\dagger,m}d^{\dagger,l}b^s d^n}(q)$ matrix, although their row and column indices are different (different collection of the non-vanishing $\mathcal{N}$). An explicit example is the $\hat{\rho}^{b^{\dagger}b}(q)$ and $\hat{\rho}^{d^{\dagger}d}(q)$ in the single particle case in section \ref{sec:singlemode}. At face value, such symmetry leads to the equal Von-Neumann entropy $S=-{\rm tr}(\rho \log{\rho})$ for R+ and R- contractions for an Unruh modes number eigenstate. 
However, the true physical implication of pseudo-thermal density matrix features on the amount of entanglement and information might need more careful calculation with well-defined metrics like negativity, relative entropy defined for QFT, etc \cite{Plenio:2007zz, Witten:2018zxz}, applied to the countably infinite dimensional density matrix representation here. We leave it to future works due to a lack of expertise on these topics.

\begin{figure}[ht]
    \centering
    \includegraphics[width=0.5\textwidth]{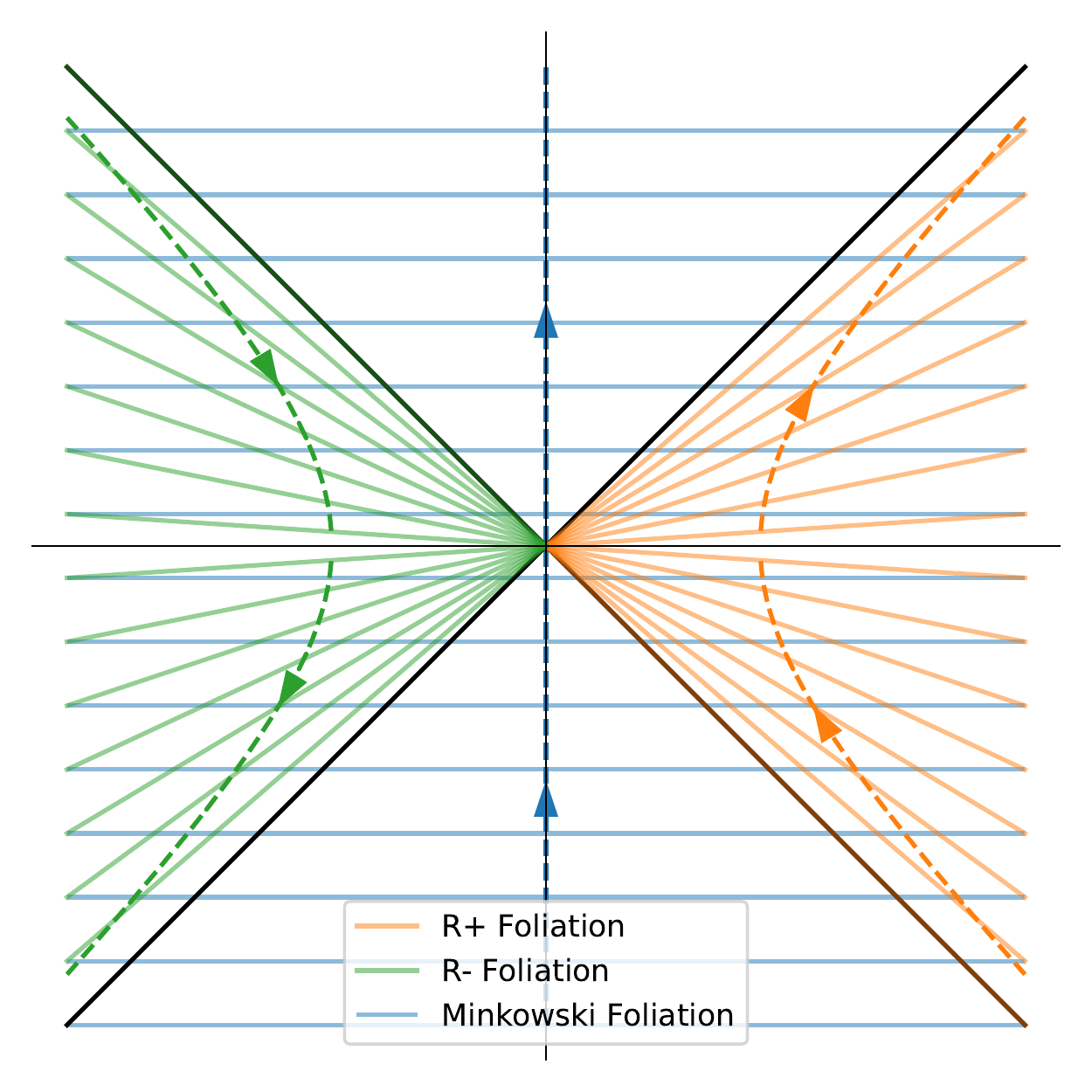}
    \caption{The foliation structure of Minkowski metric and Rindler metric. Blue horizontal lines are the equal time slices under Minkowski foliation, and the inclined straight lines going through the origin from green to orange are the equal time slices under Rindler foliation. The flow of time is forward for $R+$ and backward for $R-$ observer, with respect to the time flow for a Minkowski observer.}
    \label{fig:foliation}
\end{figure}

Given the preemptive statements about the information and entanglement metrics above, it is still useful to remind ourselves that metrics are only handy algebraic compression of the information carried by the specific quantum state or ensemble. The symmetry between partially-traced $\rho_{R+}$ and $\rho_{R-}$ of Unruh eigenstates, at least, strongly implies the equal measurability of a mode excited on maximal foliation on both partitions separated by an event horizon. Only the mode generated on a Rindler vacuum, namely the particle defined with respect to the proper time of the accelerated observer, is absolutely not detectable on the complementary Rindler wedge. The representation of a mode generated by different types of creators is summarized in table \ref{tab:visibility}. The foliation structures of the Minkowski $\leftrightarrow$ R+, R- spacetime are illustrated in figure \ref{fig:foliation}.

{\renewcommand{\arraystretch}{1.3}
\begin{table}[ht]
\centering
\caption{Creators and their visibility under different bases of the quantum field modes. Their definitions are specified in equations (\ref{eq:operatora})-(\ref{eq:operatorddagger}). Minkowski plane waves and Minkowski Unruh modes are generated on the same vacuum with different classical waveform decompositions, while Rindler observers' creator and annihilation operators are mixed in the Bogoliubov transformations from the former two. `Not visible' in the table means that a generator or annihilator on one subspace of the manifold is not affecting the states excited on the other subspace of the manifold.}
\begin{tabular}[t]{|c|c|c|c|c|}
\hline
 Creators & \makecell{Minkowski \\ plane waves} & \makecell{Minkowski \\ Unruh modes} & R+ & R- \\
\hline
$u^{\dagger}$, $v^{\dagger}$ & $a^{\dagger}$ & - & \multicolumn{2}{c|}{$(b^{\dagger},d)$ or $(d^{\dagger},b)$}\\
\hline
$a^{\dagger}$ & - & $u^{\dagger}$, $v^{\dagger}$ & \multicolumn{2}{c|}{$(b^{\dagger},d)$ or $(d^{\dagger},b)$}\\
\hline
$b^{\dagger}$ & $(a^{\dagger},a)$ & $(u^{\dagger},v)$ & - & Not Visible\\
\hline 
$d^{\dagger}$ & $(a^{\dagger},a)$ & $(u,v^{\dagger})$ & Not Visible & -\\
\hline
\end{tabular}
\label{tab:visibility}
\end{table}%
}

%\textbf{The horizon is preventing the particles generated on the foliated structured spacetime on one side to be seen by the other side. The particles generated before the formation of the horizon, namely the modes living in the full foliated structure intersecting the foliated structure on the subspace of the spacetime partitioned by the horizon, is not destructed by the formation of the horizon.}

%In the next section, we will justify the similar results about pseudo-thermal ensemble on the Schwarzschild black hole metric. The generalized Unruh effect suggests that the black hole horizon is not swallowing all the information compressed onto it during the collapsing procedure, even without any sort of fancy quantum gravity/string theory remedies. The information could be partially, even almost fully preserved throughout the non-unitary evolution of a quantized planewave mode falling in then evaporating out from the horizon. 

%Even if the bijectivity assumption is not true, and we only have injection from the Minkowski Fock space to the thermal density matrix space, we can still restore a considerable amount of information of the collapsed pure state.

\section{A Potential Solution to the Black Hole Information Paradox Without Quantum Gravity}
\label{sec:blackhole}
In this section, we will explore the applicability of the results in section \ref{sec:singlemode} and \ref{sec:beyondsingle} on the Kruskal-Schwarzschild relative non-inertial frames pair. To analogously find the mapping like \{Minkowski-Unruh Fock space density matrix $\leftrightarrow$ R$\pm$ pseudo-thermal density matrix\} in the previous section for the \{infalling, black-hole-forming star density matrix $\leftrightarrow$ Schwarzschild outside/inside pseudo-thermal density matrix\} duality, we need two conditions to be satisfied: 
\begin{itemize}
    \item \textbf{Condition 1:} A positive frequency mode under Kruskal foliation consists of a mixture of positive and negative frequency modes under Schwarzschild inner and outer foliation.
    \item \textbf{Condition 2:} The infalling positive frequency plane waves inside the collapsing shells are imprinted as Kruskal positive frequency boundary conditions on the past event horizon, which is an effective extrapolation of the collapsing shell metric toward infinite past. 
\end{itemize}

%We address the past horizon as effective here because in the scenario of a collapsing and evaporating finite black hole, the outgoing horizon is not located at infinite past. The Kruskal positive frequency boundary condition is an effective extrapolation to the infinite past.

Why do we need (but possibly, not only need) these two conditions for the purpose of extracting information from a black hole using the generalized Unruh effect? The importance of the first condition is plain to see, as it is the trigger of any generalized Unruh effect discussed in section \ref{sec:introduction}. The mixture of positive and negative frequency modes results in the mixture of creation and annihilation operators thus leading to the informative entanglement between two partitions of the manifold, R+/R- or outside/inside the event horizon. %The uniqueness of the corresponding contracted density matrix on a partition of the manifold to the maximal foliation density matrix, is ensured by the positive + negative frequency-specific combination. Imagine the case where we have two positive modes entangled across two partitions, the contracted ensemble would fall back to the perfectly thermal case by offsetting the vacuum energy with the extra modes. Only when there is mixture between positive and negative frequency modes, such shifting is not working. The annihilator on one partition of the manifold works like a seesaw, raising the counting on the other side of the partition. The event horizon of, for example, a Schwarzschild black hole is an one-sided acceleration analogy of the Rindler horizon in the strong gravity field.

The second condition echoes an existing type of view that the information of a collapsed black hole is imprinted on its event horizon, or some deformation/stretching of that surface \cite{Mathur:2009hf,Susskind:1993if}. They are largely stimulated by the holography, AdS/CFT line of thoughts \cite{Maldacena:1997re,Almheiri:2020cfm}. This work shares many common grounds with them, with one obvious difference. We take a look at the postulates of Susskind, Thorlacius, and Uglum's work \cite{Susskind:1993if} as an example. Their postulate 2 and 3, the semi-classical field equations and discrete energy level of the field living on black hole metric, are adopted in this work as well. However, we deliberately break the unitarity from their postulate 1 in the process of formation and evaporation of the black hole. The mapping we investigate is between the density matrix generated by maximal foliation eigenstates and the pseudo-thermal density matrix subset in the full density matrix set generated by the partitioned spacetime manifold eigenstates. A pure state in the former is mapped into a highly stochastic pseudo-thermal ensemble in the latter, but we will see that the information imprinted on the past horizon should still be retrievable to a certain extent due to the coherent excitation on the stochastic background. 

A chronicle way to understand the relationship between these two conditions is that {\bf Condition 2} resolves an ingoing field mode living on an evolving astrophysical black hole metric into an approximated boundary condition, at the past event horizon which is an extrapolation of collapsing shell metric to the infinite past. {\bf Condition 1} secures the solution of outgoing ensemble under the boundary condition granted by {\bf Condition 2}.

\begin{tcolorbox}[breakable]
Remark: One potential strategy to connect our result to those previous works postulating unitarity is to define the thermal ensemble modulo the temperature as the vacuum state, and any pseudo-thermal ensemble that deviates from the thermal ensemble as excited states. Such a new formalism should redefine the unitarity on the basis of a thermal and pseudo-thermal ensemble instead of the old-fashioned pure states, which could be regarded as a special case of $T=0$. More discussion towards the end of subsection \ref{subsec:isomorphismbhstates}. %Actually, the hierarchy of $ \hat{\rho}_{R+}^{b^{\dagger,l}d^{\dagger,m}b^n d^s}(q)$ is connected by acting corresponding numbers of Rindler creators and annihilators on both sides of a canonical density matrix, multiplied by a normalization factor. %If such formalism is self-consistently feasible, Page's estimation on the information curve \cite{page1993information} can be directly applied to the framework in this paper. %Alternatively, a possible extension to the subsection \ref{subsec:collapsingshell} is to treat of the initial conditions for the collapsing shell modes with detail, thus to allow the analytical calculation of $I_r = S_{\rm therm} + {\rm tr}(\rho_r \log{\rho_r})$ as a function of $S_{\rm therm}$. Again, those outlooks require further study in the definition of information metric under countably infinite Hilbert space density matrix representation.
\end{tcolorbox}

Since the two conditions are fairly standard in previous literature on the black hole information problem topic, it is an option for experienced readers to treat them as standard postulates and jump ahead to subsection \ref{subsec:isomorphismbhstates}. The following subsections \ref{subsec:kruskalpositivefreq} and \ref{subsec:collapsingshell} are dedicated to arguing the validity of the two conditions by painting more details on the original arguments made by \cite{Unruh:1976db}, which demonstrated the feasibility of approximating the collapsing shell information by the positive Kruskal frequency boundary conditions living on the past horizon of the Schwarzschild black hole. The improvements here are that our calculations will be in 4D spacetime and that we will try to amend some minor errors in \cite{Unruh:1976db} along the road.

In any case, once the two conditions above hold, the calculations and arguments for the generalized Unruh effect in Minkowski/Rindler case after equation (\ref{eq:unruhannihilator}) naturally follow. Because the explicit expression of the metric or the Klein-Gordon equation solutions is not used anywhere in the derivation about the quantum mechanical aspect of the problem, after canonical quantization.

Now let us start from the Schwarzschild and Kruskal metrics, using the conventions in \cite{Unruh:1976db} equation (2.20) and (2.25).

\subsection{Mixture of Positive and Negative Frequency Modes Across the Asymptotic Event Horizon}
\label{subsec:kruskalpositivefreq}

\begin{figure*}[htb]
    \centering
    \includegraphics[width=0.8\textwidth]{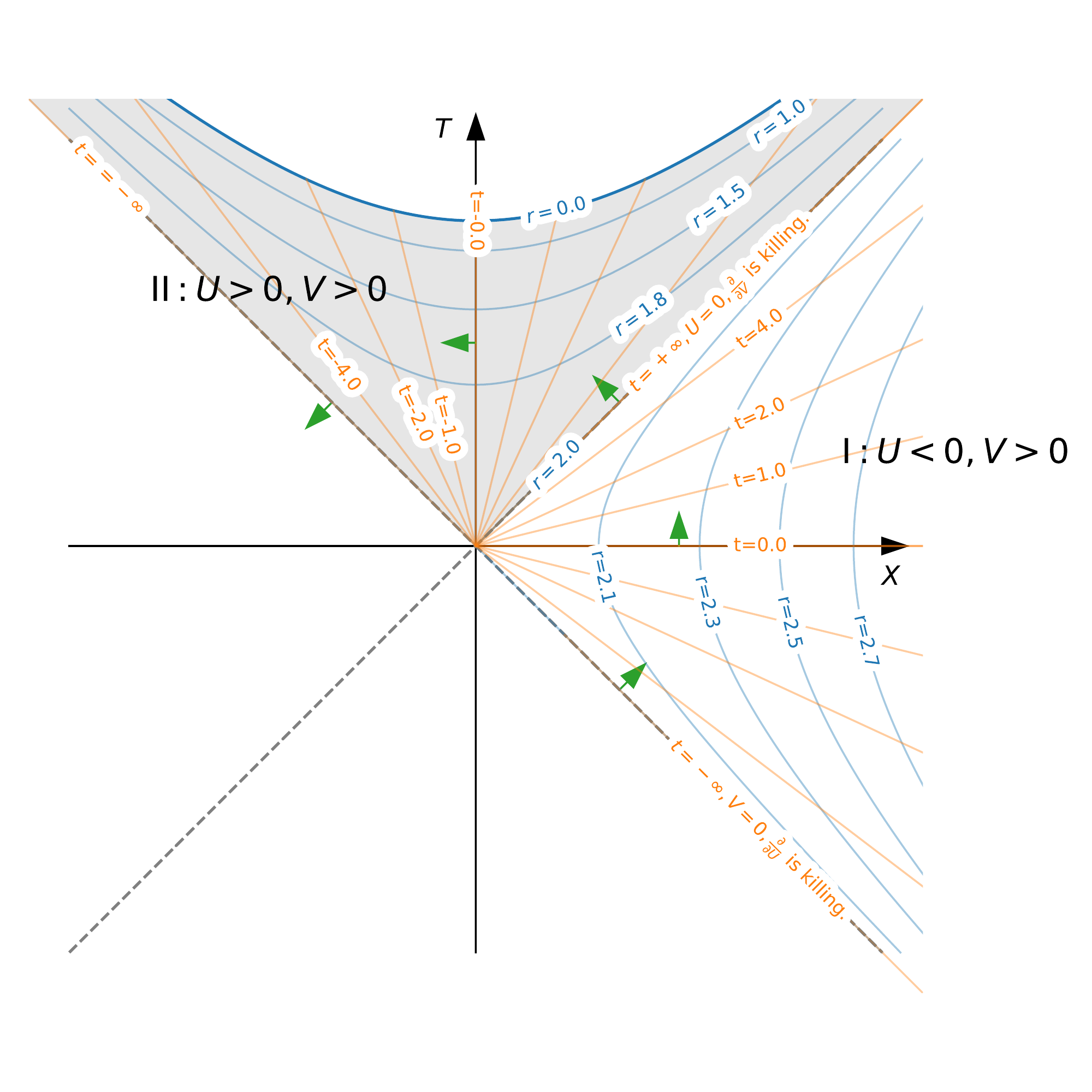}
    \caption{Kruskal metric with Schwarzschild coordinates charted. Green arrows are local timelike killing directions.}
    \label{fig:kruskal}
\end{figure*}

Schwarzschild metric:
\begin{equation}
    ds^2 = (1-2M/r)dt^2-(1-2M/r)^{-1}dr^2-r^2(d\theta^2+\sin^2\theta d\phi^2).
\end{equation}
Under a coordinate transformation, it can be written as Kruskal metric \cite{tHooft:2019xwm}:
\begin{align}
    ds^2 &= 2M \frac{e^{-r/2M}}{r}dU dV - r^2 (d\theta^2 + \sin^2\theta d\phi^2) \label{eq:kruskalmetric}\\
    U & = -4M e^{-\frac{1}{4M}(t-r-2M\ln{(\frac{r}{2M}-1)})}, {\rm for } \ r\geq 2M \label{eq:kruskalcoord0}\\
    U & = 4M e^{-\frac{1}{4M}(t-r-2M\ln{(1-\frac{r}{2M})})}, {\rm for } \ r<2M\\
    V & = 4M e^{\frac{1}{4M}(t+r+2M\ln{(\frac{r}{2M}-1)})}, {\rm for } \ r\geq 2M \\
    V & = 4M e^{\frac{1}{4M}(t+r+2M\ln{(1-\frac{r}{2M})})}, {\rm for } \ r< 2M
    \label{eq:kruskalcoord}
\end{align}
There are multiple ways to write down the Kruskal coordinates, and our expression here is consistent with the Kruskal diagram with Schwarzschild chart as shown in figure \ref{fig:kruskal}.

The relationship:
\begin{align}
    U &= T-X \\
    V & = T+X
\end{align}
is always satisfied, for region I:
\begin{align}
    T &= 4M \sqrt{\frac{r}{2M}-1}e^{r/4M}\sinh{t/4M}\\
    X & = 4M \sqrt{\frac{r}{2M}-1}e^{r/4M}\cosh{t/4M}
\end{align}
and region II:
\begin{align}
    T &= 4M \sqrt{1-\frac{r}{2M}}e^{r/4M}\cosh{t/4M}\\
    X & = 4M \sqrt{1-\frac{r}{2M}}e^{r/4M}\sinh{t/4M}
\end{align}

%The conventional way of investigating, or rather constructing, the manifold of an eternal Kruskal metric is to analytically extend region I and II into the other two quarters to cover the full Cartesian charting of $T-X$ coordinates. However, 
Here we do not extend the original Schwarzschild spacetime, instead, we glue together the past outging horizon and the future outgoing horizon so that the two regions are connected through the surface $t\rightarrow -\infty, r = 2M$. Without losing generality, it can be done by enforcing boundary condition $\phi(U\rightarrow +\infty) = \phi(U\rightarrow -\infty)$. For an astrophysical black hole, regions I and II are sufficient to describe the physics we care about. Such an eternally existing Kruskal manifold exactly expressed by equation (\ref{eq:kruskalmetric}) is only a far-field ideal approximation of the collapsing shell metric that we will consider in the next subsection, so we do not need to worry too much about the singularity at $r=0$ and the artificially exerted periodicity at infinitely far past and future outgoing event horizons. 
\begin{tcolorbox}[breakable]
    Remark: Considering the bounded distance in lightcone coordinate where asymptotic Schwarzschild approximation is valid, as an evaporating black hole only exists in finite time, the outgoing lightcone coordinate boundary conditions at far past and future outgoing horizon should be at finite extremals $\phi(U=U_{\rm max}) = \phi(U=U_{\rm min})$. The finiteness of the astrophysical black hole naturally discretizes the energy levels $\omega$ of the system. The argument here does not remove the assumption of discrete energy levels, instead, it is just a self-consistency check. 
\end{tcolorbox}

Since confining to the regions I and II of the Kruskal metric grants us only one outgoing/past horizon and ingoing/future horizon, we will use each pair of the words interchangeably in the following text.

%\begin{figure}[htb]
%    \centering
%    \includegraphics[width=0.3\textwidth]{schwarzschild1.png}
%    \includegraphics[width=0.3\textwidth]{schwarzschild2.png}
%    \includegraphics[width=0.3\textwidth]{schwarzschild3.png}
%    \caption{An illustration of how we glue the an non-extended Schwarzschild/Kruskal manifold shown in figure \ref{fig:kruskal} to avoid introducing white hole.  }
%    \label{fig:kruskalfoliation}
%\end{figure}

In Kruskal metric, the covariant vector $\frac{\partial}{\partial U}$ is Killing on the past horizon $V=0$, and $\frac{\partial}{\partial V}$ is Killing on the future horizon $U=0$. The proof is briefly shown as follows. Labeling the coordinates $(U,V,\theta,\psi)$, we can check the Killing vector conditions for covariant vectors $\xi^U_{\mu} = (1,0,0,0)$ and $\xi^V_{\mu} = (0,1,0,0)$:
\begin{align}
    \nabla_{\mu}\xi^{U}_{\nu}+\nabla_{\nu}\xi^{U}_{\mu} = -2\Gamma^0_{\mu \nu} = \delta_{\mu 0}\delta_{\nu 0} 2 g^{01}g_{10,r} \frac{\partial r}{\partial U} = 0 
\end{align}
Substituting equations (\ref{eq:kruskalcoord0}) - (\ref{eq:kruskalcoord}) in, we get \cite{enwiki:1072003828}:
\begin{equation}
    r = 2M \left(1+ W_0 \left(\frac{UV}{16M^2 e}\right)\right),
\end{equation}
where $W_0(z)$ is the positive branch of Lambert W function \cite{enwiki:1072003828}. On both future and past horizons, $UV=0$, and the derivative of Lambert W function $W'_0(0)=1$. On past horizon, $V=0$, hence $\frac{\partial r}{\partial U}=0$ and $\frac{\partial}{\partial U}$ is killing, vice versa. 

Hence on the past horizon $\mathcal{H}^-, V=0$, the solutions to the Klein-Gordon equation for a scalar field in Kruskal coordinates are featured by the modes $e^{-i\omega U}$. The positive frequency modes are those analytic and bound in the lower half complex plane of ${\rm Im}(U)<0$. Similarly, the decomposition of Kruskal modes at the future horizon is represented by $e^{-i\omega V}$ due to the killing of the timelike vector field $\frac{\partial}{\partial V}$ in that region. %The symmetry between the features of past and future horizon breaks from this point on because $V$ does not change sign across the future horizon on our non-extended Schwarzschild manifold \footnote{Actually, nor does $V$ change sign on the extended Schwarzschild metric across the future horizon. The sign change of $U,V$ never coincide with when their differentials are killing in that case, so actually the Unruh effect, thus the Hawking radiation, should not happen on the extended version of Schwarzschild manifold.}. This point should be clear in the following mathematical discussion of the positive/negative frequency modes mixing mechanism. As a result, the decision of not extending the Schwarzschild manifold is essential for multiple desired physical features, including the very existence of Unruh effect and the fact that there is only one type of Unruh mode, going out from past horizon. 

After discussing the eigenmodes on the maximal foliation in the Kruskal metric, we now look into the field equation solutions in the Schwarzschild metric. \cite{Unruh:1976db} showed that the semi-classical field solutions in Schwarzschild metric near the horizon are representable by a set of eigenmodes $e^{\pm i \omega t}e^{\pm i \omega r_*}$, where $r_* = r+2M\ln(\frac{r}{2M}-1)$ outside the horizon and $r_* = r+2M\ln(1-\frac{r}{2M})$ inside the horizon. We can denote the inner and outer near-horizon solutions by 

\begin{align}
    \phi_{\omega}^{in}(r^*,t) & = \begin{cases} & e^{2\pi M \omega} e^{\pm i \omega t}e^{\pm i \omega r_*}, {\rm \ \ for }\ r \leq 2M \\
     & 0,  {\rm \ \ for }\ r>2M,
    \end{cases} \\
    \phi_{\omega}^{out}(t,r^*) & = \begin{cases} & e^{-2\pi M \omega} e^{\pm i \omega t}e^{\pm i \omega r_*}, {\rm \ \ for }\ r \geq 2M \\
     & 0,  {\rm \ \ for }\ r<2M,
    \end{cases}
\end{align}
with the same mathematical expression modulo a normalization factor, but vanish on the complementary side. We also \textbf{assume} a normalization factor difference $e^{\pm 2\pi M \omega}$ between inside and outside modes possibly due to the volume difference between two partitions of the 4D sub-spacetime. %This normalization factor does not have concrete mathematical justification since we know formally the plane waves are not normalizable. However we will see that they fit into the puzzle of Kruskal positive frequency modes decomposition into Schwarzschild inner and outer modes. It is also important to remember, that the time-evolving part of the solution is different between $\phi^{in}$ and $\phi^{out}$. 

By observing equations (\ref{eq:kruskalcoord0}) - (\ref{eq:kruskalcoord}), we notice that:
\begin{align}
    \left(\frac{U}{4M}\right)^{i4M\omega} &= e^{-i\omega(t-r_*)},  \ \ \ \ U>0, \label{eq:boundkruskalmode0}\\
    \left(-\frac{U}{4M}\right)^{i4M\omega} &= e^{-i\omega(t-r_*)},  \ \ \ \ U<0
    \label{eq:boundkruskalmode}
\end{align}
On the real axis of $U$. Inside the horizon, $r_*$ is the timelike direction, hence the expression above implies the mixture of the positive frequency $t$ modes outside the horizon and negative frequency $r_*$ modes inside the horizon, with the same frequency amplitude. %If we check the foliation in terms of $T$, as described in the previous paragraphs, in a fixed direction in figure \ref{fig:kruskal}, we also see that the $r_*$ and $t$ slices intersecting the reference $T$ slices are always flowing in the opposite direction.

The left-hand side expression in equations (\ref{eq:boundkruskalmode0}) and (\ref{eq:boundkruskalmode}) satisfies the condition of being bound in the lower half complex plane for $U$. Next, we combine them in a way that secures analyticity along the full $U$ real axis. Notice that the Kruskal-Unruh mode $\phi_{\omega}(U)$ should be continuous across $U=0$, and the first derivative is non-zero (those Kruskal-Unruh modes are supposed to be the free-falling fields that actually travel across the horizon, i.e. non-zero flux there). A relatively opposite sign between $U>0$ and $U<0$ regimes in the above expressions is then necessary. Otherwise, they are out of phase across $U=0$. 

Combining our knowledge about the positive frequency Kruskal $U$ modes and their relation with Schwarzschild eigenmodes, we can write down the decomposition:

\begin{widetext}
\begin{equation}
    \phi_{\omega}(U) \propto  (e^{2\pi M \omega} \phi^{out}_{\omega}(t,r^*) - e^{-2\pi M\omega} \phi^{in,*}_{\omega}(r^*,t)) \propto \begin{cases} & \left(-\frac{U}{4M}\right)^{i4M\omega}, U<0 \\
    & - \left(\frac{U}{4M}\right)^{i4M\omega} , U>0
    
    \end{cases}
    \label{eq:kruskalunruh}
\end{equation}

\begin{figure}[htb]
    \centering
    \includegraphics[width=0.7\textwidth]{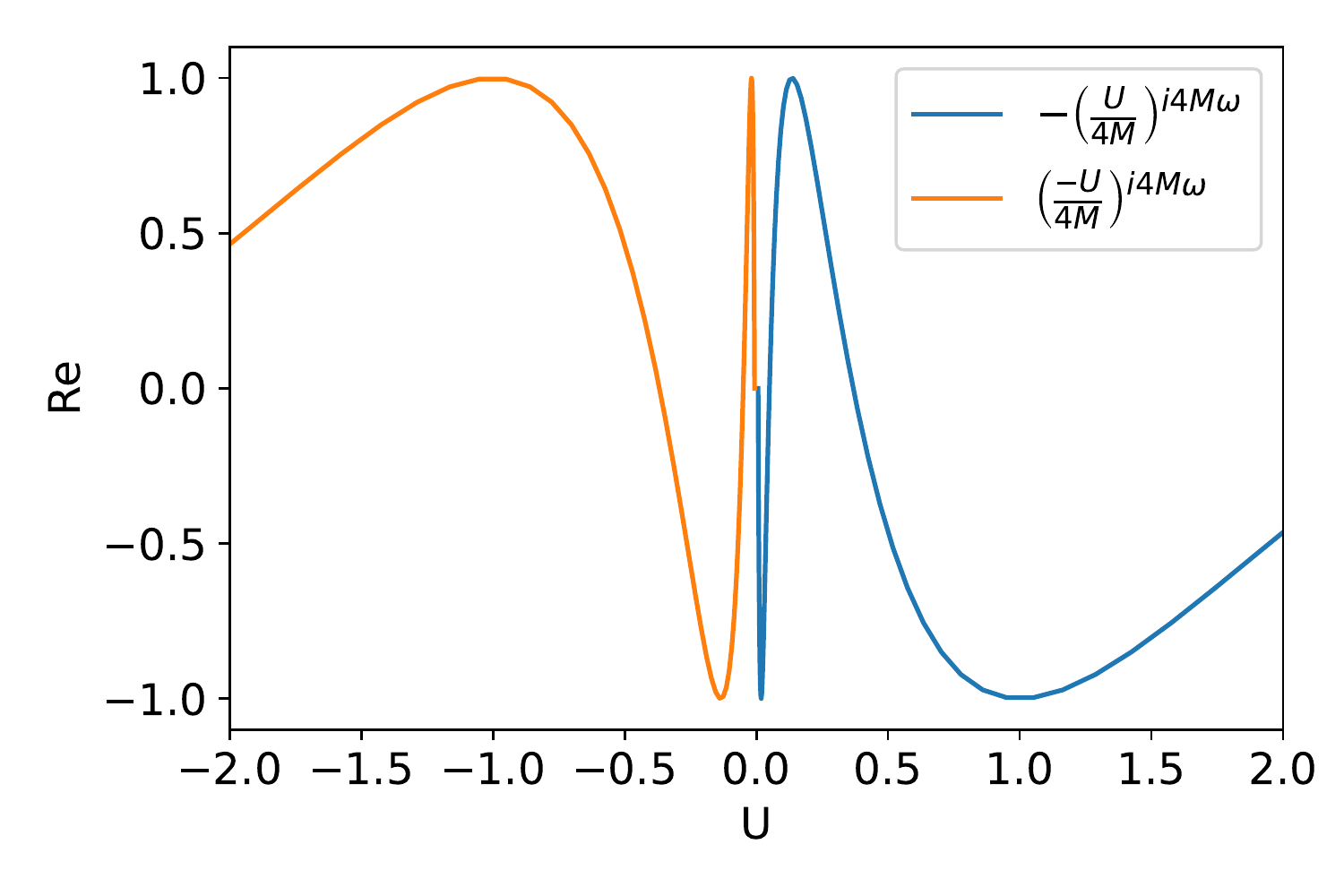}
    \caption{A Kruskal Unruh mode expressed in equation (\ref{eq:kruskalunruh}) modulo normalization factor. $\omega=\frac{5\pi}{4M \Delta \log(|U|)}$.}
    \label{fig:kruskalunruhmode}
\end{figure}
\end{widetext}
when constrained to the real axis of $U$, as illustrated in figure \ref{fig:kruskalunruhmode}. $\phi^{out/in}_{\omega}(t,r^*)$ are the positive frequency modes corresponding to each of their own time-like coordinates. $\phi^{out}_{\omega}(t,r^*)$ is outgoing from the horizon, and $\phi^{in}_{\omega}(r^*,t)$ is also departing from the horizon on the other side \footnote{Falling inward might be a confusing way to put it, although people are used to saying so. Let us not be fooled by the letter label, and remember that $r$ is timelike inside the horizon. $t$ increase with increasing $r^*$ (along time-flow) is the most accurate description for mode $\phi^{in}_{\omega}(r^*,t)$.}.

 %Gazing the function above extended to the full complex plane, we can see that the fluctuation due to the phase angle of $U$ is canceled out by the two parts of the expression, which could be weighted analytically by their complex plane phases to extend the analyticity to the lower half of the plane. We label the inner mode $\phi^{in,*}(\omega)$ with a complex conjugate because as shown in equation \ref{eq:boundkruskalmode} it is a negative frequency mode for $r_*$. 

By observing the mathematical expression of the Unruh modes in equation (\ref{eq:kruskalunruh}), we notice that the Unruh modes are basically the plane waves in $-\log{U}$ space, instead of the normal plane waves in $U$ space ($e^{-i\omega U}$). And it is these Unruh mode frequencies in $-\log{U}$ space, not the plane waves with respect to the Kruskal lightcone coordinate $U$, that determines the Schwarzschild plane wave frequencies corresponding to the $t$ foliation infinitely far way from the horizon. This point can be clearly seen from the equation (\ref{eq:boundkruskalmode}), where the equality is connecting a Kruskal-Unruh mode on the left-hand side and a Schwarzschild plane wave mode outside/inside the horizon on the right-hand side. %The Fourier transformation in $\log{x}$ space is exponentially blue-shifted from the Fourier transformation in $x$ space for a fraction of power, which should hit the UV cutoff much faster than plane waves.  %However, we do not to worry about the potential divergence due to the blue-shift in collapsing black hole case, as we will see in the next section that the plane waves inside the shell evolve directly into the Unruh modes near horizon. % This blue-shift due to the transformation into log space Fourier modes counter-acts with the well-known exponential red-shift of a mode falling into the black hole, thus leaving finite frequency spectrum features to the observer far away, measuring the evaporation radiation of a black hole.

    Before we depart from this section, we would like to stress again that the neat eternal black hole considered here is only an asymptotic approximation of the collapsing shell metric representing an astrophysical black hole, in its most condensed limit. It is well known that an observer outside a Schwarzschild radius will never witness the exact accretion of an ingoing particle onto the event horizon, in the sense that such an event is not on any of the equal-time Cauchy surfaces of the outside observer. On the other hand, Schwarzschild metric is always a valid local approximation of the metric near an observer, outside a sphere enclosing a bulk of mass. From this perspective of view, all massive objects are gravitationally thermalized and evaporating. The formation of an event horizon exactly at $r_s$ corresponding to the enclosed mass is never accomplished before the evaporation of the asymptotic black hole, from the perspective of view of any observer outside the sphere. In other words, quite intuitively, singularity does not emerge in the collapse of an astrophysical black hole, no matter how seemingly close it is to an exact event horizon; The existence of a global event horizon is equivalent to the existence of a singularity, or a defect of the spacetime manifold from the first place. 
    
    On this note, the question `what if a particle falls into the black hole event horizon' is actually a problematic, even if not completely wrong question to be asked by an observer outside the horizon, because it is incompatible with the locality of physics rules. In general relativity, the physics rules, represented by some equations of motion for the fields, are the same in different reference frames. Such consistency is known to be broken when the measurements from different reference frames are used in a single set of equations \footnote{For example, Einstein equations are bound to be broken if we use the Ricci curvature expressed Newtonian gauge and energy-momentum tensor expressed in synchronous gauge.}. %If we fully respect the locality of an observer, and presumably we do not start with a defect on the spacetime manifold, no particle can `fall into' a black hole event horizon from the perspective of an outside observer. More over, in general relativity it is forbidden to carry out algebra using the quantity measured by an outside observer and the quantity measured by a Kruskal time-flow traveler (the observer that actually travels across a horizon, but instead of a global-event one, a local-particle one that is specific to another observer). 
    We often confusingly find ourselves in the paradox of infinite redshift or diverging energy-momentum density at the black hole horizon, or information loss after reaching the exact event horizon. It is likely only because we were carrying out illegal maths that simultaneously admit the measurements by different observers. An observer that has appeared outside the horizon far past, at the exact event horizon, and outside the horizon far future does not exist.

{\bf A reasonable speculation is that an eternal black hole with an event horizon can only exist as a conceptual approximation, not in the real physical world}. It is in the stance of an infinitely extending metal plate in the electrodynamic problems when we concern about physics instead of maths. %The most subtle ignorance on one of the corner stones of the general relativity, the locality or the general principle of relativity, will lead one into a dead end. 

\begin{tcolorbox}[breakable]
Remark: In two special cases, \{Minkowski $\leftrightarrow$ Rindler\} and \{Schwarzschild $\leftrightarrow$ Kruskal\} non-inertial frame pairs, we found the opposite time-flow phenomena on a shared Cauchy surface in those diffeomorphism connected metric pairs of the same (+,-,-,-) pseudo-Riemannian manifold. Inertial frame pairs, connected by special diffeomorphisms generated by Poincar\'{e} groups, do not have opposite time-flows. To be more specific, in \{Minkowski $\leftrightarrow$ Rindler\} case, for a flat manifold with no singularity, as shown in figure \ref{fig:foliation}, on $t=\tau=0$ Cauchy surface $\left( \frac{\partial}{\partial t} ,\frac{\partial}{\partial \tau}\right)|_{R+} >0$ while $\left( \frac{\partial}{\partial t}, \frac{\partial}{\partial \tau}\right)|_{R-} <0$. Similarly, in \{Schwarzschild $\leftrightarrow$ Kruskal\} case for a flat manifold with a singularity, on the Cauchy surface, the past horizon $\mathcal{H}^-$, that is shared by both foliation structures, $\left( \frac{\partial}{\partial U} ,\frac{\partial}{\partial t}\right)|_{out} >0$ while $\left( \frac{\partial}{\partial U}, \frac{\partial}{\partial r_*}\right)|_{in} <0$. Notice that the time-flow direction is not specified by being labeled by some letter related to ``t", but is determined as the {\bf killing direction with non-negative signature}. We suspect that this kind of spacetime foliation structure is what truly underlies the Unruh effects, yet the differential manifold knowledge required to unveil the fundamentals of such phenomena in general non-inertial frame pairs is beyond our reach in this paper.

The outcome of the above opposite foliation structure after quantization manifests as follows. In non-inertial frame problems, the Hilbert spaces of advanced (one-way) time-evolving states do not coincide entirely in two ways of foliation. When we write down the exact equality between the states in two relatively accelerating frames, formally it is done in a more inclusive Hilbert space that incorporates both advanced and retarded field solutions, or states, after quantization. Those retarded states vanish soon enough into the investigation under the lower-bounded algebra we defined, $a | 0\rangle=0$.

\end{tcolorbox}

\subsection{Ingoing Modes Inside the Shell of a Collapsing Shell Metric}
\begin{figure}[htb]
    \centering
    \includegraphics[width=0.7\textwidth]{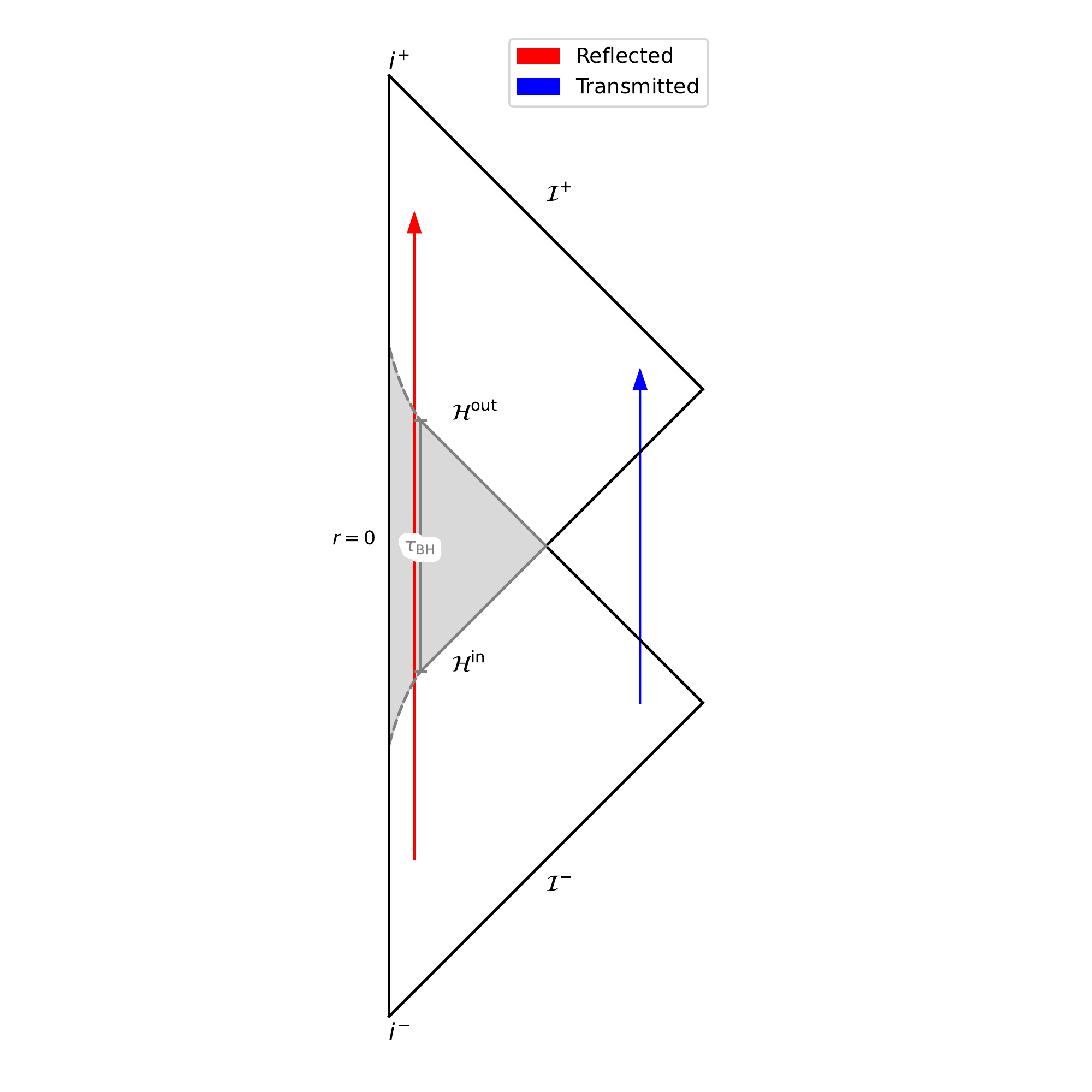}
    \caption{A finite-lifetime black hole. The blue arrow denotes the transmitted scalar field modes that have negligible amplitude near the horizon; They are the matters floating around and never get absorbed into the black hole. The red arrow denotes the modes reflected by the event horizon. }
    \label{fig:finitebh}
\end{figure}

In this section, we will move on to the discussion of astrophysical black holes that form and evaporate in a finite lifetime. The goal is to verify the \textbf{Condition 2}, which helps argue that the infalling positive frequency modes can be represented by the Kruskal outgoing horizon positive frequency modes. The Penrose diagram of such a finite-lifetime black hole is illustrated in figure \ref{fig:finitebh}.

\label{subsec:collapsingshell}

%It has been discussed in 2D by \cite{Unruh:1976db} that "the $\xi$ (Kruskal) definition of positive frequency does correspond to replacing the collapse process by boundary conditions on $H^{-}$". This section will try to demonstrate how the collapsing process information gets printed on the past, shrinking horizon of an evaporating black hole, in the spacetime we live in reality, 4D spacetime. The key of the argument lies in the centrifugal barrier, which is exerted by W. Unruh in 2D derivation, and comes naturally in with the diverging potential at $r=0$ for a 4D gravitational system.

The scalar field modes that start evolving from the past Cauchy surface $i^-$ for $\mu=0$ or $\mathcal{I}^-$ for $\mu\neq 0$ can be classified into two types, one never travels near the black hole horizon during the finite lifetime of an astrophysical black hole, the other has substantial amplitude near the horizon ($\phi(r_s<r<r_s+\delta)>\epsilon$, for $t$ during $\tau_{\rm BH}$). We call them transmitted and reflected modes respectively. The transmitted modes are those matters floating around the black hole and never get close enough to them, thus are not plagued by the black hole information problem; They unitarily evolve from the past before black hole formation to the future after black hole evaporation. The reflected modes travel through the horizon, and they are the main investigation objects in any black hole information problems.

The four dimensional collapsing shell metric, generalized from the 2D collapse from \cite{Unruh:1976db}, can be written as:

\begin{equation}
    ds^2 = \begin{cases} & d\tau^2 - dr^2, \ \ \ \ r<\hat{R}(\tau)\\
    &\left( 1- \frac{2M}{r}\right)dt^2 - \frac{dr^2}{1-2M/r}, \ \ \ \ r>\hat{R}(\tau).
    \end{cases}
    \label{eq:collapsingshell}
\end{equation}
The shell radius is given by:
\begin{equation}
    \hat{R}(\tau) = \begin{cases} R_0, \ \ \ \ \tau<0\\
    R_0-\nu\tau, \ \ \ \ \tau>0.
    \end{cases}
\end{equation}
The collapsing shell approximately describes such a physical system, where a shell of matter shrinks its size at velocity $\nu$. The collapsing of a bulky distribution of the matter, or the additional matter accreting onto an existing black hole could be decomposed by layers of collapsing shells with the in-shell metric modified from the completely flat Minkowski one. We do not consider the back-reaction of the field on the metric. Hence, strictly speaking, the retrieval of the information calculation in the following subsection based on the current subsection is only for the perturbative part of the whole chunk of collapsed mass. A complete, non-linear level analysis is not available until the puzzle of curved spacetime and quantum field theory unification is fully resolved, i.e. the inclusive formulation of gravitational and particle interactions.  %We set this motion up by hand, but in principle it is achievable through the gravitational interaction in our massive scalar field toy model. 

It is assumed that $\nu$ is fast enough $1-\nu<\frac{4M}{R_0+2M}<<1$ for simplicity in the later calculations. Using light cone coordinates:
\begin{align}
    & \bar{U} = \tau-r+R_0, \ \ \ \ \bar{V} = \tau+r-R_0 \\
    & \bar{u} = t-r_* + R_{0*},\ \ \ \ \bar{v} = t+r_* - R_{0*}
\end{align}
we can rewrite the collapsing shell metric as:

\begin{equation}
ds^2 = \begin{cases} & d\bar{U}d\bar{V} - r^2 d\Omega^2, {\rm \ \ \ \ inside\ the\ shell}\\
& (1-2M/r)d\bar{u}d\bar{v}-r^2d\Omega^2 {\rm \ \ \ \ outside\ the\ shell}

\end{cases}
\end{equation}

$\bar{u},\bar{v}, \bar{U}, \bar{V}$ are related by the condition that the $ds$ on the shell match each other expressed by in-shell or out-shell coordinates \cite{Unruh:1976db}:

\begin{equation}
    \frac{du}{dU} = \begin{cases} (1-\frac{2M}{R_0})^{-\frac{1}{2}}, &\ \bar{u}, \bar{U} <0 \\
     \frac{\hat{R}(\frac{\bar{U}}{1+\nu})}{(1+\nu)(\hat{R}(\frac{\bar{U}}{1+\nu})-2M)} &\left[ \nu+ \left( 1- \frac{2M(1-\nu^2)}{\hat{R}(\frac{\bar{U}}{1+\nu})}\right)^{1/2}\right], \\ &\ \bar{u}, \bar{U} >0
    
    \end{cases}
    \label{eq:dudu}
\end{equation}

\begin{equation}
    \frac{dv}{dV} = \begin{cases}  (1-\frac{2M}{R_0})^{-\frac{1}{2}}, & \ \bar{v}, \bar{V} <0 \\
     \frac{\hat{R}(\frac{\bar{V}}{1+\nu})}{(1+\nu)(\hat{R}(\frac{\bar{V}}{1+\nu})-2M)} & \left[ [1-\frac{2M(1-\nu^2)}{\hat{R}(\frac{\bar{V}}{1+\nu})}]^{1/2}-\nu\right], \\ &\  \bar{v}, \bar{V} >0
    
    \end{cases}
    \label{eq:dvdv}
\end{equation}

We put a bar on these lightcone coordinates to distinguish them from the Kruskal metric ones in the previous subsection.

The Klein-Gordon equation for the massive scalar in this collapsing shell coordinate becomes:
\begin{widetext}
\begin{align}
    r^2 \frac{\partial}{\partial \bar{U}}\frac{\partial}{\partial \bar{V}} \phi +r\left( \frac{\partial r}{\partial \bar{U}} \frac{\partial \phi}{\partial \bar{V}}+\frac{\partial r}{\partial \bar{V}} \frac{\partial \phi}{\partial \bar{U}}\right)-\frac{1}{\sin{\theta}}\frac{\partial}{\partial \theta} (\sin{\theta}\frac{\partial}{\partial \theta} \phi)-\frac{1}{\sin^2 \theta} \frac{\partial^2}{\partial^2 \psi} \phi-\mu^2 r^2 \phi =0 \ \ \ \ &{\rm In\ shell} \\
     r^2 (1-\frac{2M}{r})^{-1}\frac{\partial}{\partial \bar{u}}\frac{\partial}{\partial \bar{v}} \phi 
    +r(1-\frac{2M}{r})^{-1}\left( \frac{\partial r}{\partial \bar{u}} \frac{\partial \phi}{\partial \bar{v}}+\frac{\partial r}{\partial \bar{v}} \frac{\partial \phi}{\partial \bar{u}}\right) -\frac{1}{\sin{\theta}}\frac{\partial}{\partial \theta} (\sin{\theta}\frac{\partial}{\partial \theta} \phi)-\frac{1}{\sin^2 \theta} \frac{\partial^2}{\partial^2 \psi} \phi -\mu^2 r^2 \phi =0 \ \ \ \ &{\rm Out\ shell}
\end{align}

The angular part of the scalar field can be solved by spherical harmonics. The remaining light-cone coordinate dependent part of the scalar field $\phi(U,V)$ resolves into:

\begin{align}
    r^2 \frac{\partial}{\partial \bar{U}}\frac{\partial}{\partial \bar{V}} \phi +r \frac{\partial \phi}{\partial r}+ \ell (\ell+1) \phi -\mu^2 r^2 \phi=0 \ \ \ \ & {\rm In\ shell} \label{eq:kginshell}\\
     r^2 (1-\frac{2M}{r})^{-1}\frac{\partial}{\partial \bar{u}}\frac{\partial}{\partial \bar{v}} \phi 
    +r(1-\frac{2M}{r})^{-1}\left( \frac{\partial r}{\partial \bar{u}} \frac{\partial \phi}{\partial \bar{v}}+\frac{\partial r}{\partial \bar{v}} \frac{\partial \phi}{\partial \bar{u}}\right)  + \ell (\ell+1) \phi-\mu^2 r^2 \phi=0 \ \ \ \ &{\rm Out\ shell}
\end{align}

\end{widetext}

%For both in-shell and out-shell equation of motions, the angular momentum term $l(l+1) \phi$ should be negligible in most of the area, except for when $r \rightarrow 0$. This is exactly the origin of the centrifugal barrier that was implemented by hand in \cite{Unruh:1976db} for the 2D toy model. In general, $\ell$s of the collapsing modes are not exactly zero, and that leads to the $\phi(r=0)=0$ solution for the in-shell equation of motion.

Even for a non-spinning Schwarzschild black hole, the scalar field solution in general does not have $l$ exactly equal to zero. The non-exact-zero of the angular momentum for most of the scalar field modes forms a centrifugal barrier that leads to the $\phi(r=0)=0$ solution for the in-shell equation of motion (\ref{eq:kginshell}).

We are interested in the $\phi$ modes behavior near the horizon when the shell approaches the horizon. This condition specifies the final stage of the collapse $\tau \rightarrow \frac{R_0-2M}{\nu}, t \rightarrow + \infty$. The approaching to infinity of the out-shell time $t$ when $\hat{R}(\tau)$ indefinitely approaches $2M$ demonstrates that for an observer outside the shell ($\frac{dt}{d\tau}$ diverges $\sim \frac{1}{\hat{R}-2M}$), the black hole event horizon only forms asymptotically. Outside the shell, near the horizon, as $(1-2M/r)^{-1}$ diverges and $\frac{\partial r}{\partial \bar{u}} \sim \frac{\partial r}{\partial \bar{v}} \sim 0$, the first term of the equation of motion dominates, and the EoM simplifies to:
\begin{equation}
    \frac{\partial}{\partial \bar{u}}\frac{\partial}{\partial \bar{v}} \phi =0, \ \ \ \ {\rm Out\ shell,}\ t \rightarrow + \infty, r\rightarrow 2M
    \label{eq:eomoutshell}
\end{equation}

%Inside the shell, the equation of motion throughout the collapsing history is also dominated by the first term:
%\begin{equation}
%    \frac{\partial}{\partial \bar{U}}\frac{\partial}{\partial \bar{V}} \phi =0, \ \ \ \ {\rm In\ shell,} 
%    \label{eq:eominshell}
%\end{equation}

%So far, there are two assumptions for the black hole in our calculation are made here: it is non-Kerr, to ensure small angular momentum $\ell (\ell+1)$, and it has decent mass $2M >> m_{\rm pl}$. They are both reasonable assumptions for our investigated object, shell collapsed Schwarzschild black hole. Another assumption is that for the $\phi$ scalar field inside the shell, the radial distribution of $\phi$ is roughly flat. 

%Given equations \ref{eq:eominshell} and \ref{eq:eomoutshell}, the in-shell and out-shell equation of motion solutions are both of the form:

This equation is solved by:
\begin{equation}
    \phi(\bar{u},\bar{v}) = f(\bar{v})+g(\bar{u}) , \ \ \ \ {\rm Out\ shell,}\ t \rightarrow + \infty, r\rightarrow 2M.%\\
    %\phi(\bar{U},\bar{V}) &= F(\bar{V})+G(\bar{U}) , \ \ \ \ {\rm In\ shell,}\   \hat{R}(\tau)> r >> m_{\rm pl}
\end{equation}

We assume the continuity of the scalar field across the shell, and we know that according to equations (\ref{eq:dudu}) and (\ref{eq:dvdv}), $\bar{u}=\hat{U}(\bar{U})$ and $\bar{v}=\hat{V}(\bar{V})$, where $\hat{\ }$ denotes a single variable function. Namely out/in-going light-cone coordinates do not mix from the $ds$ matching at the shell. Thus we can adopt the ansatz for in-shell EoM solution near the horizon with the same form:

\begin{equation}
    \phi(\bar{U},\bar{V}) = F(\bar{V})+G(\bar{U}) , \ \ \ \ {\rm In\ shell, \ \tau \rightarrow \frac{R_0-2M}{\nu}, r\rightarrow 2M}
\end{equation}

%Note that such form of solutions are only applicable to where the approximation conditions of the equation of motions are satisfied. Beyond these regions, $u,v$ dependence of the field could be mixed. However, what we got should be sufficient to our specific interest in the infalling in-shell modes and the out-shell modes that they evolve into, as we will see in the following paragraphs.

%We only need one set of light cone coordinates $(\bar{u},\bar{v})$ or $(\bar{U}, \bar{V})$ to specify the spacetime, and the switching from one to another across the shell is just for the simplicity. The transformation between two types of light cone coordinates is set by the equality of $ds$ expression in two systems on the shell. According to equations (1.6) and (1.7), such transformation has the feature that $\bar{u}/\bar{v}$ only depends on $\bar{U}/\bar{V}$, hence in the region $\ t \rightarrow + \infty, r\rightarrow 2M$, the final collapse stage, the continuum of scalar field gives raise to relationship:
The continuum of $\phi$ across the shell near the horizon further implies:

\begin{align}
    f(\bar{v}) = F(\hat{V}(\bar{v})), g(\bar{u}) = G(\hat{U}(\bar{u})), 
    \label{eq:uvmodestransfer}
\end{align}
The $\bar{u}(\bar{U})$ and $\bar{v}(\bar{V})$ dependent functions are outgoing and ingoing modes respectively.

%Now that the in-going and outgoing modes are separable, we can focus on investigating the outgoing part $g(\bar{u}) = G(\hat{U}(\bar{u}))$. 
%Back to the problem we are trying to solve in this subsection, what kind of mode we have leaving the past horizon of a Kruskal metric. For the mode living on the past horizon of a Kruskal metric, i.e. the continuum to the final collapse stage of the collapsing shell metric, we are specifically interested in the $u$ dependent term of $\phi(\bar{u}) = f(\bar{v})+g(\bar{u})$, because we know that the $v-$ light cone coordinate saturates to a constant, so do its related mode $f(\bar{v})$. 
The centrifugal barrier effect discussed in the previous paragraphs tells us $\phi(r=0) = F(\bar{V})+G(\bar{U}) = 0$. Strictly speaking, the separable ansatz is only justified near $r=2M$, thus the possible oscillatory motion between $ 0<r<2M$ might modify this condition by a phase shift, but it is unimportant for the following derivations. Inside the shell, along $r=0$ worldline, we have the relationship $U-V = 2 R_0$. Thus the 1D functions $F(x)$ and $G(x)$ has the following relationship:
\begin{equation}
    G(x) = -F(x-2R_0).
    \label{eq:centrifugalbarrier}
\end{equation}
Combining equation (\ref{eq:uvmodestransfer}) and equation (\ref{eq:centrifugalbarrier}), we get
\begin{align}
    g(\bar{u}) = -F(\hat{U}(\bar{u})-2R_0)
\end{align}

Remember that the metric, thus the equation of motion inside the shell is simply of flat spacetime. Hence the collapsing modes whose wavefronts evolve with decreasing $r$ as $\tau$ increases could be represented by the positive frequency modes $F(\bar{V}) \sim e^{-i\Omega \bar{V}}$ just inside the shell \footnote{Whether $\mu = 0$ or not is unimportant here; The positive frequency modes solution for massive or massless particles have the same form. %Again, we should not confuse the particle picture with the field solution modes picture.
}. These modes, when leaving the past horizon of Kruskal spacetime, are reflected into $g(\bar{u})$ in the form:
\begin{align}
    g(\bar{u}) \sim -e^{-i\Omega (\hat{U}(\bar{u})-2R_0)}
\end{align}

A physics interpretation is that the infalling mode is bounced back by the centrifugal barrier at $r=0$ (possibly with a phase shift).

We use $\sim$ symbol to waive the careful treatment of normalization factors, only focusing on the spacetime coordinates dependence of a mode.

By integrating equation (\ref{eq:dudu}), at $O(1-\nu)$ order we have \cite{Unruh:1976db}:
\begin{equation}
\hat{u}(\bar{U}) = \begin{cases}  (1-2M/R_0)^{-1/2}\bar{U},& \ \ \ \ \bar{U}<0\\
  -4M\ln{(1-\nu \bar{U} /[(1+\nu)(R_0-2M)])}& \\\ \ \ \   + \bar{U} + O(1-\nu),& \ \ \ \ \bar{U}>0 
    \end{cases}
\end{equation}
Here, $\nu$ is the collapsing rate, which we assumed to be  close to $ 1$ thus a rapid collapse. In the final stage of collapse near horizon, we have $\tau \rightarrow \frac{R_0-2M}{\nu}, r\rightarrow 2M$, hence $\bar{U} \rightarrow  (R_0-2M)(\nu+1)/\nu$. %Even after the black hole starts to evaporate, immediately inside the shrinking event horizon we still have the spacetime charted in terms of the in-shell collapsing shell metric $\tau, r$, and in this immediate in-shell region we always have $\bar{U} \rightarrow  (R_0-2M)(\nu+1)/\nu + \mathcal{O}(1-\nu)$. 
In this region, $\bar{U}$ is positive, and the log term dominates over the linear term. Thus at the past horizon of Kruskal metric, we have 
\begin{equation}
    \hat{U}(\bar{u}) \approx \frac{1+\nu}{\nu}(R_0-2M)[1-e^{-\bar{u}/(4M)}]
\end{equation}

Now we proceed to connect the collapsing shell picture with the black hole picture discussed in the previous subsection. Imagine that we have an observer sitting at $r_o>>R_0$. The scalar field perturbative part has negligible back-reaction on the metric, so this observer does not need to know the exact dynamics inside the shell to figure out the evolution of the scalar field local at $r_o$. An eternal black hole and a collapsing shell have the same effective metric locally near $r_o$%and the observer is ignorant to the `true', global specetime metric \footnote{Of course, in reality, the observer cannot miss a Supernovae event if she or he is monitoring that direction closely, but we assume a (time domain) filter magically shielded these irrelevant signals that has never truly `collapsed' out.}
. Thus the evolution of the scalar field near the region of a distant observer in a collapsing shell metric is representable by the evolution of the scalar field in an eternal black hole metric. The dynamics of the scalar field happening inside the shell are imprinted as the boundary conditions at the asymptotic black hole event horizon extrapolated to $t\rightarrow -\infty$.

Matching the $r$ and $t$ coordinates of the collapsing shell metric and the imaginary eternal Schwarzschild black hole metric, the eternal black hole Kruskal metric light cone coordinate $U$ defined in equation (\ref{eq:kruskalcoord}) is related to the collapsing shell light cone coordinate outside the shell by $U = -e^{-(\bar{u}-R_0-2M\ln(R_0-2M))/4M}$. Such matching should be asymptotically exact out-shell in the far future, towards the final stage of the collapse.

Thus we see an in-shell ingoing mode $\phi(\bar{V}) \sim e^{-i\Omega \bar{V}}$ eventually evolves into
\begin{equation}
    \phi(U) \sim e^{-i\Omega \xi U}
\end{equation}
on the asymptotic event horizon, which is effectively a boundary condition for far observers. $\xi$ is a redshift factor $\xi = \frac{1+\nu}{\nu}\sqrt{R_0-2M} e^{-\frac{R_0}{4M}}$, always positive. Thus each positive frequency $e^{-i\Omega \bar{V}}$ mode is mapped into a positive frequency $e^{-i\Omega' U}$ mode after a long time of evolution, and they each could be decomposed into the Schwarzschild positive and negative frequency modes mixture outside and inside the asymptotic event horizon.

\begin{tcolorbox}[breakable]
Remark: The approach in this note could be regarded as a realization of the Horowitz-Maldacena conjecture \cite{horowitz2004black} in some sense. There, they assumed the infalling state is entangled with the vacuum (perfectly thermal after tracing out the external part) Unruh state inside the horizon. This note shows that this extra entanglement is redundant -- the Unruh state itself can be richer than a pure vacuum and sets up the boundary conditions. %However, it seems also suggest a not-so-clean conclusion that is probably hated by many theorists: a black hole embedded in the exact vacuum (thermal bath) cannot be formed from collapsing. When we talk about the black hole collapsed from a star, the boundary conditions on the past horizon of that black hole is only natural to be expected far from vacuum. 

%However, the argument above leads to another weak point of this work. We did not calculate the energy-momentum tensor from fields, and only manually assumed it to be at its maximum flow at the shell and horizon by adopting the collapsing shell metric in equation \ref{eq:collapsingshell}. It goes back to the traditional question about the back reaction of the field on the curved spacetime in the QFT curved spacetime problem. Is our collapsing shell setup with non-zero energy-momentum flow restrained on the shell a reasonable approximation, while black hole metric is a solution to the vanishing energy-momentum flow (namely, the black hole metric is always just an asymptotic and never exact in the reality)? We leave it as an open question here.
\end{tcolorbox}

%With all the reasoning stated above, this section is comparably weaker (less concrete and quantitative) than other sections of this note. For example, I did not formally illustrate how $H^-$ and the future horizon of the collapsing shell metric are glued to each other, which would cast some doubt on the redshift factor $\xi$. The topic in this section, i.e. compressing the collapse process of a black hole onto the boundary condition of the past horizon of a (approximated, in reality) eternal black hole metric, is fairly poorly discussed in the existing literature, to the best of my knowledge. A more solid argument for the positivity of the frequencies representing the collapsing process is left to future works.

\subsection{The One-to-One Mapping Between Infalling Physical States and Outside-horizon Pseudo-Thermal Physical States}
\label{subsec:isomorphismbhstates}
We have seen in section \ref{sec:beyondsingle} that, when there are two types of Unruh modes generated simultaneously in a spacetime, it was quite difficult to identify the invertible mapping between the density matrix space generated by the Unruh Fock space and the $\hat{\rho}^{lmns}$ vectors. Because when we have both the type I and the type II Unruh modes, even for a basis Fock state, there are multiple $\hat{\rho}^{lmns}$ terms due to the necessity to commute $b^{\dagger}$ and $b$ coming from $u^{\dagger}$ and $v^{\dagger}$. Things are more approachable in the Kruskal-Schwarzschild duality where we only have one type of Unruh mode moving outwards. In this section, we will try to identify the invertible mapping from an arbitrary density matrix in the Kruskal-Unruh eigenmodes Fock space to the $\hat{\rho}^{lmns}$-basis-spanned subspace of the density matrices generated by the Schwarzschild outside-horizon eigenmodes Fock space. We will be focusing on the states for a single momentum $q$ in the rest of this section, as the proof of the invertible mapping in this section can be trivially generalized given the commutativeness of the operators with different momenta. %Namely, we will go through the contraction of the states with $n(q)$ particles $q$, and the superposition of the states with different numbers of $n(q)$ here.

Canonically quantizing the equation (\ref{eq:kruskalunruh}), we get the expression of the Unruh creator in terms of the outside creator and the inside annihilator:
\begin{equation}
    u^{\dagger}_q = \frac{1}{\sqrt{2\sinh{4\pi M \omega}}} (e^{2 \pi M \omega}b^{\dagger}_q - e^{-2\pi M \omega}d_q)
\end{equation}

%We have changed some details of the notations, for example $\tilde{\omega} \rightarrow \omega$, since the Kruskal plane wave frequancies are not so useful for our discussions. 
%Note that in this section $\omega$ are the Unruh mode frequencies while $\Omega$ are the plane wave frequencies.
$u^{\dagger}$ is the Unruh creator which generates a positive frequency mode on the past horizon of Kruskal spacetime. $b^{\dagger}, d^{\dagger}$ are the Schwarzschild creators outside and inside the horizon. As usual, they each do nothing on the complementary side.%, which is a nature determined by the classical field solution they are quantized from. 

A most general Hermitian density matrix built on the Unruh Fock space with a fixed momentum $\vec{q}_0,\omega_0$ is given by:
\begin{equation}
   \rho^{K} =  \sum_{\alpha_1, \alpha_2} A_{\alpha_1 \alpha_2}| \alpha_1 \rangle \langle \alpha_2|,
\end{equation}
where $A^*_{\alpha_1 \alpha_2} = A_{\alpha_2 \alpha_1}$, ${\rm tr}(A_{\alpha_1 \alpha_2})=1$ and the matrix is positive semi-definite. $\alpha_1, \alpha_2$ are the non-negative integers, and $| \alpha \rangle$ is the normalized state with $\alpha$ Unruh particles of momentum $q_0$:
\begin{equation}
    | \alpha \rangle = \frac{1}{\sqrt{\alpha!}} u^{\dagger,\alpha} | 0 \rangle_K = \frac{1}{\sqrt{\alpha!}}e^{-2\pi M \alpha \omega}(2\sinh{4\pi M \omega})^{\alpha/2} b^{\dagger,\alpha} \hat{S} | 0 \rangle_S
    \label{eq:normalizedunruhmode}
\end{equation}
where we used the commutation relations $[b^{\dagger},d]=0$, $[d,\hat{S}] = e^{-4\pi M \omega} \hat{S} b^{\dagger}$. %, and 
%\begin{equation}
%    B(\omega_0,\alpha) = \frac{1}{\sqrt{\alpha!}}\left(2e^{-2\pi \omega_0}\tanh{2\pi \omega_0}\sinh{2\pi \omega_0}\right)^{\alpha/2}
%    \label{eq:rhonormalizefactor}
%    \end{equation}
%is a single-valued function of $\omega_0$ and $\alpha$. 
The index $K$ means Kruskal, $S$ Schwarzschild, $O$ Outside the Schwarzschild Horizon, and $I$ Inside the Schwarzschild Horizon.

The general $\rho^K$ above should be able to describe any possible infalling states/ensembles at the perturbative level. 
We contract out the states inside the Schwarzschild horizon to find the density matrix outside the horizon: 

\begin{align}
    \rho^O(A_{\alpha_1 \alpha_2}) & = \sum_{\mathcal{N}} \ _I\langle \mathcal{N}|  \rho^K | \mathcal{N} \rangle_I\\
    & = \sum_{\alpha_1 \alpha_2} A_{\alpha_1 \alpha_2} \hat{\rho}^{b^{\dagger,\alpha_1}b^{\alpha_2}},
    %C(\omega_0,\alpha_1,\alpha_2) &= A_{\alpha_1 \alpha_2} B(\omega_0, \alpha_1) B^*(\omega_0,\alpha_2)
    \label{eq:mapping}
\end{align}

where the basis matrices $\hat{\rho}^{b^{\dagger,\alpha_1}b^{\alpha_2}}$ are the special cases of equation (\ref{eq:rhogeneral}) with $l=\alpha_1, m=0, n = \alpha_2, s=0$: 

\begin{widetext}
\begin{equation}
    \hat{\rho}^{b^{\dagger,\alpha_1}b^{\alpha_2}} = Z^2 e^{-2\pi M (\alpha_1+\alpha_2)\omega} (2\sinh{4 \pi M \omega})^{\frac{\alpha_1+\alpha_2}{2}}\sum_{\mathcal{N},n_q=0}^{\infty} e^{-8\pi M E_{\rm tot}} \sqrt{C(n_q+\alpha_1,\alpha_1)C(n_q+\alpha_2,\alpha_2)} | \mathcal{N}, n_q+\alpha_1 \rangle_O \ _O \langle \mathcal{N}, n_q+\alpha_2 |
    \label{eq:pseudothermal_kruskal}
\end{equation}
\end{widetext}
$C(n_q+\alpha_1,\alpha_1) = \frac{(n_q+\alpha_1)!}{n_q! \alpha_1 !}$ is the binomial coefficient, and $Z^{-2}$ is the partition function.

As discussed in section \ref{sec:beyondsingle}, we call $\hat{\rho}^{b^{\dagger,\alpha_1}b^{\alpha_2}}$ a set of basis matrices for the outside-horizon physics, just like $\hat{\rho}^K=| \alpha_1 \rangle \langle \alpha_2|$ on the Kruskal side, because they are linearly independent of each other in terms of matrix summation algebra. The diagonal ones $\hat{\rho}^{b^{\dagger,\alpha}b^{\alpha}}$ are also properly normalized, hence a normalized, positive semi-definite, and Hermitian density matrix $\hat{\rho}^{b^{\dagger,\alpha_1}b^{\alpha_2}}$ is safely traced into a normalized, positive semi-definite and Hermitian density matrix $\rho^O$. The expression of $\rho^O=\sum_{\alpha_1 \alpha_2} A_{\alpha_1 \alpha_2}\hat{\rho}^{b^{\dagger,\alpha_1}b^{\alpha_2}}$ is unique, and we have built a one-to-one mapping between the set of arbitrary Kruskal positive frequency modes density matrices and a subset of the outside-horizon density matrices spanned by the normalized linear combination of pseudo-thermal density matrices $\hat{\rho}^{b^{\dagger,\alpha_1}b^{\alpha_2}}$. 

%It leads to a counter-intuitive confusion, that usually the partial tracing of a density matrix is a multiple-to-one mapping from a higher dimensional matrix to a lower dimensional one. We usually lose some information in such partition and contraction of the quantum systems, since the contracted out degree of freedoms become completely unknown. The exception happens when the two systems are strongly entangled in a specific way -- In our case, the positive and negative same frequency magnitude modes always mix across the horizon. An analogy is that when we already know that two spins are always prepared opposite, tracing out one of them or not does not change the information that we can extract from the other spin. In curved spacetime QFT such strong entanglement across the horizon is `prepared' by the spacetime manifold foliation structure itself.

%The positivity of the matrix $\rho^O(A_{\alpha_1 \alpha_2})$ is preserved by the positivity of $B(\omega_0,\alpha)$ and the positive coefficients of $\hat{\rho}^{b^{\dagger,\alpha_1}b^{\alpha_2}}$. As long as the creators $b^{\dagger}$ and $d^{\dagger}$ each generates a normalizable states space (which should be feasible for a simple massive scalar field), our operation of partial tracing should not break the unit trace of $\rho^O(A_{\alpha_1 \alpha_2})$ inherited from the unit trace of $A_{\alpha_1 \alpha_2}$. Hence $\rho^O(A_{\alpha_1 \alpha_2})$ is a positive, trace 1 matrix, which is a legal quantum mechanics density matrix. 
 The space of $\rho^O(A_{\alpha_1 \alpha_2})$, or the subspace of the density matrices being a normalized linear combination of the pseudo-thermal density matrices, apparently is not the full density matrix space for the physical ensembles living outside the horizon, and it should not be. For example, $\rho^O =| \alpha \rangle_O \ _O \langle \alpha| $, a pure state living outside the horizon, cannot be decomposed into $\hat{\rho}^{b^{\dagger \alpha_1}b^{\alpha_2}}$ basis at non-zero temperature $T$. %The infalling particles of the collapsing stars evolve into a special type of ensemble in late, BH evaporated universe, which is featured by the distorted thermal spectrum. The particles escaping the collapse, and living completely outside the horizon, does not belong to this subset. Combined together, they constitute the physical states outside the horizon. 
 One might wonder how we mapped a full density matrix space into a subspace of density matrices, but it is not mysterious at all for infinite element groups -- Consider how one-to-one mapping can be easily built between integers and positive integers. The surface gravity with respect to a specific observer raises the ground state of that observer by a thermal ensemble at the temperature corresponding to the surface gravity. Such gravitational thermalization could have memorizing feature, i.e. the original Kruskal density matrix $\rho^K$ could be a thermal/pseudo-thermal density matrix itself, at a different temperature. We will encounter such a scenario when approximating the collapsing astrophysical black hole by layers of collapsing shell metrics, which we leave for future works.

%Although the reflection upon the horizon seems to be mapping the full density matrix space under one metric into a subspace of the density matrix space under another non-inertial frame, it does not mean one space is larger than another nor do they have containing relationships. 
A relativistic perspective of view is helpful to understand such a phenomenon. The stance of inertial and non-inertial frames are always interchangeable. The dimensionality of their Hilbert space and density matrix space are infinity; their density matrices have this one-to-one mapping that could preserve the information from one space in the subspace at a certain temperature of the other; their Hilbert spaces are different spaces by definition, only with isomorphic representation structure.

Given the particle state/density matrix describing the collapsing history, one could analytically calculate the Page curve, i.e. the information $I_r = S_{\rm therm} + {\rm Tr}(\rho_r \log{\rho_r})$ from the density matrix at each stage of the evaporation, using the pseudo-thermal density matrices in equation (\ref{eq:pseudothermal_kruskal}). On a rather loose end, the one-to-one mapping between density matrices itself might be sufficient for us to use the degrees of freedom counting argument in the original proof of Page curve \cite{page1993information}. Without validation on the details here, we notice that a recent work \cite{JUSTC-2022-0039} starting from the same point as our paper, i.e. the Unruh effect on excited states has already presented a calculation of the Page curve in their framework. %Figure \ref{fig:entropy} shows the entropy per number counting expectation value in a single frequency band $\omega$ with decreasing black hole mass from right to left. Entanglement entropy $S = {\rm Tr} (\rho \log{\rho})$ for a canonical or pseudo-thermal ensemble is calculated from the density matrix in equation (\ref{eq:pseudothermal_kruskal}). However, a real Page curve calculation needs to take into account all the degrees of freedom of the system, namely all frequency bands, which we leave for future work. 

\begin{tcolorbox}[breakable]
Remark: As first mentioned in the remark block in section \ref{sec:blackhole}, although the mapping from a pure state density matrix to a mixed one is not unitary from a traditional quantum mechanics point of view, the unitarity could be `resumed' by re-defining the states basis matrices labeled by $\hat{\rho}^{b^{\dagger,\alpha_1}b^{\alpha_2}}$ quantum numbers modulo temperature in this supposedly non-unitary contracting operation. As long as the one-to-one mapping holds, such formalism should be feasible. %The pure-to-mixed state one-to-one mapping being an effective S-matrix in Page curve argument invites 
The speculation is that we should probably change the definition of the vacuum in our relativistic QFT theory. We suspect that in an advanced representation of the physical states living on certain curved spacetime manifold, the vacuum should be defined as degenerating with any thermal ensemble, and by acting creators on the traditional vacuum or a thermal ensemble density matrix, instead of state vector representation, we get a spectrum of excitations. Actually, simple algebra shows that the hierarchy of $ \hat{\rho}^{b^{\dagger,\alpha_1}b^{\alpha_2}}$ is connected by acting corresponding numbers of creators and annihilators on both sides of a canonical density matrix, multiplied by a normalization factor, in the high-frequency/low-temperature limit. What is more, by acting annihilator $b$ on the ground state in this formalism, the canonical ensemble, we obtain the first excitation on the opposite side $ \hat{\rho}^{d^{\dagger}d}$.

\end{tcolorbox}%One immediate problem of this proposal is that by acting annihilation operators on the vacuum defined this way we have no lower bound of the energy in this spectrum. Such issue does not necessarily lead us to a dead end, though it does hint that there is still a long way to go to fully realize the speculation here. %\footnote{One more comment on the energy bound of a system: at least we know the energy of a spacetime manifold is bound by its occupation of black hole.}. 

\section{Case Study: Spectrum Under Different Basis}
\label{sec:spectrum}

%\textbf{Assumption: infinite dimensional POVM.}

We intuitively have a rough picture of the physics that the pseudo-thermal density matrix represents. As briefly mentioned in section \ref{sec:beyondsingle}, $\hat{\rho}$ has a bump around specific frequency sitting on top of a thermal ensemble. However, it is still very elucidating to see the detailed number counting and energy spectrum that an observer would measure in the asymptotically flat regime. We will go through the detailed calculation in this section. %It contains all the possible configurations of a Hilbert space generated by Fock states, like a stochastic ensemble, yet the way each of them is weighted makes such density matrix non-trivial. 
%Like briefly mentioned in section \ref{sec:beyondsingle}, $\hat{\rho}$ has a bump around specific frequency sitting on top of a thermal spectrum. In this section we will explicitly plot the number density spectrum calculated from Equation \ref{eq:pseudothermal_kruskal}.

Before we start, we would like to first specify several presumptions and approximations. The first thing to stress is again, that we are going to apply some quantum information and computation techniques well-established for finite-dimensional Hilbert space on the infinite-dimensional Hilbert space. Specifically, in this section, we need to use the concept of positive operator-valued measures (POVM) for number/energy measurements. Secondly, we ignore the grey-body factor, which causes a frequency dependent $<1$ transmission rate across the potential barrier extended in the intermediate region away from the black hole horizon. The grey-body factor is well-studied in the particle packet perspective of view for Hawking flux \cite{Gray:2015xig}, and in our scalar field eigenmode scenario, this effect originates from the complicated form of the eigenmodes in the transition region between the near horizon and asymptotically flat regions. Recall that (only) in those two extreme regimes the Klein-Gordon equation solutions have simple plane-wave-like expressions in terms of $t, r $ or $r_*$, and they have $\leq 1$ transmission rates in between captured by the grey-body factor. We argue that since the grey-body factor is related to the physics not in the immediate vicinity of the horizon, it could be mounted separately later in the standard way from previous literature. The result we present in this section are not considering the grey-body factor and assume the transmission rate $T=1$.

To find the energy spectrum of a density matrix, we need to calculate the expectation value $\langle E(\omega)\rangle$ as a function of the frequency $\omega$. In quantum mechanics, the expectation value of a positive operator-valued measure (POVM) for a density matrix can be calculated as \cite{peres2002quantum}:
\begin{equation}
    \langle \bold{E} \rangle  =  {\rm Tr}(\rho \bold{E}) = {\rm Tr}(\sum_{\alpha}A_{\alpha}{\rho}_{\alpha} \bold{E}) = \sum_{\alpha} A_{\alpha} \langle \bold{E} \rangle_{\alpha}
    \label{eq:expectationvalue}
\end{equation}
where $\sum_{\alpha} A_{\alpha}=1$ and ${\rm Tr}(\rho_{\alpha}) = 1$. The energy in a certain frequency $\bold{E}_{\omega}$ is a POVM of our system, and we calculate the expectation value using the above equation, as a function of frequency, to obtain the energy spectrum. %We can thus calculate the expectation value, i.e. the energy spectrum in our case, of a POVM for each normalized density matrix $\rho_{\alpha}$, then calculate their weighted average.

%We are going to specify the physical system under investigation in this section, by formally present the density matrix the POVM in the following paragraph. 

We start with a wavepacket of Unruh modes living on the outgoing horizon of the Schwarzschild black hole. We express this Unruh wavepacket of the quantized real scalar field as follows:
\begin{equation}
    |\Gamma \rangle_K = \int_0^{\infty} d\omega g(\omega)u^{\dagger}_{\omega} |0\rangle_K
    \label{eq:unruhwavepacket}
\end{equation}
where $g(\omega)$ is the shape of an arbitrary wavepacket, which is normalized to unity:

\begin{equation}
    \int_0^{\infty} d\omega |g(\omega)|^2 = 1
\end{equation}

%and we take it to be Gaussian function $g(\omega) = \frac{1}{(2\pi \sigma^2)^{1/4}}e^{-\frac{(\omega-\omega_0)^2}{4 \sigma^2}}$. Equation \ref{eq:unruhwavepacket} is a pure state with excitation number 1 of certain wave-packet. The choice of Gaussian function is only an example, and the convolution of the pseudo-thermal spectrum with the Unruh wavepacket is fairly straightforward as we will see later. 

The density matrix living outside the Schwarzschild black hole horizon contracting out the inner physical states is given by:
\begin{align}
    \rho^O = &\sum_{\mathcal{N}} \ _{I}\langle \mathcal{N}| \Gamma\rangle_K \ _K\langle \Gamma | \mathcal{N} \rangle_I\\
     = & \int_0^{\infty} \int_0^{\infty} d\omega_1 d \omega_2 g(\omega_1) g^*(\omega_2) \hat{\rho}^{11}(\omega_1,\omega_2)
\end{align}
%where the density matrix component normalizing factor $B(\omega,\alpha)$ is specified in equation \ref{eq:rhonormalizefactor}, and $\hat{\rho}^{11}(\omega_1,\omega_2)$ is the general $\hat{\rho}$ defined in equation \ref{eq:rhogeneral} with $l=n=1, m=s=0$, and $b^{\dagger}_{\omega_1}$, $b_{\omega_2}$ are with different frequencies. 
where $\hat{\rho}^{11}(\omega_1,\omega_2)$ is as in the equation (\ref{eq:pseudothermal_kruskal}).

Among all the $\hat{\rho}^{11}(\omega_1,\omega_2)$ terms, only the terms with equal frequency $\omega_1=\omega_2=\omega$ are contributing to the diagonal terms in density matrix $\rho$ -- `diagonal' means that a component is expressed as the direct product of two identical states $|\mathcal{N} \rangle \langle \mathcal{N}|$. With $\omega_1 \neq \omega_2$, the two states of a density matrix component, $|\mathcal{N}, n_{\omega_1}+1\rangle$ and $|\mathcal{N}, n_{\omega_2}+1\rangle$, are never identical throughout all the configurations $\mathcal{N}$. %Hence we have a nice feature (like any quantum mechanical density matrix should have) that the off-diagonal terms in $\rho^{K}$ under Unruh modes basis contributes and only contributes to the off-diagonal terms in partitioned and contracted density matrix $\rho^O$. Same for the diagonal terms. 
For the purpose of calculating the expectation value of a POVM, we focus on the diagonal terms in the density matrix from now on. 
\begin{equation}
    {\rm diag}(\rho^O) =  \int_0^{\infty}  d\omega |g(\omega)|^2 \hat{\rho}^{11}(\omega)
\end{equation}
$\hat{\rho}^{11}(\omega)$ are the normalized density matrix components, with ${\rm Tr}(\hat{\rho}^{11}(\omega))=1$. Thus, for the measurement of energy distributed to a specific frequency,
\begin{equation}
  \bold{E}_{\omega} | \mathcal{N} \rangle = n_{\omega} \omega ,
\end{equation}
the energy expectation value over the full density $\rho^O$ in $d\omega$ band can be decomposed into:
\begin{align}
    \langle \bold{E}_{\omega}\rangle & = {\rm Tr} ( \bold{E}_{\omega}\rho^O) \\ &=  \int_0^{\infty}  d\omega' |g(\omega')|^2 \langle \bold{E}_{\omega} (\omega')\rangle
    \label{eq:Eexpectation}
\end{align}
where
\begin{equation}
    \langle \bold{E}_{\omega} (\omega')\rangle = {\rm Tr} (\bold{E}_{\omega} \hat{\rho}^{11}(\omega'))
\end{equation}
is the energy spectrum for a specific single Unruh mode contracted normalized pseudo-thermal density matrix $\hat{\rho}^{11}(\omega')$. 

Next, we refer to the traditional derivation of the Planck formula to calculate $\langle \bold{E}_{\omega} (\omega')\rangle$ \cite{planckformula, pathria2016statistical}. Because $\hat{\rho}^{11}(\omega')$ is normalized and the distribution of the occupation number of a frequency is uncorrelated with other frequencies (multiplicative coefficients for each frequency in all the density matrix elements), we can calculate the expectation of the energy of a frequency $\omega$ as:
\begin{equation}
    \langle \bold{E}_{\omega} \rangle_{\omega'} = \frac{\sum_{n_{\omega}} p_{\omega'}(\omega,n_{\omega}) \epsilon_{\omega}}{\sum_{n_{\omega}} p_{\omega'}(\omega,n_{\omega})}
\end{equation}
where $p_{\omega'}(\omega,n_{\omega})$ is the probability of we finding $n_{\omega}$ of a frequency in the configuration $\mathcal{N}$, in a system prepared with density matrix $\hat{\rho}^{11}(\omega')$. $n_{\omega}$ always run from $0$ to $\infty$. %We will see later that it does not always mean that there are exactly $n_{\omega}$ particle counting in the actual physical state.

When $\omega \neq \omega'$, the distribution of $\langle \bold{E}_{\omega} (\omega')\rangle$ is plainly Bose-Einstein for our massive scalar field, as $p_{\omega'}(\omega,n_{\omega}) \propto e^{-n_{\omega}\omega/T}$, $\epsilon_{\omega} = n_{\omega}\omega$, where $T=\frac{1}{8\pi M}$.
\begin{align}
    \langle \bold{E}_{\omega} (\omega' \neq \omega)\rangle  = & \frac{\sum^{\infty}_{n = 0} n\omega e^{-n\omega/T}}{\sum^{\infty}_{n = 0} e^{-n\omega/T}}\\
     = & \frac{\omega \sum^{\infty}_{n=0}\frac{de^{-n\omega/T}}{d(-\omega/T)}}{\sum^{\infty}_{n=0}e^{-n\omega/T}}\\
      = &\frac{\omega \frac{d}{d(-\omega/T)} (1-e^{-\omega/T})^{-1}}{(1-e^{-\omega/T})^{-1}}\\
      =& \frac{\omega}{e^{\omega/T}-1}
      \label{eq:omegaprime}
\end{align}

For $\omega = \omega'$, according to equation (\ref{eq:pseudothermal_kruskal}) with $\alpha_1=\alpha_2=1$, $p_{\omega'}(\omega',n_{\omega'}) \propto (n_{\omega'}+1)e^{-n_{\omega'}\omega'/T}$, $\epsilon_{\omega'} = (n_{\omega'}+1)\omega'$. This distribution is manifestly different from the canonical one for other frequencies, especially with the number counting in frequency $\omega = \omega'$ starting from 1, instead of 0.  %The full particle configuration $\mathcal{N}$ in equation (\ref{eq:rhobb}) runs over all possible combinations of $n_{\omega_i}$, each going through $0,1,2... \infty$. However, in physical state $|\mathcal{N}, n_{\omega}+1 \rangle$, the eigenvalue of the number counting operator for $\omega$ is not $n_{\omega}$, but $n_{\omega}+1$. Also, the coefficient determining the probability of $|\mathcal{N}, n_{\omega}+1 \rangle$ is not simply $\propto e^{-n_{\omega}\omega/T}$ (like all other frequencies), but $\propto e^{-n_{\omega}\omega/T}(n_{\omega}+1)$. 
It is this relative difference, of $\omega = \omega'$ compared with other canonically occupied frequencies, recording the information of $\omega'$ corresponding to the original Unruh mode. Such difference is non-trivial, i.e. cannot be eliminated by some sort of relabeling of the number counting. %This fact can be easily understood by looking at the lowest possible physical counting of the $\omega=\omega'$ quanta: others have ground state starting from $n=0$, while $\omega=\omega'$ has ground state starting from $n=1$ \footnote{This is just a quick way of understanding results for $\hat{\rho}^{bb}$; Actually, in $\hat{\rho}^{dd}$ where $n_{\omega=\omega'}$ does start counting from $0$, because of a factor of $(n_{\omega=\omega'}+1)^2$ larger probability of measuring the special frequency, $\omega=\omega'$ still stands out in the energy or number counting spectrum, just with a slightly lower amplitude.}. 
The energy expectation value of $\omega=\omega'$ frequency is given by:
\begin{widetext}
\begin{align}
    \langle \bold{E}_{\omega} (\omega'=\omega)\rangle = & \frac{\sum^{\infty}_{n = 0} \omega (n+1)^2 e^{-n\omega/T}}{\sum^{\infty}_{n = 0} (n+1) e^{-n\omega/T}}\\
     = & \omega \frac{\sum^{\infty}_{n = 0}(n+1)n e^{-n\omega/T}+\sum^{\infty}_{n = 0}(n+1) e^{-n\omega/T}}{\sum^{\infty}_{n = 0} \frac{d}{d(e^{-\omega/T})}e^{-(n+1)\omega/T}}\\
     =& \omega \frac{e^{-\omega/T}\sum^{\infty}_{n = 0}\frac{d^2}{d(e^{-\omega/T})^2}e^{-(n+2)\omega/T}+\frac{d}{d(e^{-\omega/T})}((1-e^{-\omega/T})^{-1}-1)}{\frac{d}{d(e^{-\omega/T})}((1-e^{-\omega/T})^{-1}-1)}\\
     & = \omega \frac{e^{-\omega/T}\frac{d^2}{d(e^{-\omega/T})^2}((1-e^{-\omega/T})^{-1}-1-e^{-\omega/T})+(1-e^{-\omega/T})^{-2}}{(1-e^{-\omega/T})^{-2}}\\
     =& \omega \frac{2e^{-\omega/T}(1-e^{-\omega/T})^{-3}+(1-e^{-\omega/T})^{-2}}{(1-e^{-\omega/T})^{-2}}\\
     =& \omega \frac{e^{\omega/T}+1}{e^{\omega/T}-1}
\end{align}
\end{widetext}

Supposing that the discrete energy levels labeled by $\omega$ and $\omega'$ are distinguishable at discretization unit $h_0$, we can combine the expression of $\langle \bold{E}_{\omega} (\omega' \neq \omega)\rangle$ and $\langle \bold{E}_{\omega} (\omega'=\omega)\rangle$ into a delta function:
\begin{equation}
    \langle \bold{E}_{\omega} (\omega')\rangle = \frac{\omega}{e^{\omega/T}-1}+\delta\left(\omega'-\omega\right)h_0 \omega'\frac{e^{\omega'/T}}{e^{\omega'/T}-1}
\end{equation}
Similarly, the number counting spectrum:
\begin{equation}
    \langle \bold{n}_{\omega} (\omega')\rangle = \frac{1}{e^{\omega/T}-1}+\delta(\omega'-\omega)h_0\frac{e^{\omega'/T}}{e^{\omega'/T}-1}
\end{equation}

Substituting $\langle \bold{E}_{\omega} (\omega')\rangle$ back to equation (\ref{eq:Eexpectation}), the integration gives us:
\begin{equation}
     \langle \bold{E}_{\omega} \rangle = \frac{\omega}{e^{\omega/T}-1} + h_0|g(\omega)|^2  \frac{\omega e^{\omega/T}}{e^{\omega/T}-1}
     \label{eq:pseudospec}
\end{equation}

The delta function $\delta(\omega'-\omega)$ for each individual $\hat{\rho}^{11}_{\omega'}$ ensures us the simple convolution with the Unruh wave-packet in the pseudo-thermal spectrum. The first term gives us a black body radiation spectrum, while the second term enhances the power by a factor of $h_0|g(\omega)|^2 e^{\omega/T}$. This result echoes the stimulated Hawking radiation by J\"{u}rgen Audretsch and Rainer M\"{u}ller in 1992 \cite{audretsch1992amplification}, with a focus on showing how the deviation of the Hawking radiation from a perfect black-body spectrum incorporates the information of the waveform of the specific in-going excitation. Notice that the low-temperature limit $T \rightarrow 0$ of equation (\ref{eq:pseudospec}) goes back to the inertial frame result:
\begin{equation}
    \langle \bold{E}_{\omega} \rangle|_{T \rightarrow 0} = |g(\omega)|^2 h_0 \omega
\end{equation}

\begin{figure*}[ht]
    \centering
    \includegraphics[width=0.49\textwidth]{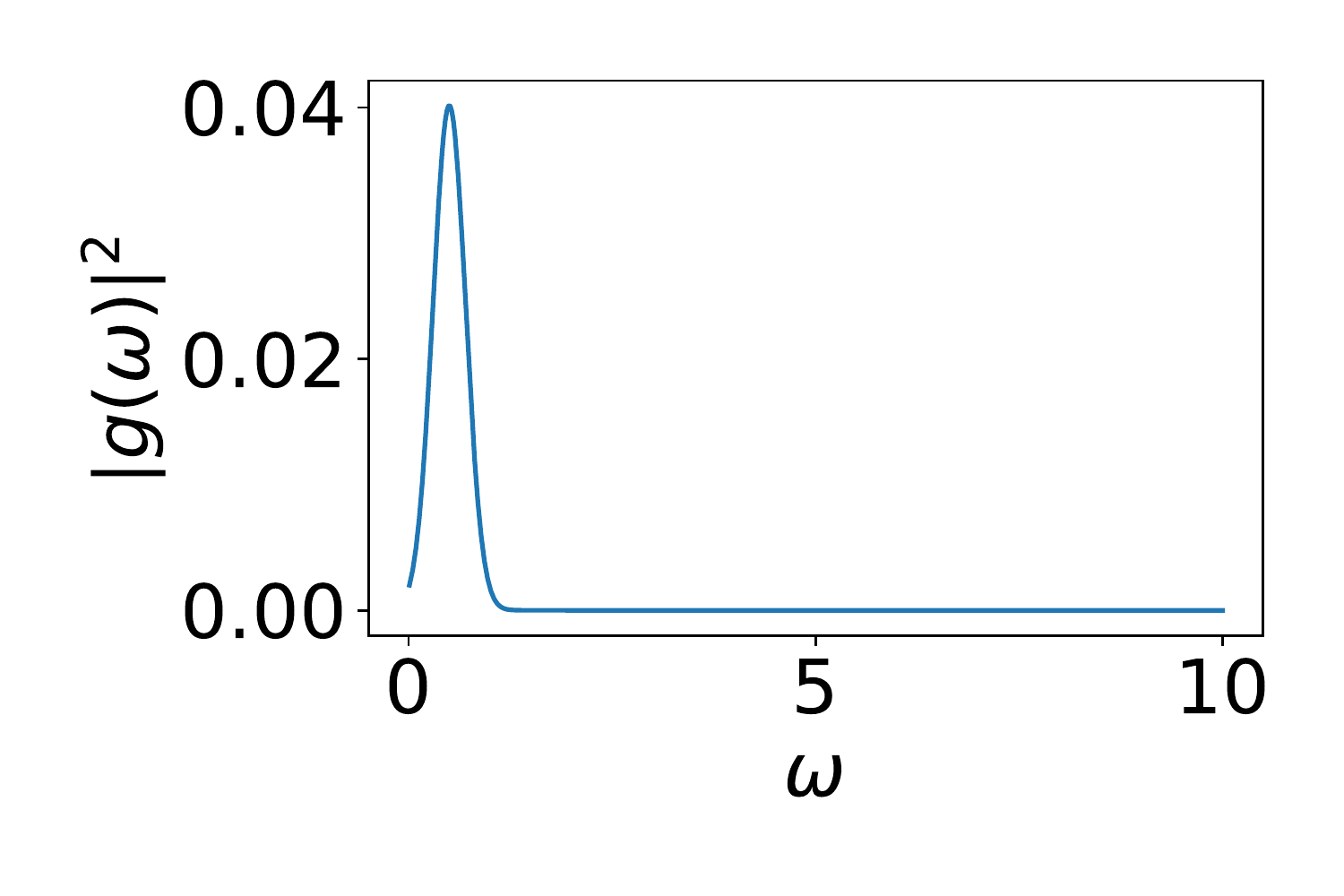}
    \includegraphics[width=0.49\textwidth]{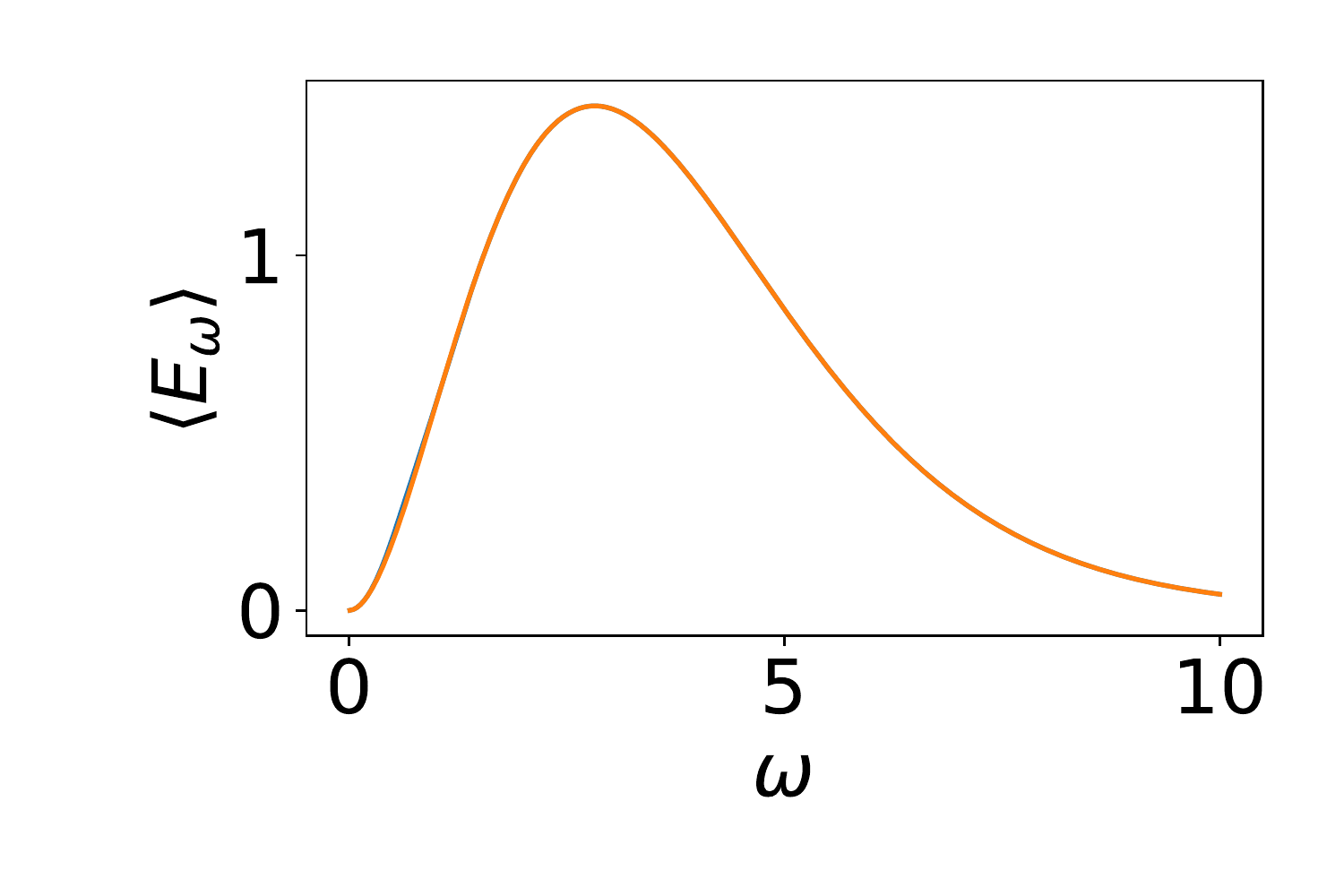}
    \includegraphics[width=0.49\textwidth]{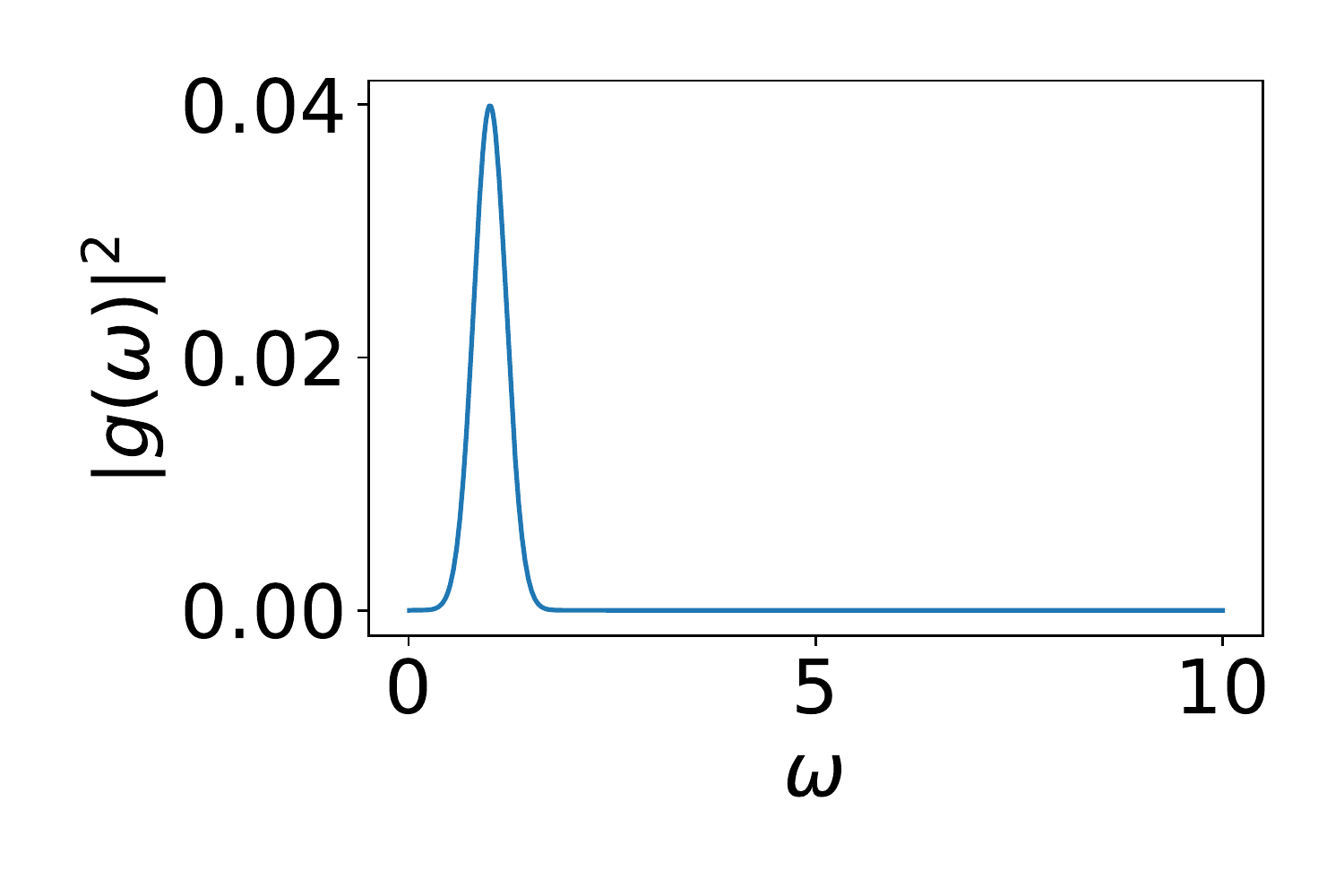}
    \includegraphics[width=0.49\textwidth]{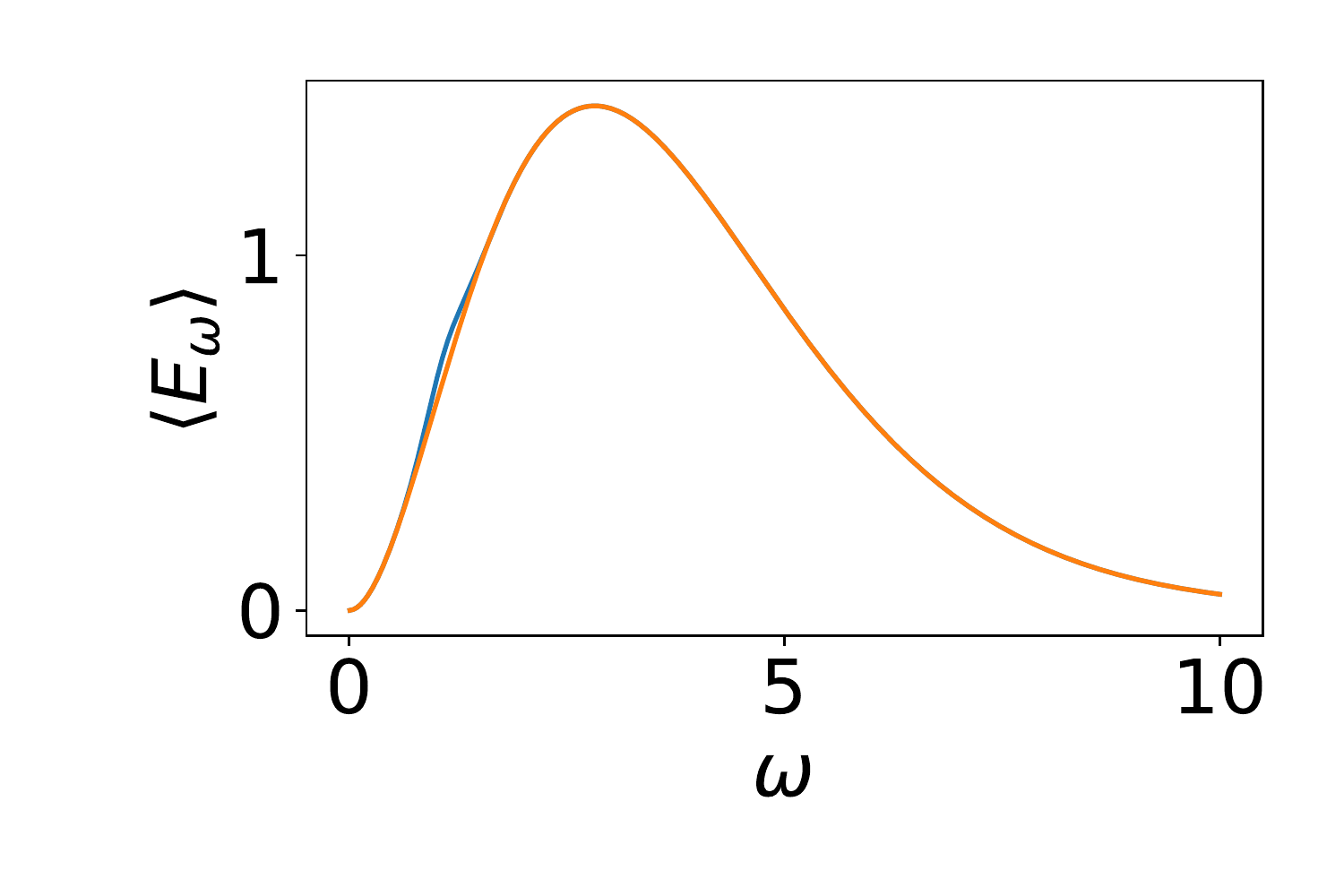}
    \includegraphics[width=0.49\textwidth]{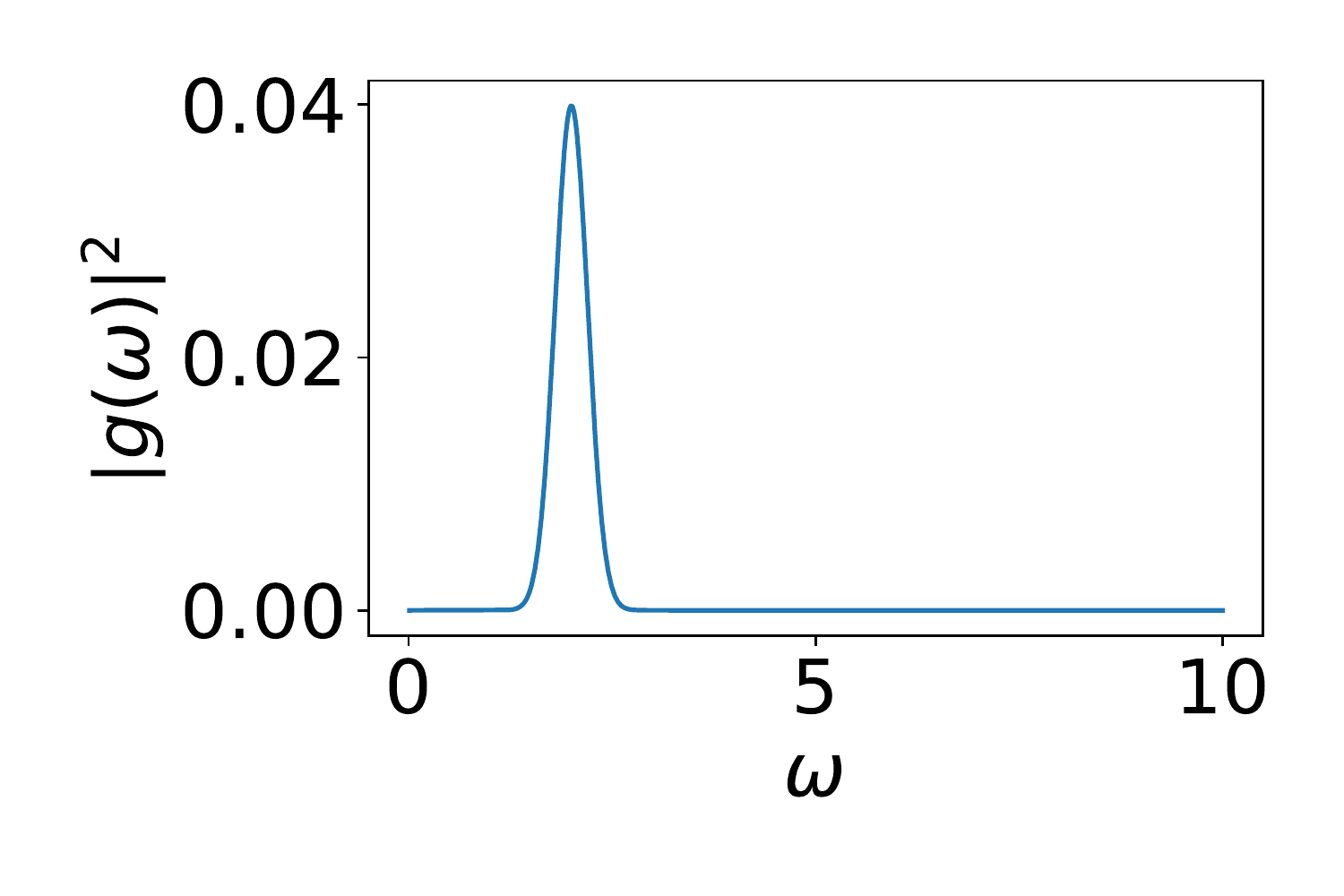}
    \includegraphics[width=0.49\textwidth]{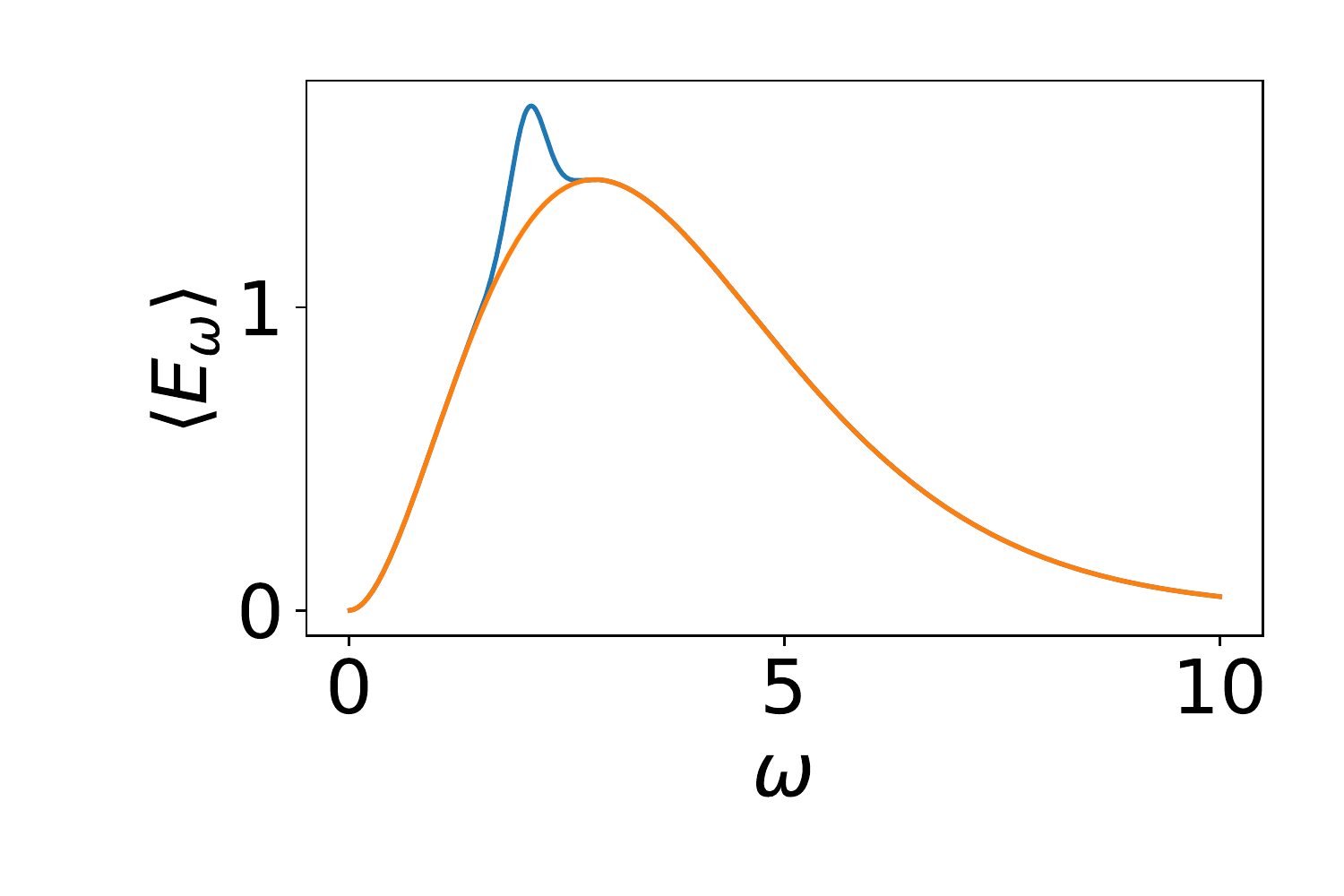}
    \includegraphics[width=0.49\textwidth]{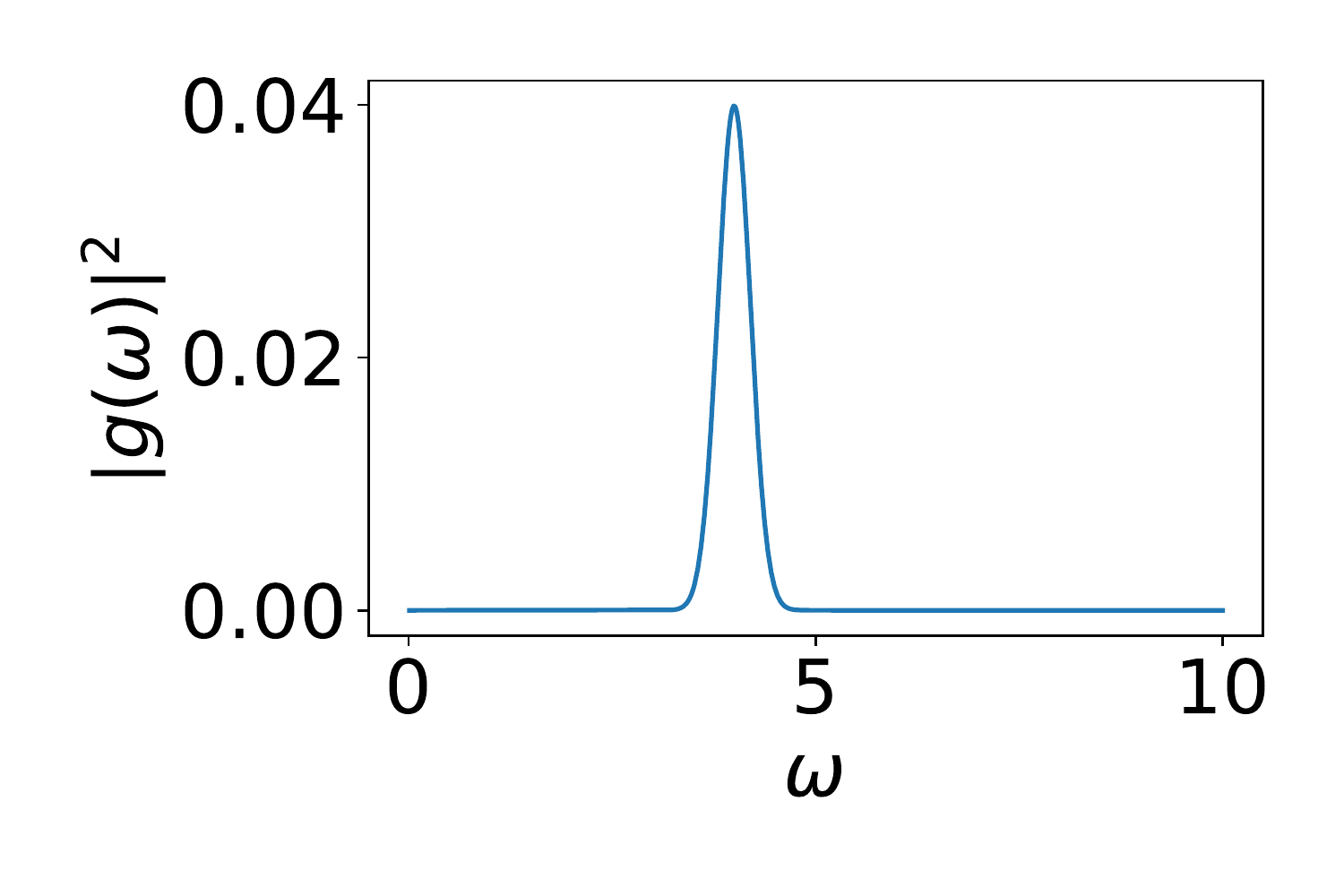}
    \includegraphics[width=0.49\textwidth]{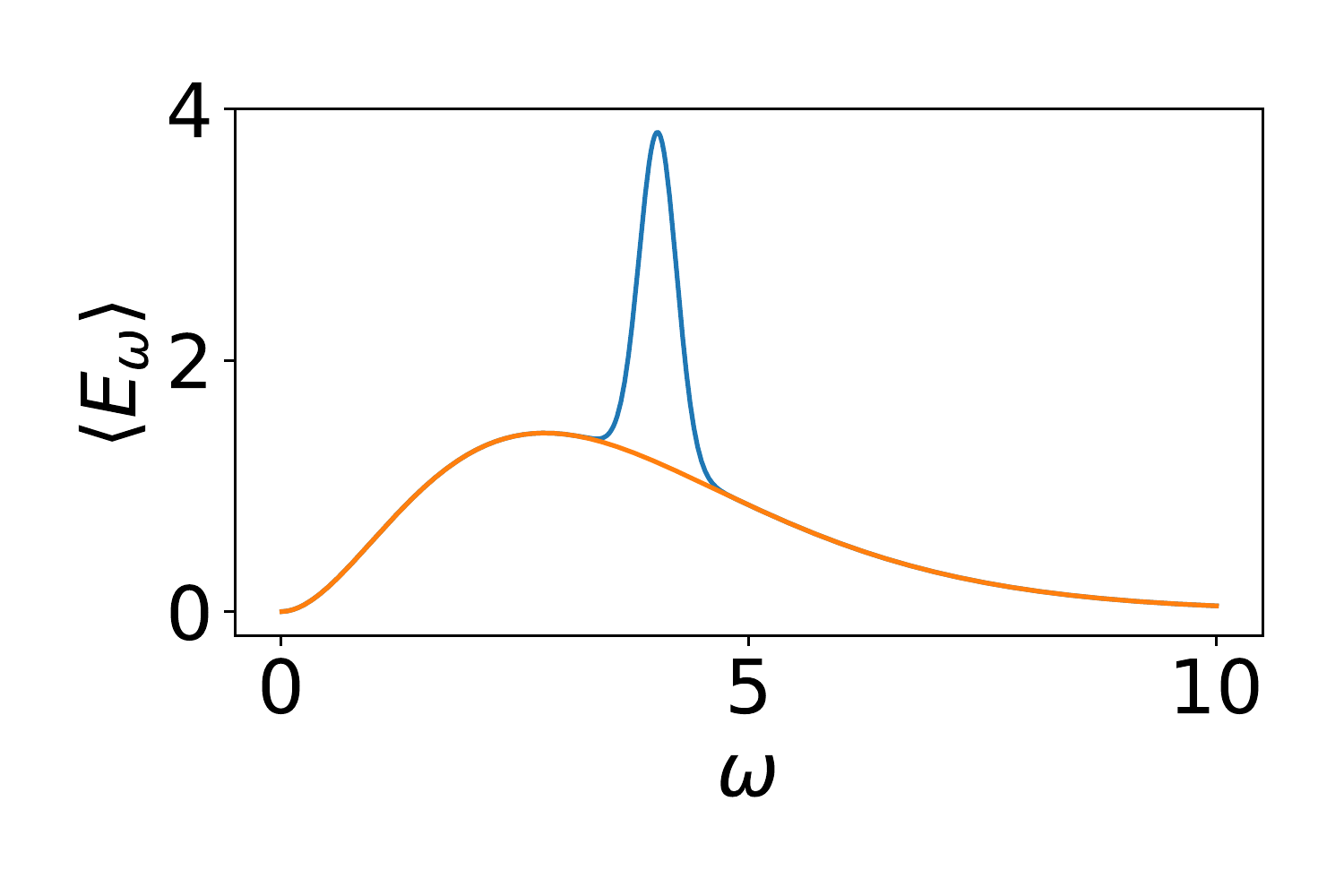}
    \caption{Left panels: the Gaussian wavepackets of type-I Unruh modes. Right panels: the energy spectrum on R+ (or outside the horizon), after tracing out the states on R- (or inside the horizon).}
    \label{fig:spectra}
\end{figure*}

In Figure \ref{fig:spectra} we illustrate several Gaussian $g(\omega) = \frac{1}{(2\pi \sigma^2)^{1/4}}e^{-\frac{(\omega-\omega_0)^2}{4 \sigma^2}}$ wavepackets and their corresponding energy spectrum $\langle \bold{E}_{\omega} \rangle$ measured by an observer infinitely far away outside the horizon. We multiply equation (\ref{eq:pseudospec}) by $\omega^2$, which is achievable by assuming a uniform angular modes distribution, to compare with the black body radiation spectrum depicted by the orange curves. We set the frequency discretization constant $h_0=1$ for illustration purposes. Actually, changing the excitation number in equation (\ref{eq:unruhwavepacket}) to $\alpha$ multiplies the second term in equation (\ref{eq:pseudospec}) by $\alpha$, so Figure \ref{fig:spectra} could be physically achieved by a large number of excitations setting $\alpha h_0=1$. The subtlety is that by contributing a nontrivial amount of energy into the energy spectrum, the Unruh modes at the horizon presumed to be perturbative might break the assumption of no backreaction. Such complexity can be ignored for now as the illustration here is only dedicated to showing that there will be a featured deviation from the black body spectrum in Hawking radiation, considering the generalized Unruh effect. 

The applicability of the above calculation of the pseudo-thermal ensemble spectrum is wide, for example, we could calculate the spectrum for higher number excitation, realistic Unruh wavepackets at the horizon, angular distribution, fermion generalized Unruh effect, etc. The technique in this section can be fairly straightforwardly applied those further investigations into the collapsing history of a black hole. In any case, hopefully, the amazing bottom line has been illustrated by those plots in figure \ref{fig:spectra} clearly, that by observing the offset of the Hawking radiation spectrum from the black body spectrum, we could infer the information of the particles that have fallen into the black hole. Again, our result is only valid at the perturbative level. 

Although we assumed a uniform angular distribution of the infalling modes in our illustration, it is useful to notice that the angular distribution of the infalling modes will be preserved when they propagate out as Hawking radiations for non-rotating Schwarzschild black holes. Because the scalar field eigenmodes in Kruskal and Schwarzschild metrics have the same angular dependent part. In reality, the Hawking radiation for cosmological black holes is at temperatures so low that it has never been observed. The results obtained here might need to be tested in some other laboratories like asteroidal mass primordial black holes, which are still in the open window to take all the dark matter \cite{Carr:2020gox}, or at the particle horizons that could be produced by accelerations or simulated systems in the lab. The inflation and reheating research can also potentially investigate the effect of the particle horizon formed from the rapid expansion of the universe using the generalized Unruh effect as a tool. 

%the energy spectrum seen by an observer sitting infinitely far outside the horizon corresponding to arbitrary positive Kruskal frequency boundary conditions on $H^-$ can be obtained. \textbf{Namely, in principle it is possible to reconstruct the collapsing procedure of a black hole by measuring its radiating spectrum, which in general is not just a vanilla black body radiation spectrum.} As long as we did treat the density matrix normalization right in our calculation in this section, it is impossible to prepare the pseudo-thermal density matrix as a linear combination of a perfectly thermal density matrix and a pure state density matrix. Namely, in principle the black hole pseudo-thermal radiation is distinct from a particle production outside the black hole plus a thermal, vacuum Hawking radiation background. 

%Although we assumed a perfectly uniform angular distribution of the infalling Unruh modes in our figure, to facilitate the comparison to black body power spectrum, a very handy outcome of our result is actually that the angular distribution of the infalling modes will be preserved when they propagate out as Hawking radiations for non-rotating Schwarzschild black holes. Because the compressed 2D spherical coordinates for Kruskal and Schwarzschild metric are trivially mapped into each other. In the rotating black hole case, things might be more complicated, and we will leave it for future discussions. 

\begin{figure*}[ht]
    \centering
    \includegraphics[width=0.49\textwidth]{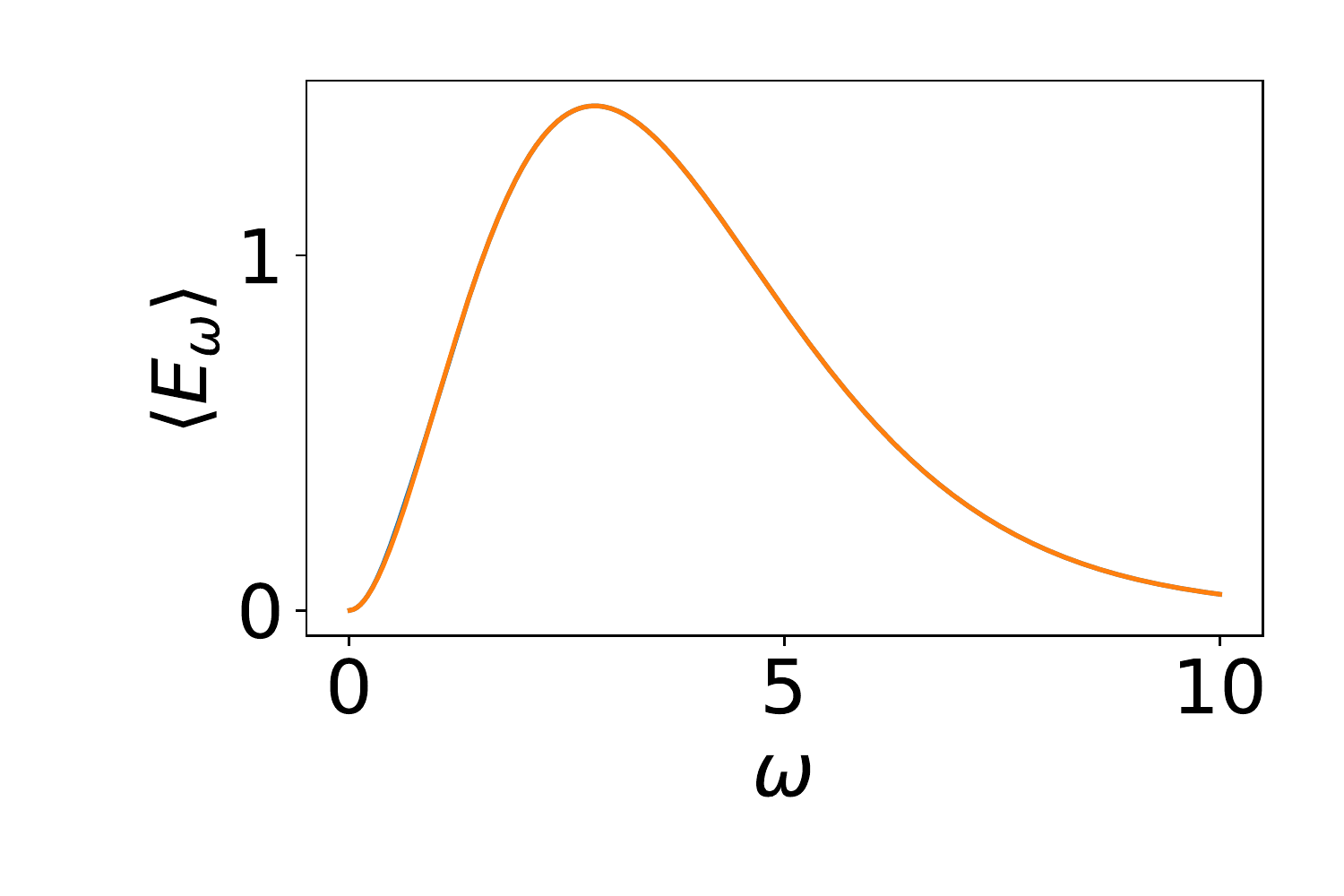}
    \includegraphics[width=0.49\textwidth]{pseudothermal_spec0_5.pdf}
    \includegraphics[width=0.49\textwidth]{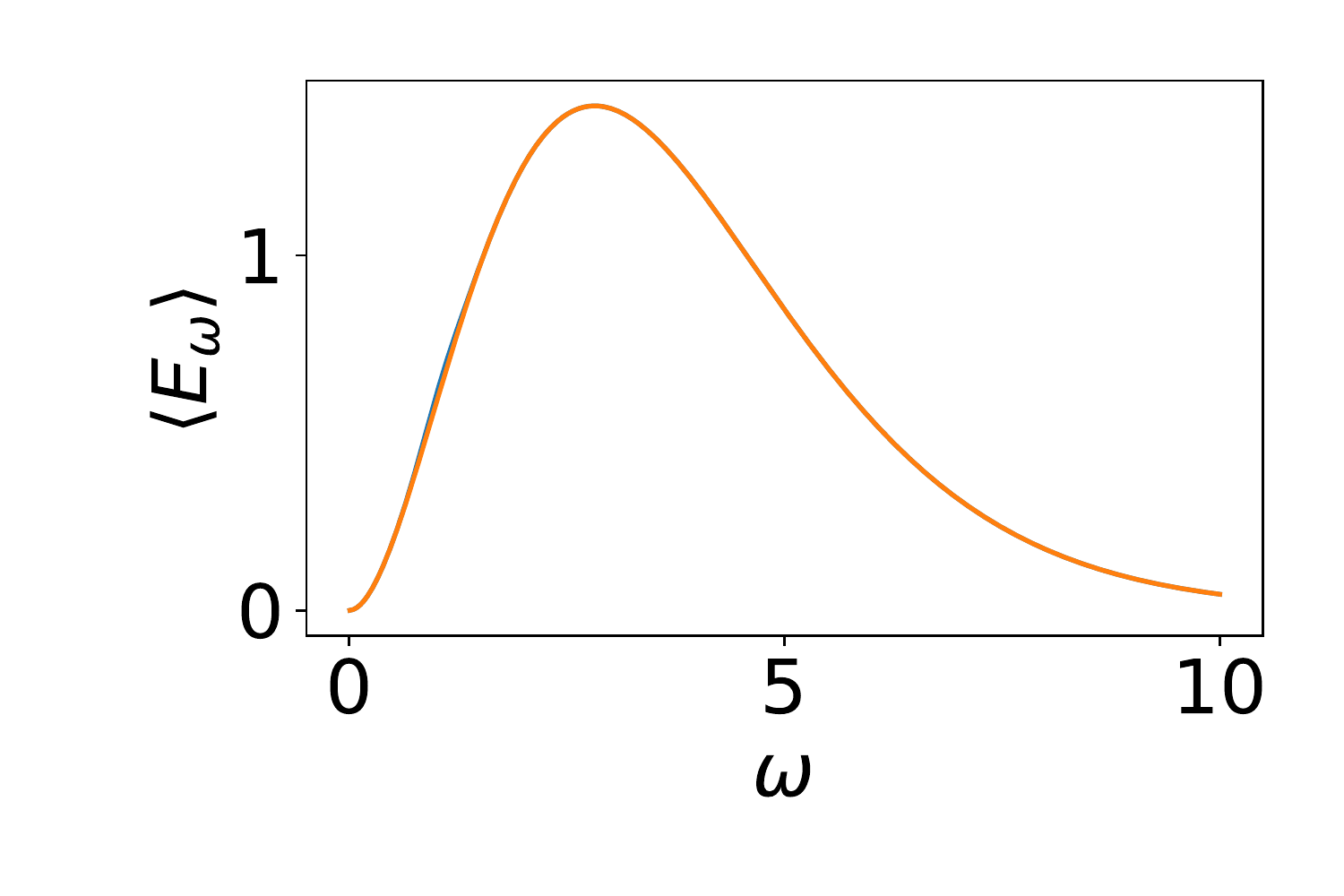}
    \includegraphics[width=0.49\textwidth]{pseudothermal_spec1_0.pdf}
    \includegraphics[width=0.49\textwidth]{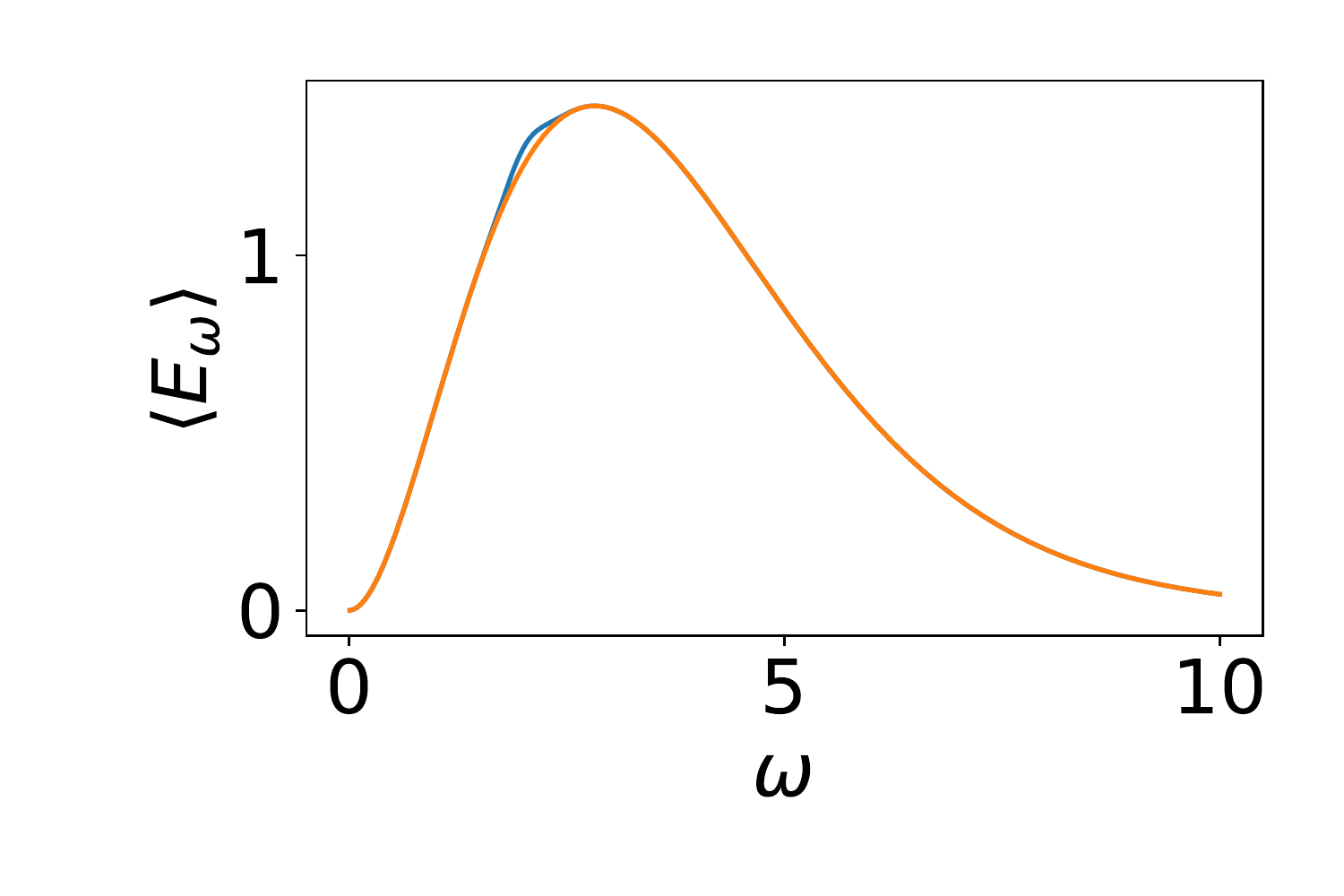}
    \includegraphics[width=0.49\textwidth]{pseudothermal_spec2_0.pdf}
    \includegraphics[width=0.49\textwidth]{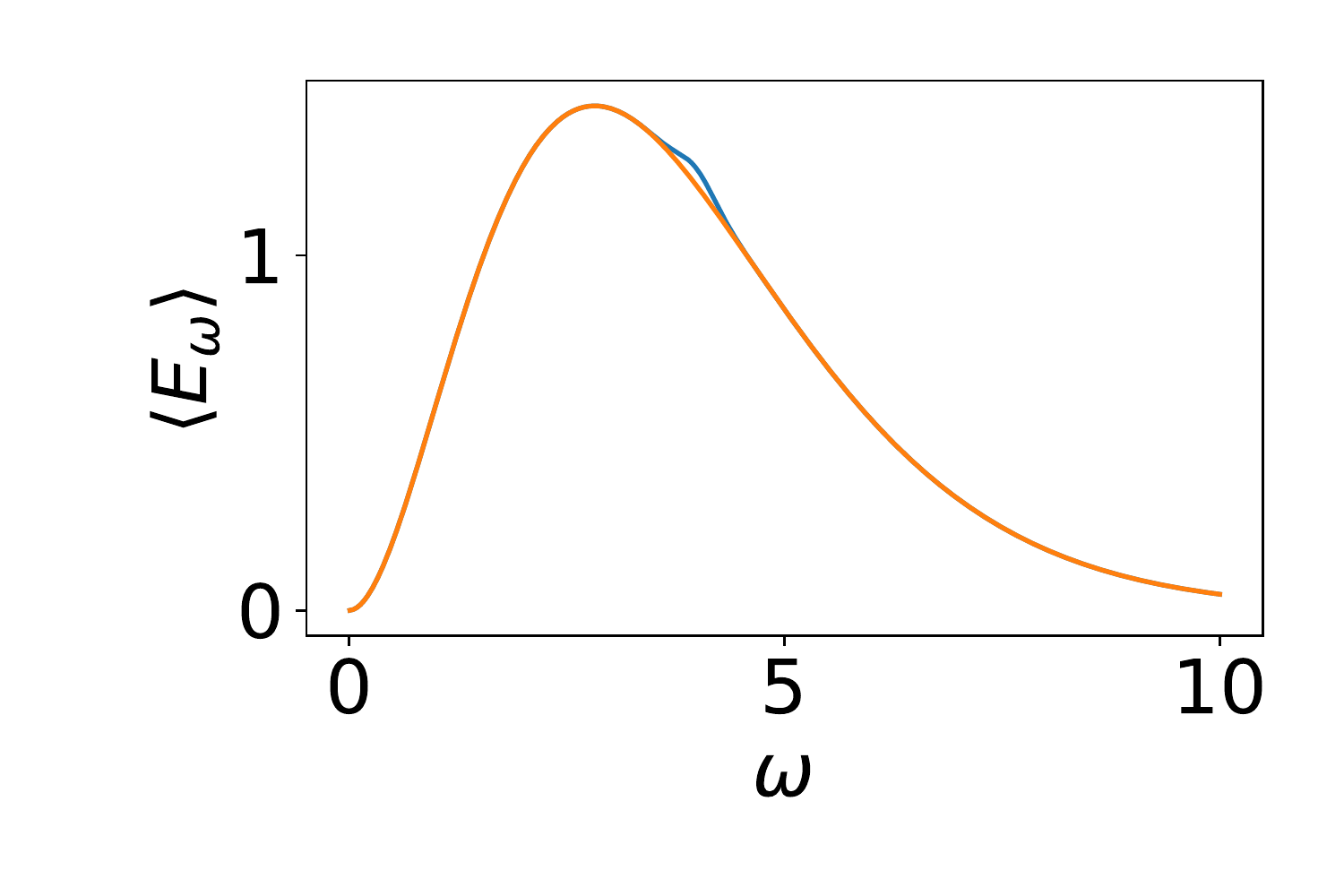}
    \includegraphics[width=0.49\textwidth]{pseudothermal_spec4_0.pdf}
    \caption{Left panels: the energy spectrum on R- (or inside the horizon), after tracing out the states on R+ (or outside the horizon). Right panels: the energy spectrum on R+ (or outside the horizon), after tracing out the states on R- (or inside the horizon). The parity is broken by the asymptotic direction of the initial Unruh mode.}
    \label{fig:spectra_dd}
\end{figure*}

\begin{tcolorbox}[breakable]
Remark: Back to the Rindler-Minkowski case. The results shown by figure \ref{fig:spectra} are also representable for the $\hat{\rho}^{bb}$ on $R+$ wedge starting from an excitation on the Minkowski vacuum. The energy-frequency spectrum is very intuitive here: The observer on R+ measures a peak on top of the thermal bath whose temperature is proportional to the acceleration; when the particle has very high energy compared to the background temperature, the observation goes back to Minkowski case; And when the acceleration is high, the low-energy particle is buried under the hot thermal bath. This serves as a decent sanity check of our result. The implication for an R- observer coming after that is rather counter-intuitive. The spectrum of $\hat{\rho}^{dd}$ calculated in the similar way as $\hat{\rho}^{bb}$ is shown in figure \ref{fig:spectra_dd}. Although at a much lower amplitude, the same frequency bump caused by $C(n+1,1)= n+1$ factor is always present in the spectrum observed by an $R-$ observer, just as implied by the equal Von-Neumann entropy of $\rho^{bb}$ and $\rho^{dd}$. Remember that the Unruh mode we start from is a purely type-I mode in this case, whose propagation asymptotically align with the trajectory of an R+ observer. At the theoretical level, it seems that the information of this oppositely traveling particle is not completely lost on the $R-$ wedge. The engraving of a piece of information on a horizon seems to be equally observable by both sides no matter which direction it is traveling. 
\end{tcolorbox}

\section{Discussions}
\label{sec:discussions}
\subsection{Assumptions and Caveats}
We would like to preemptively summarize the price for the neat implication of this work. We could not exhaust the potential problems, but we try our best to list the assumptions and caveats that come in at different steps of our derivation in this paper here.

\begin{enumerate}
    \item \textbf{The infinite dimensionality of the Hilbert space of a quantum field theory system.} By postulating the scalar field energy levels (EoM solution eigenvalues) to be discrete, we obtain an infinite but countable dimension of the Hilbert space, thus the density matrices. However, as far as we know, many of the quantum information techniques under the density matrix formalism used in this paper have only been well-established for finite-dimensional quantum systems. Since most of the pseudo-thermal density matrices are well-behaved (diagonalizable), the main concern is the normalizability of the density matrix. We verify the normalization of equation (\ref{eq:pseudothermal_kruskal}) with $\alpha_1=\alpha_2$ by explicitly calculating the series summation backstage, and speculate that the consistent physics interpretation should not be spoiled for no reason. But truth be told, strictly speaking, the normalization of pseudo-thermal density matrices $\hat{\rho}$ is speculation at this point. Even if the normalization has no problem, the direct applicability of quantum information concepts like Von-Neumann entropy and POVM on infinite dimensional Hilbert space is still an open question.
    \item \textbf{The transformation between plane wave modes and Unruh modes.} As mentioned before, it takes a transformation from $e^{-i\omega x}$ and $x^{-i\omega}$ basis to go from one way of quantization to another. The exponential relationships between the Fourier transform variable here could cause severe blue- and redshifts. When considering the discrete eigenvalues, such log-space transformation poses a question of the energy level structures that can truly span mutually complete basis modes. %\footnote{This factor is not spoiling the results obtained in collapsing shell metric though. Because as shown in section \ref{sec:blackhole}, a plane wave mode inside the shell is reflected into a Kruskal Unruh mode with finite red-shift. A Minkowski excitation of the plane wave mode might need more delicate discussion.}
    
    \item \textbf{The practical measurability of the pseudo-thermal spectrum.} The one-to-one mapping between an arbitrary Unruh mode density matrix and the pseudo-thermal density matrix vector only ensures the conceptual invertibility of the operation. Due to the high stochasticity of the pseudo-thermal ensemble, the story could be stated in another way that the information will indeed be buried under the thermal bath. Even though there is coherent excitation on the whole ensemble, it might be too weak to be detected. %In the pseudo-thermal ensemble, the information is recorded in the eigenvalue differences for example $\langle n_{\omega}-n_{\omega_{\rm ref}} \rangle$, instead of the absolute number counting of a frequency $\langle n_{\omega} \rangle$. Repetitive measurements need to be done to reduce the variance. Extensive measurements are needed when $\omega/T$ is small, hence whether the information problem is completely solved by the one-to-one mapping between the density matrices is debatable.
    
    \item \textbf{Enhanced emission from the black holes.} An obvious risk presents for the generally enhanced Hawking radiation. In extreme cases, excitation peaks could even potentially make a black hole no longer black, when $g(\omega)$ spectrum deposits its energy in high frequencies. However, there is a possibility that high-frequency emission spectra are indeed allowed by the realistic black holes depending on their collapsing history: We could have already observed them but could not distinguish them from the bright accretion disc emissions, or have attributed them to the unexplained diffuse cosmic ray excess \cite{Knodlseder:2005yq,Fermi-LAT:2015sau}. The enhanced Hawking radiation will also accelerate the evaporation, which brings in problems or opportunities depending on the masses of the primordial black holes. % What is more, in our current toy model of massive scalar the evaporation of black hole is always accelerated. On the other hand, we could also say that this theory benefits from such implications, in the sense that it could be potentially tested against our available observations in the foreseeable future. We will leave the in-depth investigation to the future, but give two qualitative argument that the Unruh modes spectrum $g(\omega)$ should favor restoring its power in lower frequency bands. 

%First, the frequency $\omega$ of the Unruh mode as boundary condition on $H^-$ comes from a plane wave mode living inside the shell. The in-shell coordinate system is essentially a comoving frame of the coherently infalling modes, so in such comoving frame both the time and space oscillation of the field should not be very drastic. Secondly, the redshift factor $\xi = \frac{1+\nu}{\nu} \sqrt{R_0-2M}e^{-\frac{R_0}{4M}}$ is exponentially suppressed by the duration of the collapsing procedure -- the higher the compression ratio $R_0/r_s$ is, the stronger these Unruh modes are redshifted. 

%Whether these two factors or other mechanisms could suppress the Unruh modes frequencies to low enough level comparing to the Hawking radiation temperature is yet to be investigated. There is also possibility that certain level of high frequency emission lines, as long as they are not too strong that they could cause fast decay rate of the black holes, are indeed allowed by the realistic black holes depending on their collapsing history, and we have already observed them but could not distinguish them from the bright accretion disc emissions, or have attributed them to the unexplained galactic center excess lines \cite{Knodlseder:2005yq,Fermi-LAT:2015sau}. 
\end{enumerate}

\subsection{Generalized Unruh Effect on Other Particle Horizons}
%In sections \ref{sec:introduction} to \ref{sec:beyondsingle}, this paper derived the generalized Unruh effect for arbitrary excited state in the Fock space of a massive scalar field. In section \ref{sec:blackhole}, we applied the results obtained from the generalized Unruh effect globally to a mass shell collapsing into a black hole. The `global' here is defined in contrast with `local', and the `globalness' throughout section \ref{sec:blackhole} is reflected in the fact that we charted the spacetime manifold in an unbounded region, $0<r<\infty$. We get a conclusion that the Hawking radiation of the evaporating black hole encodes the information of the collapsing history in its spectrum features deviating from black body radiation spectrum.

The general Unurh effect technique is in principle applicable to any particle horizon, regardless of the cause of it being the acceleration, massive objects, or inflation. %Beyond our application of the generalized Unruh effect on the black hole system, we believe that the Bogoliubov transformation, the premise of our derivation, is a powerful tool for us to further the understanding of QFT in curved spacetime. It could even lead to the unification of general relativity and QFT through a different and/or complementary path than the most popular attempts in this direction, the quantization of the gravity. 

The Bogoliubov transformation approach has been explored for particle production during inflation for a long time. A recent paper by K. Kaneta, S. M. Lee, and K. Oda \cite{kaneta2022boltzmann} have carried out an instance of comparing Boltzmann and Bogoliubov ways of dark matter production during inflation, and they reached a great agreement. It is an encouraging work unveiling the alternativeness of the gravitational thermalization picture and the direct quantization of the graviton picture. One major merit of the Bogoliubov over the Boltzmann approach though, is that it can make analytical predictions at very low-$k$ regimes, which is not something that can be done with the Boltzmann approach by carrying out conventional particle theory calculation at the perturbative level for the quantized gravitons. The reason is that at the low-$k$ regime, the scalar field stops being localized enough and loses its particle property. It is in analogy to the stretch of the waveform near the asymptotic Schwarzschild black hole in our work. 

Despite the advantage at the low-$k$ regime, the Bogoliubov approach has an obvious disadvantage in the traditional vacuum initial state treatment. The method to incorporate other non-gravitational particle productions has been unknown in the Bogoliubov approach. The generalized Unruh effect working with the excitation at the horizon as initial conditions could fill this gap and give an inclusive description of different particle production mechanisms through a differentiated Bogoliubov approach. %In practice, it might be computationally much less economical than the Boltzmann approach, but it could be the only option for those large-scale modes beyond the Compton wavelength of a particle.

%\subsection{Implication of the Generalized Unruh Effect on the Thermodynamic Aspects of the Gravity}
%The application of generalized Unruh effect to a macroscopic system like black hole, as shown in this paper, has the potential to solve the information paradox. However, the power of our Generalized Unruh effect results from the first half of this paper, section \ref{sec:introduction} to \ref{sec:beyondsingle} is far beyond the application to this single problem. The trigger of the generalized Unruh effect, mixture of positive and negative frequency on a horizon, could happen locally. 

Asymptotic horizons are always formed when an observer accelerates with respect to the surrounding system. %Those horizons are only conceptually defined to facilitate the investigation of a realistic physical system, and work by approximating a field as boundary condition extrapolated to distant regime along time axis., we mean the effective event horizon that could be extrapolated from the instantaneous and local spacetime curvature the observer feels. 
We could break down a smooth geodesic into a series of short hyperbolas, then use the Minkowski $\leftrightarrow$ Rindler generalized Unruh effect to investigate how a bunch of information restored in local field operators $\phi(O)$ in an element volume $O$ is passed from upstream to downstream of a bundle of geodesics. Of course, there is arbitrariness in the definition of upstream/downstream, i.e. the direction of the time flow. Suppose at a certain world point $P_i$ the geodesic is represented by the local hyperbola segment with acceleration $\alpha_i$ between two successive moments. %. The acceleration is with respect to the local Minkowski metric at $P_i$ (basically, the local Minkowski and local Rindler coordinates correspond to the first order and second order time derivative expansions of the geodesics). At quantum level, what is happening is that 
The local field operators $\phi(P_i)$ passed its information to the local field operators at the next moment $\phi(P_{i+1})$, with a slight thermalization due to $\alpha_i$, according to generalized Unruh effect.

We remember that there is arbitrariness in the definition of the time arrow in the first place. However, going in any arbitrary direction of time flow always add a thermal ensemble to the initial state, thus presumably increasing the entropy. We do not know which one leads to the other, but from the argument above it seems that there is a connection between the second law of thermodynamics and the simple law of causality, which forbids looping geodesics. %If we interpret $dt \equiv dS$, a looping time-like geodesic suggests a source of entropy in the system circled by the geodesic loop.

\subsection{Emergent Gravity from the Stochasticity of Quantum Fields}

The ubiquitousness of the local particle horizon and thus the local Unruh effects as described in the previous subsection intrigues us with this conceptual speculation:
%This one-way increasing thermalization is reflected in the fact that we cannot turn back after choosing a certain time direction of a bundle of geodesics. It should be clarified that `no turning back' is not an artificial choice, it is due to the existence of the local Rindler horizon generated at each moment on the geodesic -- Because of the existence of horizon (that bound to be created even momentarily when there is acceleration), the observer can only choose to be on one side of the Rindler wedge, and tracing out the other due to (again, even momentary) the inaccessibility from causality. Whichever the wedge the observer happened to stay on \textbf{defines} the continued, fixed direction of the time flow. 

%In summary, the story above could be regarded as a relativistic QFT explanation of the second law of thermodynamics: \textbf{The growth of entropy along the time arrow is caused by gravitational thermalization. It roots in the inaccessibility of spacetime vincinities around a geodesic, and it is a result of causality strictly obeyed at each moment.}

%By far the readers can clearly see where this discussion is leading to. We are proposing to reconsider the necessity of quantizing the gravity to serve for the purpose of uniting general relativity and quantum field theory. We are suggesting that the thermodynamics interpretation of gravity for quantum fields living on 4D Lorentzian, or even 3D Euclidean manifold could be a faster routine to this grand purpose. Specifically, we would like to postulate one more equivalence after Einstein's:
\begin{widetext}
\begin{equation*}
    {\rm Gravity} \equiv {\rm Spacetime\ curvature} \equiv {\rm Temperature\ of\ the\ quantum\ fields}
\end{equation*}
\end{widetext}

Each of these equivalences is supposedly revertible, and those quantities are completely local. By this proposed relationship we could derive the former from the latter or the other way around. Namely, it falls into the emergent gravity category of ideas. To the best of our knowledge, the thermodynamics interpretation of general relativity was first discussed in \cite{Jacobson:1995ab}, proposing the connections like $T \leftrightarrow \kappa$, $dt \leftrightarrow dS$ that have been repetitively appearing throughout our paper. People have been reluctant to take it further than an interpretation since then. We think the way to go a step forward along their path could be to come up with a formal way to replace Ricci curvature in the Einstein-Hilbert action with a heat reservoir of the fields. By solving for the temperature of the fields from this refined action with a heat reservoir, if feasible, we can accordingly solve the metric of the spacetime. %In this way we may surpass the biggest problem of QFT in curved spacetime calculation so far, that we always need to presume a metric to start with. 

We are going to briefly comment on these thought experiments that are dedicated to proving the quantum nature of gravity as a closing remark to the far-fetched discussion in this section. To the best of our knowledge, most, if not all, of those quantum gravity experiments' setups, are indistinguishable between a quantized gravitational field (metric) and the local application of the Unruh effect, namely the gravity emergent from the stochasticity of quantum fields -- they both render the gravitational phenomena quantum natures. Usually, such experiments intend to prove the quantum nature of gravity by measuring the entanglement introduced or mediated by gravitational interaction. For example, the Gedankenexperiment, which is essentially a gravitational version of the Stern-Gerlach experiment \cite{Belenchia:2018szb} and Marletto \& Vedral's proposal of double mass interferometer experiment \cite{Marletto:2017kzi}. What lies as the fundamental to these experiments is usually a perturbed gravitational field whose propagation is semi-classically describable, and the experiments are designed to show that such perturbed gravitational fields serve as the quantum entanglement messenger between two already well-established quantum systems. However, in a stochastic quantum field description of gravity, we do not need to introduce this additional quantum of graviton to mediate the entanglement -- the ubiquitous quantum field of the two presumed quantum systems spread out in spacetime is already a mediator of the entanglement. The entanglement is passed on by local Unruh effects. In principle, no experiment can distinguish between the picture of `The graviton $h_{\mu \nu}$ mediated the entanglement between $Q_1$ and $Q_2$' and `$Q_1$ interact with $Q_i$ distributed ubiquitously in the spacetime spanned until $Q_2$, thus the two eventually brought into entanglement'. They are just two alternative ways of telling the same story. 

For the reasons presented in this section, we figure that emergent gravity from the stochasticity of the quantum fields might be a competitive way toward the incorporation of gravity into the quantum picture. This approach, as illustrated in the calculation in this paper, has the potential to produce {\bf calculable and testable} predictions.

\section{Conclusions}
\label{sec:conclusions}
%In sections \ref{sec:introduction} - \ref{sec:beyondsingle}, this work use concrete commutator algebra done under QFT in curved spacetime framework to generalize the Unruh effect from vacuum state to an arbitrary density matrix of the states in a Minkowski vacuum excited Fock space. In section \ref{sec:spectrum}, we calculate the energy spectrum on a spacetime partitioned by event horizon given an initial condition on the fully accessible spacetime before partition. We apply the results on Kruskal/Schwarzschild black hole metric in section \ref{sec:blackhole}, and suggested that the generalized Unruh effect could be a potential solution to the black hole information paradox. Section \ref{sec:discussions} argue that our findings in this work further suggest the stochastic quantum field theory as gravity serve as a promising path towards the unification of gravity and quantum field theory.

We practice a straightforward idea, that generalizes the Unruh effect applied usually on the vacuum state to arbitrary excited states in this paper. The result shows that the positive frequency excitation at the horizon induces a coherent excitation on each of the configurations of a canonical ensemble measured by an accelerating observer, thus illustrating itself as a featured peak on top of the featureless black body radiation spectrum. We call such an ensemble pseudo-thermal, which in principle (in terms of density matrix representation) is distinguishable from the linear combination of a canonical ensemble and a pure excitation state. 

We apply our generalized Unruh effect result on a system with massive real scalar field in the collapsing shell metric, asymptotically a Schwarzschild black hole. We find that the particles inside the shell could be represented by the boundary condition of the Kruskal positive frequency modes living on the outgoing horizon, thus we could retrieve that information, at least partially, through the strong entanglement across the horizon. The Hawking radiation has a featured enhancement based on the collapsing history and the initial excitation inside the shell of the black hole. 

There are numerous open questions extending both bottom-ward and up-ward from this work, as discussed in section \ref{sec:discussions}. Both theoretically and observationally, the generalized Unruh effect implies intriguing conclusions and speculations to be confronted in the future.

\section*{Acknowledgements}
This work was supported by World Premier International Research Center Initiative (WPI), MEXT, Japan.

This paper has come into being through iterative discussions with multiple persons: Misao Sasaki, Elisa Ferreira, Naritaka Oshida, Masahito Yamazaki, Valeri Vardanyan, Vicharit Yingcharoenrat, Sam Pasaglia, Kaloian Lozanov and many other cosmology group members at IPMU. They provided valuable comments on and asked difficult questions about my idea. I tried my best to address most of them in the text or the remark blocks of this paper. I greatly appreciate the support provided by my Ph.D. adviser Dragan Huterer, Eiichiro Komatsu, Masahiro Takada, my friends, and my family. I would like to acknowledge the great work done by William. G. Unruh, and the information provided by him that this work very likely has never been done yet in our brief exchange. 
%Let us stress again that the Hilbert spaces we are working in through out this work are the states of a quantum field, quantized with respect to the modes as equation of motion solutions under classical general relativity and classical fields. As long as the Lagrangian remains the same, under the principle of relativity, those modes are just two representations of the same Hilbert space. The vacuum in one coordinate is not vacuum for the other, and this vacuum state is like a slice cutting the full Hilbert space into the part accessible by the positive frequencies in another basis and the part that is not accessible by the positive frequencies. 

\bibliographystyle{ieeetr}
\bibliography{generalized_unruh_effect}
\end{document}